\documentclass{aa}
\usepackage{txfonts}
\usepackage{graphicx}	
\usepackage{amsmath}	
\usepackage{amssymb}	
\usepackage{caption,subcaption}
\usepackage{textgreek}
\usepackage{verbatim}
\usepackage{gensymb}
\usepackage{xcolor}
\usepackage{placeins}

\usepackage{xcolor}
\usepackage{ulem}
\usepackage[export]{adjustbox}

\usepackage[colorlinks=true, citecolor=blue, linkcolor=blue]{hyperref}
\begin{document} 

   \title{The MeerKAT Absorption Line Survey: Homogeneous continuum catalogues towards a measurement of the cosmic radio dipole}

   \author{J. D. Wagenveld\inst{1}
          \and H.-R.~Kl\"{o}ckner\inst{1}
          \and N.~Gupta\inst{2}
          \and P.~P.~Deka\inst{2}
          \and P.~Jagannathan\inst{3}
          \and S.~Sekhar\inst{4,3}
          \and S.~A.~Balashev\inst{5,6}
          \and E.~Boettcher\inst{7,8,9}
          \and F.~Combes\inst{10}
          \and K.~L.~Emig\inst{11}
          \and M.~Hilton\inst{12,13}
          \and G.~I.~G.~J\'{o}zsa\inst{1,14}
          \and P.~Kamphuis\inst{15}
          \and D.~Y.~Klutse\inst{12}
          \and K.~Knowles\inst{14}
          \and J.-K.~Krogager\inst{15}
          \and A.~Mohapatra\inst{2}
          \and E.~Momjian\inst{3}
          \and K.~Moodley\inst{12}
          \and S.~Muller\inst{16}
          \and P.~Petitjean\inst{17}
          \and P.~Salas\inst{18}
          \and S.~Sikhosana\inst{12}
          \and R.~Srianand\inst{2}
          }

   \institute{Max-Planck Institut f\"{u}r Radioastronomie, 
              Auf dem H\"{u}gel 69, 
              53121 Bonn, Germany
        \and Inter-University Centre for Astronomy and Astrophysics,
            Post Bag 4, Ganeshkhind, 
            Pune 411 007, India
        \and National Radio Astronomy Observatory, 
             Socorro, NM 87801, USA
        \and The Inter-University Institute for Data Intensive Astronomy,         Department of Astronomy, and University of Cape Town,
             Private Bag X3, Rondebosch, 7701, South Africa
        \and Ioffe Institute, 26 Politeknicheskaya st., St. Petersburg, 194021, Russia
        \and HSE University, Saint Petersburg, Russia
        \and Department of Astronomy, University of Maryland, College park, MD 20742, USA
        \and X-ray Astrophysics Laboratory, NASA/GSFC, Greenbelt, MD 20771, USA
        \and Center for Research and Exploration in Space Science and Technology,
NASA/GSFC, Greenbelt, MD 20771, USA
        \and Observatoire de Paris, Coll\`{e}ge de France, 
             PSL University, Sorbonne University, 
             CNRS, LERMA, Paris, France
        \and Jansky Fellow of the National Radio Astronomy Observatory, 520 Edgemont Road, Charlottesville, VA 22903, USA
        \and Astrophysics Research Centre and School of Mathematics,              Statistics and Computer Science, University of KwaZulu-Natal, Durban 4041, South Africa
        \and Wits Centre for Astrophysics, School of Physics, University of the           Witwatersrand, 2050, Johannesburg, South Africa
        \and Department of Physics and Electronics, Rhodes University, 
            P.O. Box 94 Makhanda 6140, South Africa
        \and Ruhr University Bochum, Faculty of Physics and Astronomy,            Astronomical Institute, 44780 Bochum, Germany
        \and Department of Space, Earth and Environment, 
            Chalmers University of Technology, 43992 Onsala, Sweden
        \and Institut d’astrophysique de Paris, UMR 7095,
             CNRS-SU, 98bis bd Arago, 75014 Paris, France
        \and Green Bank Observatory, Green Bank, 
             WV 24944, USA
            }


 
    \abstract{The number counts of homogeneous samples of radio sources are a tried and true method of probing the large scale structure of the Universe, as most radio sources outside the galactic plane are at cosmological distances. As such they are expected to trace the cosmic radio dipole, an anisotropy analogous to the dipole seen in the cosmic microwave background (CMB). Results have shown that although the cosmic radio dipole matches the direction of the CMB dipole, it has a significantly larger amplitude. This unexplained result challenges our assumption of the Universe being isotropic, which can have large repercussions for the current cosmological paradigm. Though significant measurements have been made, sensitivity to the radio dipole is generally hampered by systematic effects that can cause large biases in the measurement. Here we assess these systematics with data from the MeerKAT Absorption Line Survey (MALS), a blind search for absorption lines with pointings centred on bright radio sources. With the sensitivity and field of view of MeerKAT, thousands of sources are observed in each pointing¸ allowing for the possibility of measuring the cosmic radio dipole given enough pointings. We present the analysis of ten MALS pointings, focusing on systematic effects that could lead to an inhomogeneous catalogue. We describe the calibration and creation of full band continuum images and catalogues, producing a combined catalogue containing 16,313 sources and covering 37.5 square degrees of sky down to a sensitivity of 10~\textmu Jy/beam. We measure the completeness, purity, and flux recovery statistics for these catalogues using simulated data. We investigate different source populations in the catalogues by looking at flux densities and spectral indices, and how they might influence source counts. Using the noise characteristics of the pointings, we find global measures that can be used to correct for the incompleteness of the catalogue, producing corrected number counts down to 100~-~200~\textmu Jy. We show that we can homogenise the catalogues and properly account for systematic effects. We determine that we can measure the dipole to $3\sigma$ significance with 100 MALS pointings.}


\titlerunning{MALS: Homogeneous continuum catalogues towards a measurement of the cosmic radio dipole}
\maketitle
%

\section{Introduction}

The vast majority of sources seen at radio wavelengths outside of the galactic plane are known to be at cosmologically significant distances \citep[$<z>\sim0.8$, e.g.][]{Longair1966,Condon2016}. This makes homogeneous samples of radio sources ideal to study the local luminosity function, along with the large scale structure and evolution of the universe \citep{Longair1966}. The number counts of radio sources were used as evidence against a static Euclidean universe \citep{Ryle1955}, providing a strong argument in favour of a strongly evolving universe even before the discovery of the Cosmic Microwave Background \citep[CMB,][]{Penzias1965}. As radio sources trace the large scale structure of the universe, they are expected to abide by the cosmological principle, which asserts that the universe is homogeneous and isotropic. However, there is an anisotropy expected in the number counts of radio sources, caused by the velocity of the Solar system with respect to the cosmological background. This expresses itself as a dipole, and is the dominant anisotropy observed in the CMB \citep{Lineweaver1997}. The movement of the observer induces Doppler boosting and relativistic aberration that cause the apparent luminosity and position of radio sources to shift, resulting in a dipole in the number counts of radio sources. A measurement for the cosmic radio dipole was first proposed by \citet{Ellis1984} who showed that $2\times10^5$ sources, adequately distributed along the axis of the dipole, are required for a $3\sigma$ measurement of the radio dipole, assuming the Solar system velocity derived from CMB measurements.

Using data from the National Radio Astronomy Observatory (NRAO) VLA Sky Survey \citep[NVSS,][]{Condon1998}, \citet{Blake2002} made the first significant measurement of the dipole, with a direction and amplitude that corresponds to that of the CMB. Subsequent studies were performed with the NVSS and other radio surveys, such as the Westerbork Northern Sky Survey \citep[WENSS,][]{Rengelink1997}, the Sydney University Molonglo Sky Survey \citep[SUMSS,][]{Mauch2003}, and the Tata Institute for Fundamental Research (TIFR) Giant Metrewave Radio Telescope (GMRT) Sky Surveys first alternative data release \citep[TGSS ADR1,][]{Intema2017}. It was found that the cosmic radio dipole, while being consistent in direction, significantly differs from that of the CMB in amplitude \citep[e.g.][]{Singal2011,Rubart2013,Tiwari2013,Tiwari2015,Colin2017}. These early dipole measurements found that survey-wide systematic effects, which cause varying source densities, can greatly bias dipole estimates. This is usually remedied by strict cuts in flux density, dramatically decreasing the number of usable sources. Even with these flux density cuts some surveys such as TGSS yield anomalous dipole results which have been attributed to systematics due to problems with flux calibration \citep[e.g.][]{Singal2019,Bengaly2018,Siewert2021}. While results differ depending on survey and estimator used, the amplitude of the radio dipole is consistently larger (by a factor of 2-6) than the amplitude of the CMB dipole \citep[see][for an overview]{Siewert2021}, while the direction of the dipole remained consistent. 
With similar results found using the number counts of Active Galactic Nuclei (AGN) at infrared wavelengths \citep{Secrest2021,Singal2021,Secrest2022}, it becomes increasingly difficult to explain them with systematic effects or faulty analysis. Only in a recent analysis by \citet{Darling2022} a dipole consistent with the CMB in both direction and amplitude was found by combining the VLA Sky Survey \citep[VLASS,][]{Lacy2020} and the Rapid Australian Square Kilometre Array Pathfinder (ASKAP) Continuum Survey \citep[RACS,][]{McConnell2020}, though it presents only one counterpoint to the many works finding an increased dipole amplitude. Considering a purely kinematic origin of the dipole, the cosmic radio dipole and CMB dipole are in obvious tension with each other. The excess dipole found in the radio therefore must be a result of a different process, which could have large implications for cosmology. As radio galaxies trace the underlying matter distribution, a dipole in their distribution would break with isotropy, one of the fundamental assumptions of cosmology. The assumption of isotropy and homogeneity is founded on the notion that we as observers do not occupy a special place in the Universe, these results suggest that there is some flaw in this assessment. 

Working towards an independent measurement of the radio dipole, we will utilise the MeerKAT Absorption Line Survey \citep[MALS,][]{Gupta2016}, a deep radio survey with pointings centred on bright radio sources. MALS is carrying out a dust-unbiased search for neutral hydrogen (H{\sc i}, 21 cm) and hydroxyl (OH, 18 cm) absorption lines at redshifts $0< z <2$ in order to unravel the processes driving the steep evolution of star formation rate density. As a blind search for absorption lines, every MALS pointing is centred on a bright ($>$200 mJy at 1.4 GHz) AGN. The targets have been chosen from the NVSS and SUMSS catalogues, and are cross-checked with Wide-field Infrared Survey Explorer (WISE) data in order to build a dust-unbiased sample of AGN \citep{Gupta2022}. Early results show that MALS is able to attain unprecedented sensitivity to absorption lines in these bright AGN \citep{Gupta2021,Combes2021}. Besides the search for absorption lines, the data taken will be sensitive enough to produce deep continuum images, down to 10~\textmu Jy/beam. With a full width at half maximum (FWHM) field of view of a degree at L-band (1.27 GHz), each MeerKAT pointing presents a few square degrees and potentially thousands of sources. With 391 pointings currently observed in L-band, the full survey will provide thousands of square degrees of deep continuum sky and hundreds of thousands of sources.

Though the expected MALS number counts are sufficient for a dipole measurement, a dipole estimate requires a homogeneous catalogue. Systematic effects influencing the sensitivity of surveys are common, and usually dealt with by making conservative cuts in the data to avoid biasing the dipole estimate. Instead, in this work we present a thorough analysis of ten MALS pointings, aiming to fully understand the systematics present in the survey data. This will allow us to account for these systematics when measuring the radio dipole using hundreds of MALS pointings. The nature of the survey provides additional challenges for this type of measurement. Previously, measurements of the dipole have been performed with contiguous surveys such as the NVSS, whereas MALS will be more sparse, sampling the sky in many different directions. However, compared to these surveys MeerKAT has a much higher sensitivity (10~\textmu Jy/beam), which allows us to probe deeper into the population of faint radio sources. Furthermore, past dipole measurements from contiguous sky surveys have been performed post factum, with little knowledge of the internal processing and therefore present systematics of these surveys beyond what is described in the literature. In this paper, we will study the first ten continuum images of MALS in-depth, in order to assess their quality. We investigate the systematics in calibration, imaging, and source finding on image quality and source counts, and extrapolate our findings to the rest of the survey.

This paper is organised as follows. In Section \ref{sec:data} we describe the MALS data. Initial creation of source catalogues and completeness measures are described in Section \ref{sec:sourcefinding}. In Section \ref{sec:final_catalog} we describe results from the full catalogue of sources. We investigate how different source populations affect the catalogues in Section \ref{sec:science}. In Section \ref{sec:dipole} we assess the prospects for a dipole measurement with MALS using the results in this paper. Finally, in Section \ref{sec:conclusion}, we summarise the findings of this paper.

\section{Data - MALS}
\label{sec:data}

\begin{figure}
    \centering
    \includegraphics[width=\hsize]{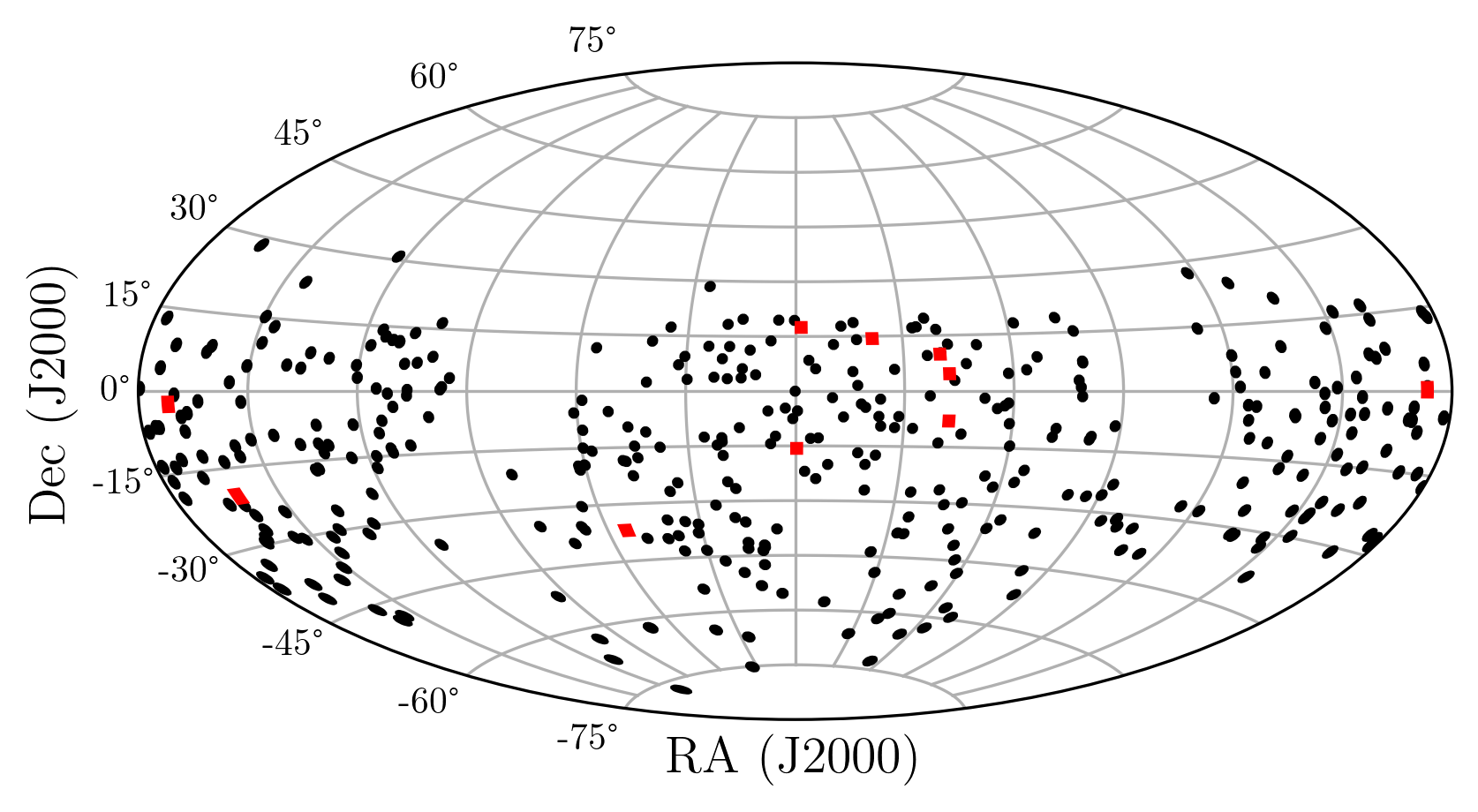}
    \caption{Sky distribution of the first 391 observed pointings of MALS. The galactic plane is largely avoided, and since 89\% of the pointings are selected directly from NVSS, the vast majority of pointings are above declination of -40 degrees. The pointings used in this analysis are highlighted in red.}
    \label{fig:mals_distribution}
\end{figure}

\begin{table*}[t]
    \centering
    \caption{Calibration details of the pointings presented in this paper, grouped by observation runs. The flux densities of the calibrators are reported by the CASA \texttt{fluxscale} task, which changes reference frequency based on whether the source is a flux calibrator or gain calibrator.}
    \resizebox{\textwidth}{!}{%
    \begin{tabular}{c c || c | c c c c c c}
    Flux cal & Flux density  & Target  & Gain cal & Flux density & Reference flux$^\mathrm{a}$ & Spectral index$^\mathrm{a}$ & Distance from target \\
             & 870 MHz &     &         & 1365 MHz & 1400 MHz  & &     \\
             & (Jy)    &     &         & (Jy)     & (Jy)      & & (degrees) \\
    \hline \hline
    J1331+3030 & 18.829 & J2023-3655 & J2052-3640 & $1.45\pm0.005$ & $1.367\pm0.017$ & $-1.258\pm0.02$ & 5.8 \\
    J1939-6342 & 14.095 &            & & \\ \hline 
    J0408-6545 & 27.027 & J0126+1420 & J0108+0134 & $3.24\pm0.01$ & $3.113\pm0.070$ & $-0.273\pm0.02$ & 13.5\\
    J1939-6342 & 14.095 &            && \\ \hline
    J1331+3030 & 18.829 & J1133+0015 & J1150-0023 & $2.86\pm0.005$ & 2.9$^\mathrm{b}$ & -- & 4.2\\
    J1939-6342 & 14.095 & J1232-0224 & J1256-0547 & $10.66\pm0.05$ & $9.82\pm0.120$ & $-0.490\pm0.05$ & 6.9 \\
    &                   & J1312-2026 & J1311-2216 & $5.5\pm0.01$ & $4.857\pm0.060$ & $-1.281\pm0.04$ & 1.8 \\ \hline
    J0408-6545 & 27.027 & J0001-1540 & J2357-1125 & $2.12\pm0.009$ & 1.8$^\mathrm{b}$ & -- & 4.4 \\
    J1939-6342 & 14.095 & J0006+1728 & J2253+1608 & $15.5\pm0.02$ & $16.199\pm0.198$ & $-0.193\pm0.03$ & 17.5 \\ \hline
    J0408-6545 & 27.027 & J0240+0957 & J0238+1636 & $0.61\pm0.002$ & $0.528\pm0.014$ & $-0.246\pm0.05$ & 6.7 \\
    J1939-6342 & 14.095 & J0249-0759 & J0240-2309 & $6.15\pm0.003$ & $5.938\pm0.131$ & $-0.154\pm0.03$ & 31.2 \\
    &                   & J0249+0440 & J0323+0534 & $2.85\pm0.002$ & $2.766\pm0.062$ & $-0.920\pm0.01$ & 8.5\\ \hline
    \end{tabular}}
    \tablefoot{$^\mathrm{a}$ \url{https://skaafrica.atlassian.net/wiki/spaces/ESDKB/pages/1452146701/L-band+gain+calibrators} for properties of MeerKAT L-band calibrators. $^\mathrm{b}$ Value from the old list of calibrators, no longer publicly available.}
    \label{tab:calibration_table}
\end{table*}

The distribution of the first 391 observed pointings of MALS is shown in Figure \ref{fig:mals_distribution}. In order to assess the data quality of the individual MALS pointings and the impact for the dipole estimates an initial set of ten pointings, shown in Figure~\ref{fig:mals_distribution} in red, has been selected out of five observing runs to probe different ranges of right ascension, declination, and central source flux density.

\subsection{Observations and Calibration}
\label{sec:obs_cal}

The general setup of a single MALS observation includes observations of three science targets and corresponding calibrators. The observation is scheduled with a flux calibrator observed for 10 minutes at the start and end of each observing run. Each target is observed for 20 minutes at a time, cycling through all targets three times for a total observing time of an hour per source. Before and after each target observation, a nearby gain calibrator is observed for one minute. Cycling between targets like this maximises the UV-coverage with minimal increase in overhead. Observing multiple targets in a single run is not only convenient in terms of processing, but is also critical in taking stock of systematic effects, such as flux density scale or phase errors, potentially introduced during observation or calibration. All observations have a correlator integration time of 8~s, with observations carried out in 32K mode, providing 32,768~channels with a channel width of 26.123~kHz. With a frequency range of 856 -- 1712~MHz, the total bandwidth is 856~MHz, with a central frequency of 1.285~GHz. 

The MeerKAT data are shipped to IUCAA and processed by the Automated Radio Telescope Imaging Pipeline (\textsc{ARTIP}). The complete deployment of \textsc{ARTIP} in MALS is described in \citet{Gupta2021}. \textsc{ARTIP} presents an environment where data can be processed according to user specifications and is based on the Common Astronomy Software Applications (\textsc{CASA}) tasks \citep{CASA2022}. Each dataset undergoes a round of basic flagging, removing known radio frequency interference (RFI) frequencies. This is followed by flux calibration, bandpass calibration, and gain calibration, each step having the possibility of additional automated flagging. The final target visibilities used for the imaging process are produced by applying the flags and calibration solutions.

As part of the overall evaluation of the individual pointings, all the available information is assessed automatically with an evaluation scheme that has been developed to trace errors of the calibration process by searching through the logging information of \textsc{ARTIP}. This scheme also extracts relevant information from the logs, such as the flux densities of the calibrator sources. An overview of the targets and calibrators of the ten selected pointings, organised by observation block, is shown in Table~\ref{tab:calibration_table}. For the gain calibrators, both the flux density determined during calibration and from a reference catalogue is listed.

\subsection{Self-calibration and continuum imaging}
\label{sec:cont_imaging}

For the purposes of continuum imaging, the data are averaged over 32 channels and divided into 16 spectral windows (SPWs), resulting in 64 channels per SPW. Once again, frequencies with known strong RFI are flagged \citep[see also Fig. 2 of][]{Gupta2021}. The resulting dataset has a total of 960 channels, a bandwidth of 802.5 MHz (869.3 --  1671.8 MHz), and a central frequency of 1.27 GHz as a result of the edges of the band being flagged.

As each field contains a strong point source at its centre, both phase and amplitude self-calibration can be performed \citep{Cornwell1989}. In total, three phase and one amplitude calibration steps are performed, with imaging each step to improve the local sky model (LSM). As is common with self-calibration in \textsc{CASA}, we use the clean components created in \verb|tclean| as the LSM for calibration. We iterate on the model by creating masks for \verb|tclean| using the Python Blob Detection and Source Finder \citep[\textsc{PyBDSF},][]{Mohan2015}. Starting with a mask containing only the central source, after a set number of iterations \textsc{PyBDSF} is used on the image to create the mask for the full field, initially with a high S/N threshold and lowering the threshold for subsequent runs to gradually expand the model. Creating the clean masks in such a way ensures that cleaning is mostly limited to real emission, while also speeding up the imaging by limiting the cleaning area. 

Though the self-calibration can be a significant improvement on the image it can also be potentially unstable. To monitor the stability of solutions, a diagnostic tool for self calibration produces a report on the variation of relevant statistics such as noise and central source flux density in different steps of calibration. As with calibration, the logs are evaluated for errors and warnings during the self-calibration process and information relevant to assessing the quality of the products, such as percentage of flagged data and theoretical noise limit, is extracted. 

Imaging is performed using \textit{Multi-term Multi-Frequency Synthesis} \citep[MTMFS,][]{Rau2011} deconvolution with four pixel scales (0, 2, 3, and 5 pixels) to model extended emission in the images and two Taylor terms to account for the spectral shape of the sources. This produces two Taylor term images, which describe the spectral shape of the emission to zeroth and first order, respectively. As such, the zeroth order Taylor term $I_0$ represents the continuum flux density of the field at the reference frequency of 1.27 GHz, while the first order Taylor term $I_1$ describes the spectral index,
\begin{equation}
    I_0 = I_{\nu_0}^{sky} ; I_1 = \alpha I_{\nu_0}^{sky}.
    \label{eq:taylor_terms}
\end{equation}
To maintain a balance between sensitivity and resolution in the images, visibilities are weighted using Briggs weighting \citep{Briggs1995} with robust value of 0. Because we are imaging with a large field of view, we use W-projection \citep{Cornwell2005} with 128 projection planes to correct for the fact that our baselines are non-coplanar. The final data products consist of the restored, model, residual, sum-of-weights, and point spread function (PSF) images for both Taylor terms. Furthermore, spectral index, spectral index error, and mask images are also produced. The continuum images have a pixel size of 2\arcsec and a size of 6000~x~6000~pixels. This results in a square image of 3.3~degrees on a side. Though individual pointings have different beams, as detailed in Table~\ref{tab:pointing_table}, they are on average aligned in the north-south direction, with a mean major axis of 9.3\arcsec and mean minor axis of 6.5\arcsec.

\subsection{Spectral index images}
\label{sec:alpha_images}

The L-band of MeerKAT has a bandwidth of 802.5~MHz, which is large enough to be sensitive to the spectral shape of the radio emission within the band. If this is not taken into account when imaging the full band, this incurs a large uncertainty in flux density. The general solution for this is MTMFS deconvolution, which models the frequency dependence of the emission with a Taylor expansion. In our case, as mentioned in Section~\ref{sec:cont_imaging}, we model the frequency dependence of the emission in the pointings to first order in $\nu$. With this we can create maps describing the spectral index $\alpha$, defined by the relation between flux density $S$ and frequency $S\propto\nu^{\alpha}$, of the emission in the image.

Although MTMFS imaging also produces a spectral index image, pixels below 5 times the peak residual are masked in this image. To retain flexibility, we therefore choose to produce the spectral index images from the Taylor term images ourselves. From the definition of the Taylor term images in Equation \ref{eq:taylor_terms}, a spectral index image can be obtained using $\alpha = I_1/I_0$, from which we will be able to measure the spectral indices of sources. To keep values in the spectral index image from diverging, pixels are masked where values in the Stokes I image are below 10 \textmu Jy $\mathrm{beam}^{-1}$. When measuring the spectral index in some region of the image, usually defined by the extent of a source, we assign a spectral index as the intensity weighted mean of the measured pixels in the spectral index image, with intensity weighted standard deviation as the error, 
\begin{align}
    \overline{\alpha} &= \frac{\sum_{i=1}^n I_{0,i} \alpha_i}{\sum_{i=1}^n I_{0,i}}, \\
    \sigma_{\alpha} &= \sqrt{\frac{\sum_{i=1}^n I_{0,i} (\alpha_i - \overline{\alpha})^2}{\frac{n-1}{n} \sum_{i=1}^n I_{0,i}}}.
\end{align}
In case more than half of the measured pixels in a region are invalid in the spectral index image, this carries over to the measured spectral index and uncertainty by assigning a masked value.

\subsection{Primary beam correction}
\label{sec:pb_correction}

Due to the primary beam response of the MeerKAT antennas, sources away from the pointing centre appear fainter than they are in reality. As this effect is not corrected for in the imaging stage, resulting continuum images will have accurate flux densities at the pointing centre but attenuated flux densities that become fainter the further from the pointing centre they are located. A simplified model of the primary beam is described in \citet{Mauch2020}, which assumes the primary beam of MeerKAT as directionally symmetric, describing it with a cosine-tapered illumination function
\begin{equation}
    P(\rho,\theta_{pb}) = \left[\frac{\cos(1.189\pi\rho/\theta_{pb})}{1-4(1.189\rho/\theta_{pb})^2}\right]^2.
    \label{eq:pb_correction}
\end{equation}
Here $\rho$ is the distance from the pointing centre and $\theta_{pb}$ represents the angular size of the FWHM of the primary beam, a quantity which is dependent on the observing frequency $\nu$,
\begin{equation}
    \theta_{pb}(\nu) = 57'.5\left(\frac{\nu}{1.5\mathrm{GHz}}\right)^{-1}.
\end{equation}
At the central frequency of our continuum images of 1.27~GHz, the FWHM of the primary beam is $\theta_{pb} = 67'$. This simplified model is implemented in the \verb|katbeam|\footnote{\url{https://github.com/ska-sa/katbeam}} \textsc{python} package. As the primary beam is frequency dependent it affects the spectral index images, increasing the measured spectral index away from the pointing centre. The spectral index change induced by the primary beam can be approximated by
\begin{equation}
    P_{\alpha}(\rho,\nu) = -8\log(2)\left(\frac{\rho}{\theta_{pb}}\right)^2\left(\frac{\nu}{\nu_0}\right)^2,
    \label{eq:alpha_corr}
\end{equation}
Again, we assume the frequency $\nu$ to be equal to the central frequency $\nu_0 = 1.27$~GHz.

In reality, the MeerKAT primary beam at L-band is more complicated and cannot be completely described by a directionally symmetric model. \citet{Villiers2022} present and analyse holographic measurements of the MeerKAT primary beam, showing the directional asymmetries present due to variations between individual antennas. For an accurate model of the primary beam, we use these holographic measurements to correct our images. As we utilise the full 802.5~MHz bandwidth of L-band for these images, a primary beam correction must take this into account. Though a wideband primary beam correction is implemented in the \textsc{CASA} task \verb|widebandpbcor|, there are no models of the MeerKAT beams available. As such, we implement the wideband primary beam correction ourselves using the same basic recipe, which consists of creating a primary beam with a frequency structure matching that of the image, in this case creating a primary beam model for each of the 16 spectral windows of the continuum data. As in the imaging step, we model the multi-frequency primary beam with two Taylor terms. The primary beam corrected Taylor term images are then defined as follows:
\begin{subequations}
\begin{align}
    I_0' &= P_0^{-1}I_0, \\
    I_1' &= P_0^{-1}\left(I_1-\frac{P_1}{P_0}I_0\right).
\end{align}
\label{eq:wb_pb}
\end{subequations}
Here, $P_0$ and $P_1$ represent the zeroth and first order Taylor term primary beams, respectively, where $P_0/P_1$ should be equal to $\alpha_{pb}$ as specified in Equation \ref{eq:alpha_corr}. 

While we use the holographic wideband primary beam corrections described in Equation \ref{eq:wb_pb} for the main results of this work, we will also briefly explore the simplified corrections of Equations \ref{eq:pb_correction} and \ref{eq:alpha_corr} and see how they compare to the wideband corrections. At applying the primary beam corrections, the image is cut off at the 5\% level of the primary beam (at the central frequency of 1.27~GHz), which leaves us with a circular image with a diameter of approximately 4000~pixels, or 2.2~degrees. As a result of reduced sensitivity towards the edges of the image, the noise is increased there.

\begin{table*}[]
    \centering
    \caption{Details on all ten pointings after complete processing and source finding.}
    \resizebox{\textwidth}{!}{%
    \begin{tabular}{c|c c c c c | c c c c c c}
    Target & RA & Dec & Flux density & Flux density & Spectral index & PSF maj & PSF min & PSF PA & Counts & $\sigma_{20}$ & Demerit \\
           &    &     & NVSS         & MALS   &     & & & & & & \\
            &   &   & (mJy)     & (mJy) & & $''$ & $''$ & $\degree$ & & (\textmu Jy $\mathrm{beam}^{-1}$) & (mJy) \\
    \hline \hline
    J0001-1540 & $00^h01^m41^s.57$ & $-15\degree40'40''.60$ & 436 & $513.6\pm1.0$ & $-0.85\pm0.04$ & 7.7 & 6.3 & -8.9  & 2132 & 26 & 14.5 \\
    J0006+1728 & $00^h06^m47^s.35$ & $+17\degree28'15''.40$ & 226 & $220.1\pm0.7$ & $-0.28\pm0.08$ & 11.4 & 6.3 & -7.1 & 1378 & 29 & 10.1 \\
    J0126+1420 & $01^h26^m13^s.24$ & $+14\degree20'13''.10$ & 577 & $685.6\pm0.7$ & $-0.95\pm0.09$ & 10.5 & 6.3 & -2.5 & 1591 & 33 & 8.6 \\
    J0240+0957 & $02^h40^m27^s.19$ & $+09\degree57'13''.00$ & 521 & $589.3\pm0.7$ & $-1.11\pm0.09$ & 10.2 & 6.6 & 9.1 & 986 & 48 & 18.7 \\
    J0249-0759 & $02^h49^m35^s.41$ & $-07\degree59'21''.00$ & 646 & $711.0\pm0.6$ & $-0.97\pm0.09$ & 9.2 & 6.6 & -1 & 2619 & 19 & 7.7 \\
    J0249+0440 & $02^h49^m39^s.93$ & $+04\degree40'28''.90$ & 420 & $472.6\pm0.3$ & $-0.80\pm0.09$ & 8.1 & 6.7 & -7.6 & 1558 & 29 & 6.3 \\
    J1133+0015 & $11^h33^m03^s.12$ & $+00\degree15'48''.90$ & 233 & $377.7\pm0.9$ & $-0.01\pm0.07$ & 8.9 & 6.7 & -15.3 & 803 & 52 & 15.9 \\
    J1232-0224 & $12^h32^m00^s.13$ & $-02\degree24'04''.10$ & 1647 & $1823.4\pm5.2$ & $-0.31\pm0.09$ & 8.5 & 6.7 & -9 & 611 & 73 & 20.8 \\
    J1312-2026 & $13^h12^m07^s.86$ & $-20\degree26'52''.40$ & 778 & $851.4\pm0.4$ & $-0.83\pm0.09$ & 7.7 & 6.3 & -9.1 & 2431 & 21 & 12.2 \\
    J2023-3655 & $20^h23^m46^s.21$ & $-36\degree55'21''.20$ & 436 & $406.6\pm0.5$ & $-0.11\pm0.12$ & 10.9 & 6.8 & -88 & 2160 & 22 & 5.8 \\
 \hline
    \end{tabular}}
    \label{tab:pointing_table}
\end{table*}

\subsection{Assessment of calibration}
\label{sec:calibration_assessment}

\begin{figure}
    \centering
    \includegraphics[width=\hsize]{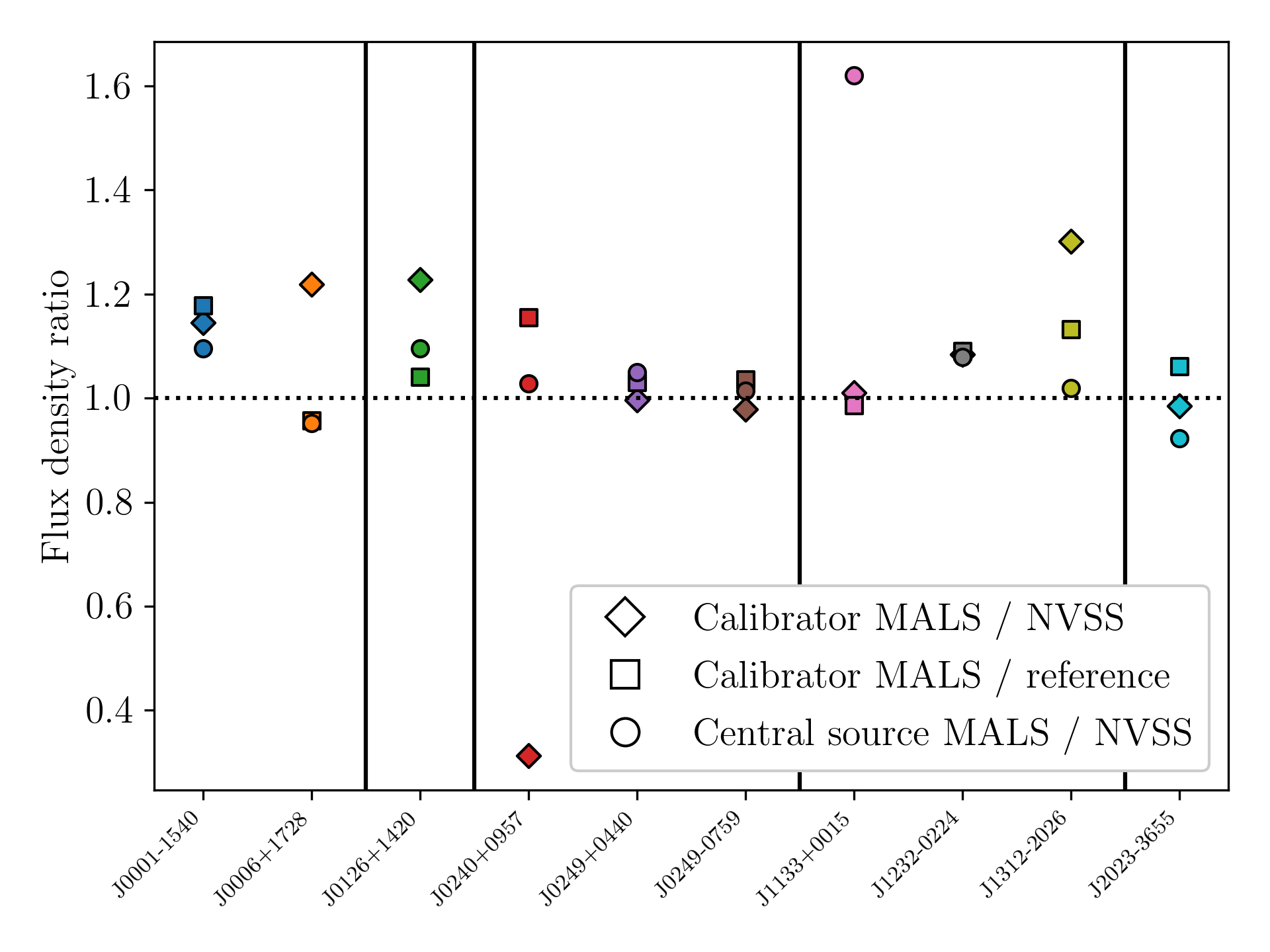}
    \caption{Flux offsets of the central sources and gain calibrators per field. Gain calibrators are compared to their reference flux density (circles) as specified in Table \ref{tab:calibration_table} and with their NVSS counterparts (diamonds). Central sources are compared to their NVSS counterparts (squares), NVSS flux densities are all converted to rest frequency assuming $\alpha=-0.75$. Sources are ordered by observing run, separated by vertical black lines.}
    \label{fig:flux_offset}
\end{figure}

Processing the raw data to the final scientific data products can introduce errors , affecting the flux density scale. A first order estimation of the flux density scale can be obtained by comparing the flux densities of the gain calibrators with their literature values. As discussed in Section~\ref{sec:obs_cal}, we evaluate the automated process of the data calibration by generating diagnostic reports and automatically evaluating logged information in order to determine problems in the data processing. This evaluation singles out errors and warnings present in the logs, allowing direct insight into any problems that might have occurred during the calibration process. Furthermore, it extracts information we can use to assess the quality of calibration from the logs, such as the flux densities of calibrator sources.

Table~\ref{tab:calibration_table} summarises the observation and calibration details, showing the targets and their associated calibrator sources. The flux densities of the flux- and gain calibrators are extracted from the logs and the flux densities of gain calibrators are compared to the MeerKAT reference catalogue \citep{Taylor2021}. We extend this to a broader assessment of the flux density scale in Figure \ref{fig:flux_offset}, where we show the flux density offsets of the gain calibrators and central sources of the individual pointings. Along with the comparison in Table~\ref{tab:calibration_table}, both gain calibrators and central sources (see Table~\ref{tab:pointing_table}) are compared to their NVSS counterparts. Flux densities are corrected for frequency using the spectral index from the reference catalogue if available, assuming $\alpha=-0.75$ otherwise. Combining the measurements from the ten pointings, the mean flux density ratios are $1.03\pm0.26$ between the gain calibrators and their NVSS counterparts, $1.07\pm0.07$ between gain calibrators and their reference values, and $1.08\pm0.19$ between central sources and their NVSS counterparts. We note that the absolute amplitude calibration of NVSS is based on \citet{Baars1977} and has an uncertainty of up to 12\% with respect to the here used \citet{Perley2017} scale, depending on the calibrator used. 

We note that the SUMSS and NVSS measurements were taken with different instruments at different times, so some variation is to be expected. The current assessment does not include astrometric precision, as calibrators are not imaged. We assess this aspect along with the another flux density scale assessment by cross-matching the full catalogue of sources with other surveys in Section~\ref{sec:cross_match}.  

\subsection{Assessment of image quality}
\label{sec:noise}

\begin{figure}
    \centering
    \includegraphics[width=\hsize]{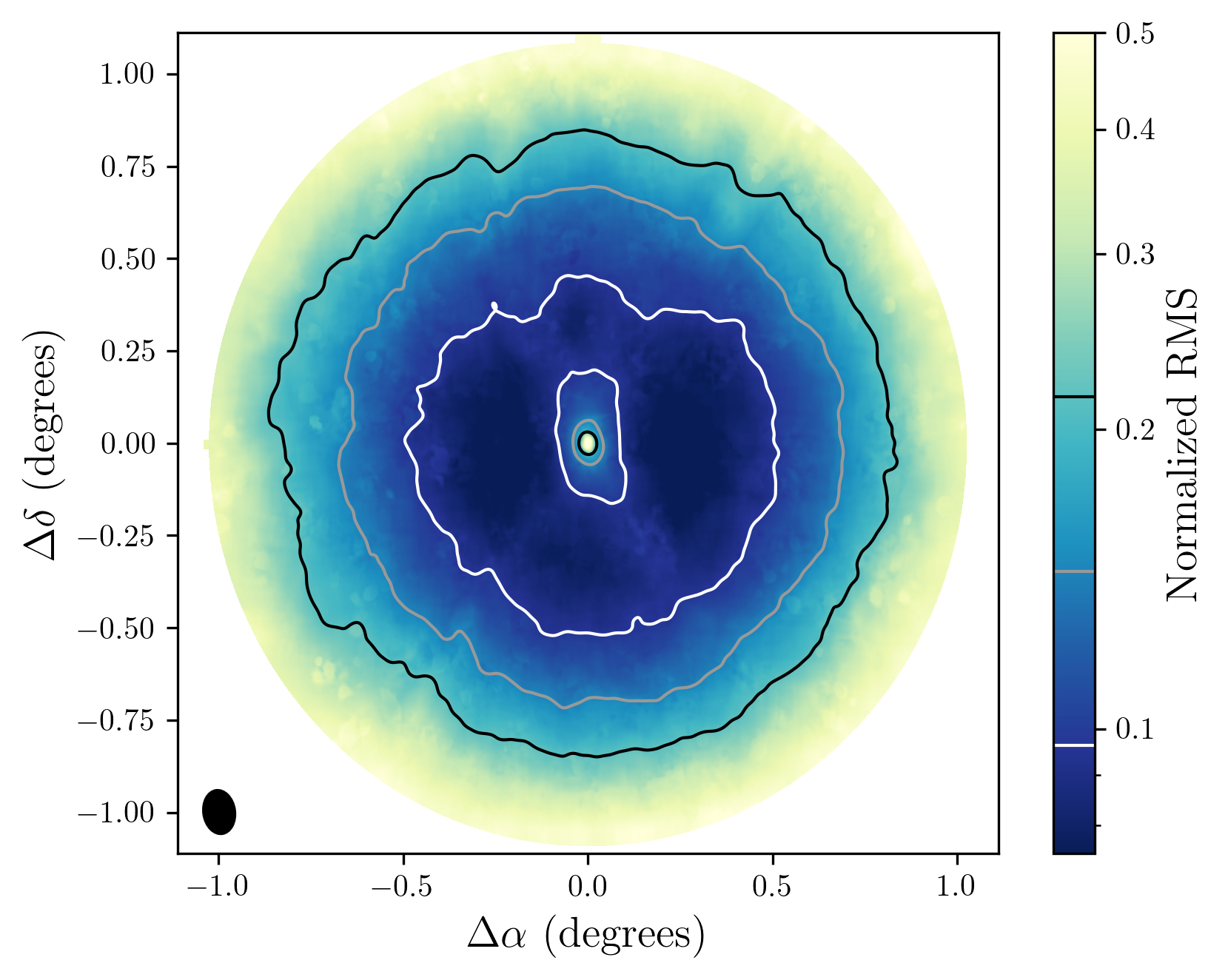}
    \caption{The median stacked pixel values of the RMS noise images of all ten pointings. As primary beam correction is applied, the noise goes up towards the edges of the image. Since a strong central source is always present, the noise is always higher in the centre as well. The given contours from white to black are 20, 40, and 60\% RMS noise coverage. The stacked beam (50x increased in size) of all the pointings is shown in the lower left corner, matching the elongated structure in the centre.}
    \label{fig:median_rms}
\end{figure}

\begin{figure}
    \centering
    \includegraphics[width=\hsize]{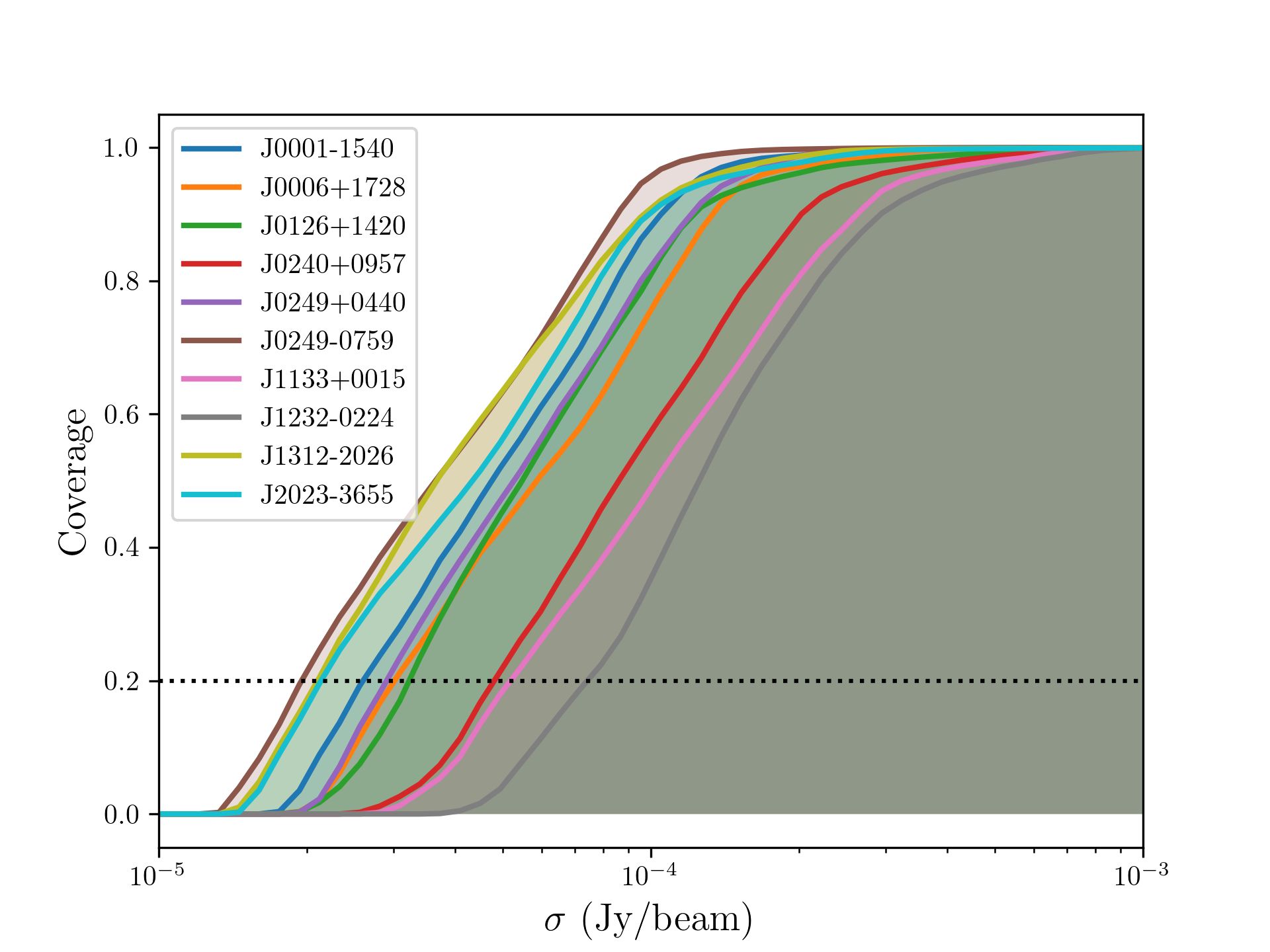}
    \caption{RMS noise coverage for all ten pointings. The dotted line indicates the 20\% coverage level, which is used to define $\sigma_{20}$. Noise varies appreciably between the pointings, however the overall structure of the RMS noise coverage curves remains consistent, indicating that $\sigma_{20}$ is a good zeroth order measure of the noise scale.}
    \label{fig:rms_coverage}
\end{figure}

With any radio image, there is a great number of variables that can influence the quality of the image, both related to intrinsic properties of the pointing and to the process of calibration and self-calibration. As discussed in Section~\ref{sec:cont_imaging}, a report is generated that monitors image statistics such as noise and central source flux during different self-calibration steps. Furthermore, the logs are automatically evaluated for possible errors and warnings and information relevant to the quality of the self-calibration and imaging is extracted. To evaluate the final image product, the image quality of the individual ten pointings is assessed by using the Root Mean Square (RMS) noise maps that are automatically produced during the source finding procedure by PyBDSF (see Section~\ref{sec:sourcefinding}). In particular, we investigate the overall noise characteristics by evaluating the sky coverage with respect to the RMS noise. A direct measurement of the noise allows us then to easily correlate image quality with other characteristics of the pointings.

We create a smoothed representation of the ten pointings by median stacking their normalised RMS noise images which is shown in Figure~\ref{fig:median_rms}. As all pointings have a strong source at their centre, the noise is increased at the pointing centre and increases towards the edges of the pointing as a consequence of the primary beam response. Figure~\ref{fig:median_rms} shows that some directional effects are left in the image. Notably, there is an elongated noise structure in the centre, associated with the bright central source, aligned in the north-south direction. The stacked beam included in the figure aligns well with the elongated structure, indicating that the most prominent structures are a result of the shape of the stacked PSF of the images. The imprint of the stacked PSF is also the most likely cause of the cross-like structure seen in the stacked image. Though we have the wideband primary beam correction based on holographic images which take into account the asymmetries present in the primary beam, pointings are observed for three separate blocks of 20 minutes in an observing night, which smears out the asymmetries in the primary beam\footnote{MeerKAT antennas have Alt-Az mounts, such that the sky rotates with respect to the dish while observing}. This effect cannot be easily corrected for in the image plane, but could be taken into account during imaging using A-projection \citep{Bhatnagar2008}. Though present, the asymmetries here are small and dominated by the other noise structures in the image. 

The usual method of determining RMS noise in an image relies on measuring RMS noise in an area close enough to the pointing centre to not be affected by the primary beam and far enough from strong sources to be unaffected by artefacts. Due to the number and structure of MALS pointings, this cannot be reliably done in an automated fashion. Instead, we investigate the differences in noise level between individual pointings by using the RMS noise images to assess the RMS noise coverage, measuring the cumulative distribution of noise levels across an image. Figure~\ref{fig:rms_coverage} shows the RMS noise coverage curves for the individual fields, and it can be seen that RMS noise coverage curves are similar in structure but offset from each other. To quantify this offset and thus characterise the noise in the individual pointings, we define $\sigma_{20}$ at 20\% RMS noise coverage, representing the noise in the central portion of the image (see Figure~\ref{fig:median_rms}). We will see that $\sigma_{20}$ excellently serves as a normalisation factor to account for the differences in noise levels between the pointings, and can be used to unify the assessment of individual pointings and extend them to the full survey.

There are several factors which can contribute to the overall noise level in an image, not all of which are easily quantifiable. However, an important aspect to consider is the shape of the synthesised beam or PSF, determined by the UV-coverage of the observation, which in turn is determined by the array configuration, observing time, and elevation of the target at the time of observation. There are two aspects to the PSF that influence image noise. A measurement of RMS noise in Jy/beam will be influenced by the shape of the beam\footnote{The clean beam of an image is determined during imaging by fitting a 2D Gaussian to the central lobe of the PSF.}, and very bright sources can have persistent and bright sidelobes from the shape of the PSF that are difficult to clean completely and as a result push up the noise in an image. To quantify this last effect we calculate the demerit score detailed in \citet{Mauch2020} to estimate the contributions of bright sources to RMS noise in the image. We calculate the independent source contributions to the errors in the image using all sources that have an unattenuated flux density of more than 100 mJy. The demerit score $d$ is then defined as
\begin{equation}
    d = \left[\sum_{S>100\ \mathrm{mJy}}^i \left\{\left(\frac{8\ln(2)\rho\sigma_p}{\theta_{pb}^2} + \sigma_g\right) S_{a,i}\right\}^2 \right]^{1/2},
\end{equation}
where the first term represents the contribution of pointing error $\sigma_p$ scaling with distance from the pointing centre $\rho$, and the second term is the receiver gain error $\sigma_g$. The contribution of each source comes in the form of their attenuated flux density $S_a$. Appropriate values for MeerKAT L-band are detailed in \citet{Mauch2020}, which we also use ($\theta_{pb} = 67'$, $\sigma_p = 30''$, $\sigma_g = 0.01$). The demerit scores of all pointings are included in Table~\ref{tab:pointing_table}. A correlation is present between demerit score and $\sigma_{20}$, and especially pointings with high $\sigma_{20}$ show increased demerit scores. Though pointings with lower $\sigma_{20}$ show more scatter in their demerit scores, this nonetheless shows demerit score as a first order estimate of pointing quality, which we can utilise as a predictive measure.

\section{Source finding}
\label{sec:sourcefinding}

With thousands of sources expected to be detected in every MALS pointing, we require an automated source finding algorithm to find and characterise these sources. A small number of these are suitable for radio images, and perform comparatively similar \citep{Hale2019}. Of these, \textsc{PyBDSF} has been used in several recent data releases of large-scale surveys, such as the LOFAR Two-Metre Sky Survey \citep[LoTSS,][]{Shimwell2019} and the Rapid ASKAP Continuum Survey \citep[RACS,][]{Hale2021}. \textsc{PyBDSF} stands out in its ability to model extended emission with its wavelet decomposition module, and provides easy ways to compile source catalogues and assess the quality of the fields. Besides generating catalogues, \textsc{PyBDSF} provides  output maps related to the input image, such as the RMS noise images we used in Section~\ref{sec:noise}, and mean and residual images. Once RMS noise and mean maps are obtained \textsc{PyBDSF} allows these maps to be used as input to ensure source finding is performed with the exact same parameters. For MALS, we thus make use of \textsc{PyBDSF}, both for creating clean masks during self-calibration as detailed in Section \ref{sec:cont_imaging}, and integrating \textsc{PyBDSF} into the workflow to automatically carry out source finding, cataloguing, cross-matching and combining catalogues, using \textsc{python}-based scripts developed by the authors\footnote{\url{https://github.com/JonahDW/Image-processing}}.

In order to understand the impact of the individual pointings to a general catalogue, we evaluate the source finding procedure for each pointing. We investigate completeness (what fraction of sources do we detect) and purity (what fraction of sources is real) in source counts with respect to signal-to-noise ratio, flux density, and source size, as well as \textsc{PyBDSF}'s capability to accurately recover flux densities. 

\subsection{Stokes I Catalogues}
\label{sec:catalogs}

\begin{figure*}
    \centering
    \includegraphics[width=\textwidth]{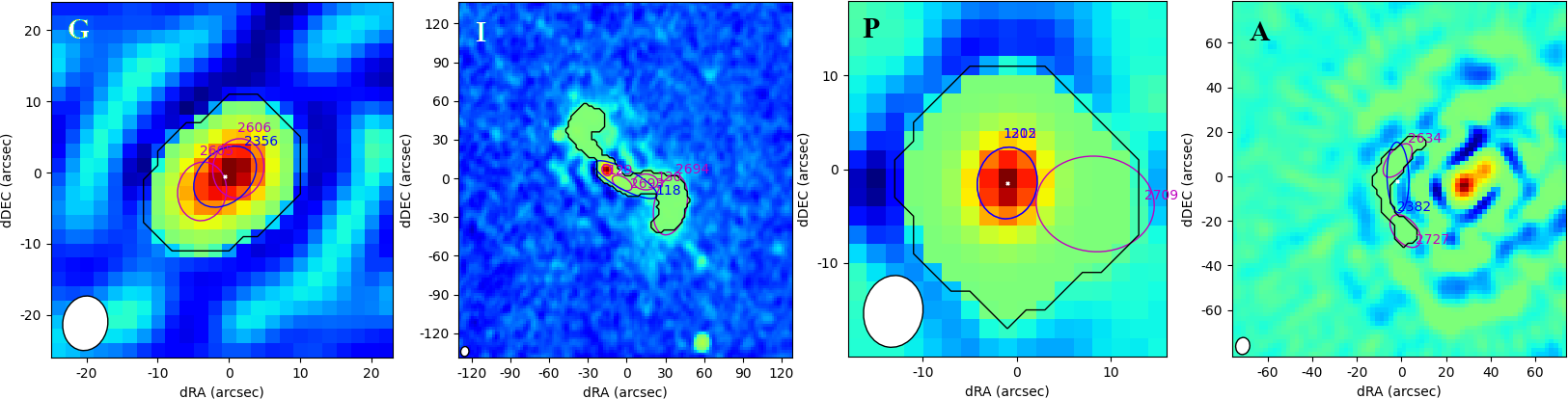}
    \caption{Examples of possible source classes. The black outline shows the island threshold, the magenta ellipses show the individual Gaussian components fit to to the source, and the blue ellipses show the the combined Gaussian describing the source. From left to right: An elongated source fit by two Gaussian components, the combined Gaussian describes the source adequately, it has been assigned the `G' class. A likely FRI source with complex structure, better described by the island than the Gaussian components, it has been assigned the `I' class. A point source with an additional noise peak that has been fitted with a Gaussian component, it has been assigned the `P' class. An artefact caused by a nearby bright source, it has been assigned the `A' class.}
    \label{fig:cutout_examples}
\end{figure*}

\begin{figure}
    \centering
    \includegraphics[width=\hsize]{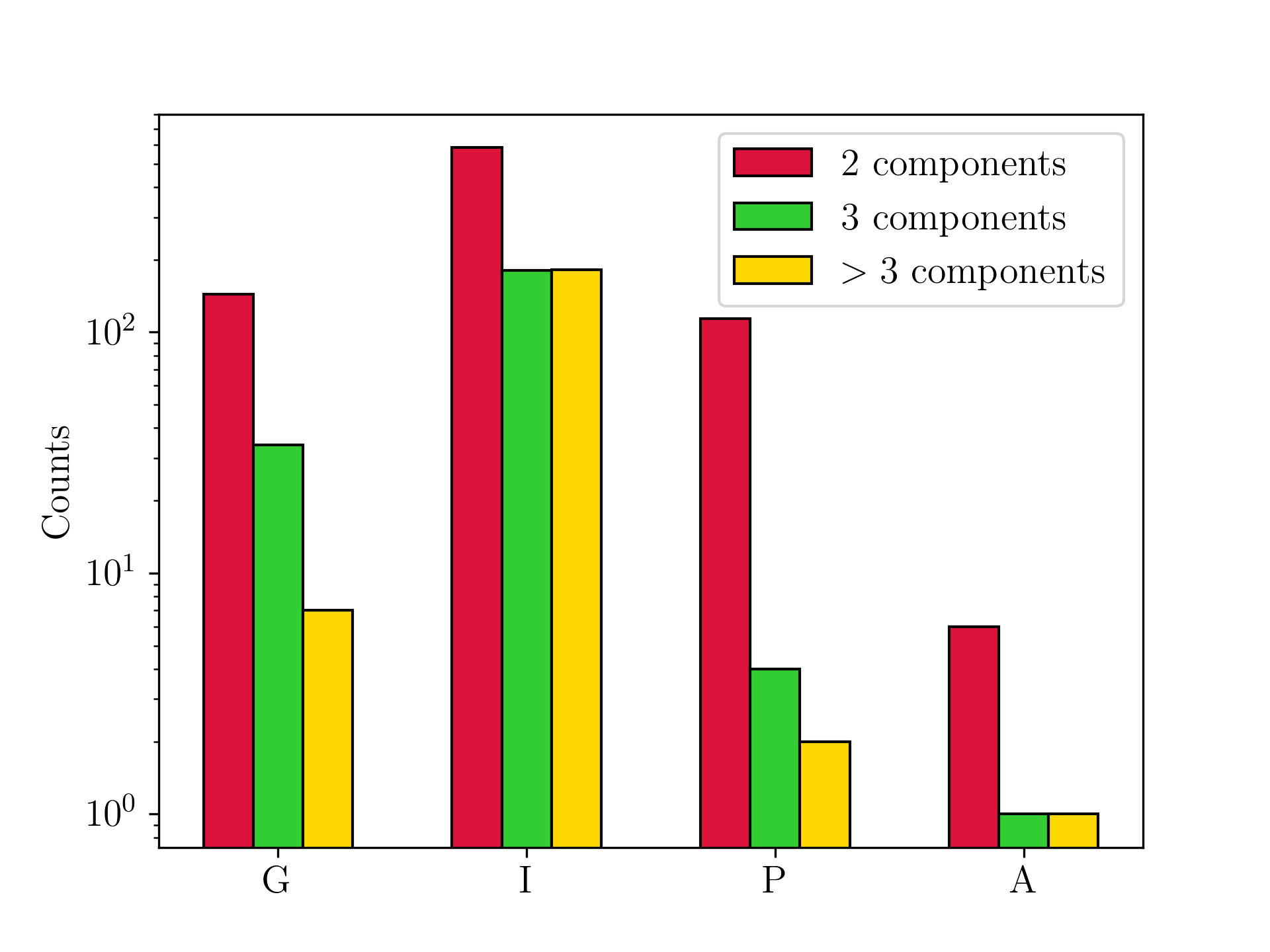}
    \caption{Assigned flags to sources with multiple Gaussians, separated by the number of Gaussian components fit to them by \textsc{PyBDSF}. Keeping in mind that most of these sources are fit with two Gaussian components, the `I' class is preferred for sources with three or more components, while the `P' class consists almost exclusively of sources with two components.}
    \label{fig:cutout_classes}
\end{figure}

In order to compile a source catalogue from \textsc{PyBDSF}, various steps are needed which depend on the initial setup of \textsc{PyBDSF}. \textsc{PyBDSF} identifies islands of emission which are brighter than the island threshold. Within these islands \textsc{PyBDSF} finds emission peaks above a corresponding pixel threshold, and for each peak found fits a 2D Gaussian to the peak and surrounding emission. Performing source finding on our MALS images, we impose an island threshold of 3\textsigma, and a pixel threshold of 5\textsigma. Individual Gaussian components are combined into sources in a way that can be specified by the user, and we elect to combine Gaussian components that occupy the same island into a single source. The RMS noise in the images is determined by a sliding box, and we decrease the size of the sliding RMS box near bright sources to avoid spurious detections of artefacts around these sources as much as possible. Furthermore, to improve fitting of extended sources in the field, we enable \`a trous wavelet decomposition \citep{Holschneider1989}. The \textsc{PyBDSF} settings can be summarised as follows:
\begin{verbatim}
    thresh_isl = 3.0
    thresh_pix = 5.0
    rms_box = (150,30)
    adaptive_rms_box = True
    adaptive_tresh = 100
    rms_box_bright = (40,15)
    group_by_isl = True
    atrous_do = True
    atrous_orig_isl = True
    atrous_jmax = 3
\end{verbatim}
For the purposes of analysing and building the final catalogue, we require a number of output products from \textsc{PyBDSF}. The output from source finding includes both a catalogue of sources and of individual Gaussian components. Furthermore, a background RMS noise and background mean image are produced, as well as a residual image. 

For a single pointing, we run source finding using \textsc{PyBDSF}, and modify the output source catalogues by adding to the existing columns the ID of the MALS pointing, a source name following IAU convention, and distance of the source to the pointing centre. As \textsc{PyBDSF} does not calculate spectral indices unless an image has multiple channels, we measure the spectral index of the sources in our fields from the spectral index images as described in Section \ref{sec:alpha_images}, using the extent of the Gaussian (major axis, minor axis, position angle) of the source to define the region in the image. 

Though \textsc{PyBDSF} is configured to avoid spurious detections as much as possible, it is unavoidable that some artefacts are falsely identified as sources. We identify artefacts around the 10 brightest sources in each image by flagging sources within 5 times the major axis of the beam that have less than 10\% of the peak flux density of the bright source. This is largely motivated by the shape of the PSF, which can have sidelobes with a strength of up to 10\% of the maximum. Though this does not get rid of all false detections in the image (see Section \ref{sec:purity}), it flags the most prominent imaging artefacts.

To assess the quality of the Gaussian fitting by \textsc{PyBDSF}, we perform visual inspection on select sources. We create cutouts from the images and perform visual inspection, which is implemented in a separate module based on \textsc{python} and \textsc{CASA}\footnote{\url{https://github.com/JonahDW/CASA-Poststamp}}.
\textsc{PyBDSF} assigns each source a flag indicating whether the source is fit by a single Gaussian (`S'), multiple Gaussian components (`M') or Gaussian component(s) on an island with other sources (`C'). Since all Gaussian components that occupy the same island are always combined into one source, the `C' flag is not present in our catalogues. For visual inspection we consider all sources made up of multiple Gaussian components. As such, all sources that carry the `M' flag ---which make up 8\% of the all sources found in the fields--- are flagged for visual inspection. Through visual inspection we then assign an additional flag indicating the nature of the source, and how well it is described by the \textsc{PyBDSF} model.
\begin{itemize}
    \item[G:] Sources that are well described by the Gaussian model.
    \item[I:] Complex sources that are not adequately described by the Gaussian components fit to them. The flux density of these sources is better described by the integrated flux of the island, and their position by the flux weighted mean position of the island.
    \item[P:] Sources fit with multiple Gaussian components where only one is required to adequately describe the source. Other Gaussian components are likely fit to noise fluctuations coinciding with the source.
    \item[A:] Artefacts that will be flagged as such in the catalogue.
\end{itemize}
Figure \ref{fig:cutout_examples} shows an example for each of the cutout classes, and how we identify the different possible cases. To aid in visual inspection, in the source cutout we plot the individual Gaussian components, the combined source Gaussian, and the island threshold. Therefore in this step we use both the source catalogues and the Gaussian component catalogues. Along with the cutout classes, additional columns are added to the table that describe the sources. In the cutouts we measure the integrated flux density of the island, the spectral index of the island, the intensity weighted mean position of the source, and a flag indicating if these measures are valid. Additionally, the number of Gaussian components is recorded for each source, as the initial \textsc{PyBDSF} catalogue only indicates whether a source has been fit with multiple Gaussian components or not.

Figure \ref{fig:cutout_classes} shows the classification of all sources in the ten pointings that have been fit with multiple Gaussian components, 1259 in total. We see that almost all 120 sources assigned with the `P' class have two Gaussian components, and a relatively large percentage of sources with the `I' class have more than two Gaussian components assigned. Around 185 (15\%) of these sources were considered to be adequately described by their Gaussian components, while 946 (75\%) are more complex and better described by their island attributes. Only 8 sources are flagged as obvious artefacts. 

\subsection{Evaluation of individual pointings}
\label{sec:individual_assessment}

In order to determine the reliability of the source finding routine and to assess how detection of sources is affected by their properties, we measure the completeness, purity, and flux recovery statistics of the catalogues. Here we assess these qualities for individual pointings to see how characteristics of the pointings such as central source flux density and noise level affect these quantities.

To measure completeness and flux recovery, we require complete knowledge of the intrinsic flux densities and shapes of the sources that are present in the image. To that end, we use realistic samples of simulated extragalactic radio sources from the \citet{Wilman2008} simulation of the SKA Design Study (SKADS). Though more recent simulations such as the Tiered Radio Extragalactic Continuum Simulation \citep[T-RECS,][]{Bonaldi2019} are available, the SKADS catalogues include morphology details of all sources and source components, which is necessary information when injecting sources into the data. From the SKADS simulations we create mock catalogues with 5000 sources uniformly distributed in flux density that have a flux density above 10~\textmu Jy, which is equal to the limit of thermal noise (10~\textmu Jy/beam) for an unresolved source. With this we allow for the possibility of noise fluctuations to push sources above the detection threshold. We inject sources from the mock catalogue uniformly distributed into the residual images produced by \textsc{PyBDSF}, which are devoid of sources but share the noise characteristics of the original images. We then perform the source finding routine again on these images, using the same mean and RMS noise maps determined by \textsc{PyBDSF} from the original image. This ensures that source finding is performed in the exact same way as the original image. We consider a source recovered if it is detected within the FWHM of the major axis of the clean beam from the original position. In order to reach a more robust measure, this process is repeated 50 times for each pointing, separately for point sources and resolved sources.

\subsubsection{Completeness}
\label{sec:completeness}

\begin{figure*}
    \centering
    \includegraphics[width=\textwidth]{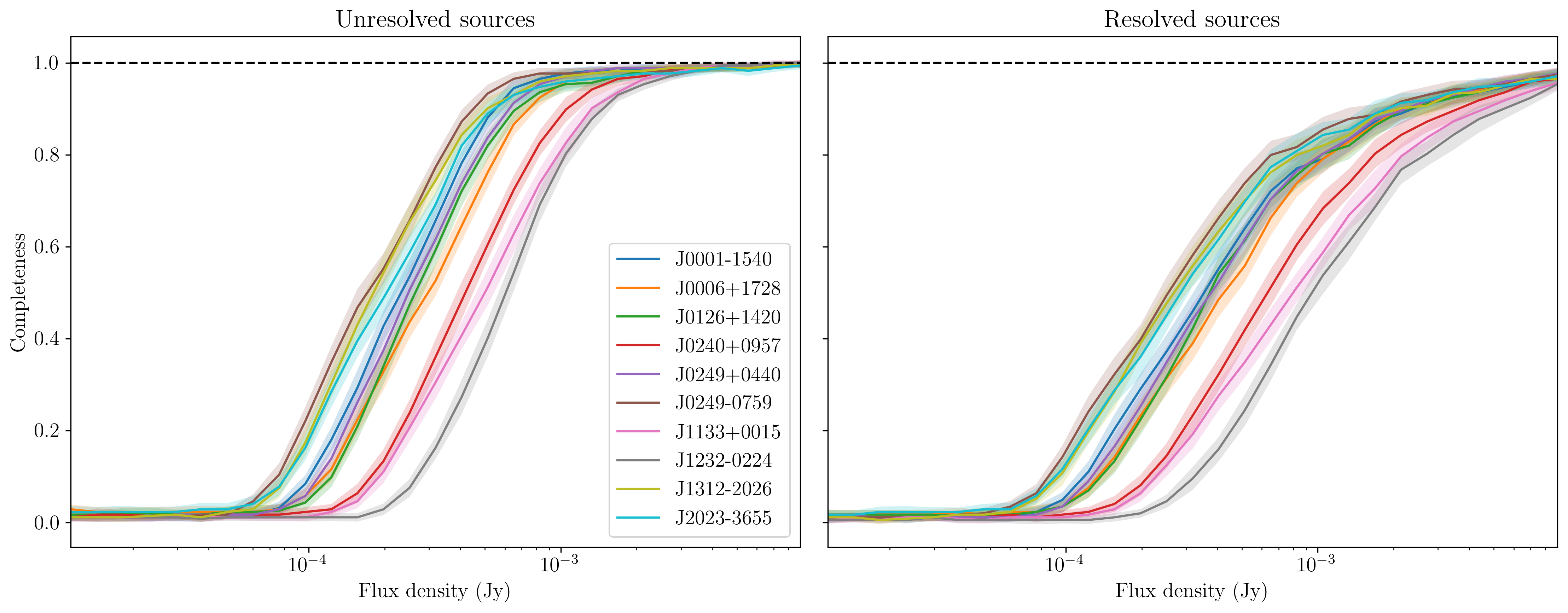}
    \caption{Completeness for unresolved sources (left) and resolved sources (right) for the different fields and their associated uncertainties as a function of flux density. There are larges differences between the pointings, where pointings with higher noise levels have lower overall completeness. Completeness is lower for resolved sources as well.}
    \label{fig:pointing_completeness}
\end{figure*}

\begin{figure*}
    \centering
    \includegraphics[width=\hsize]{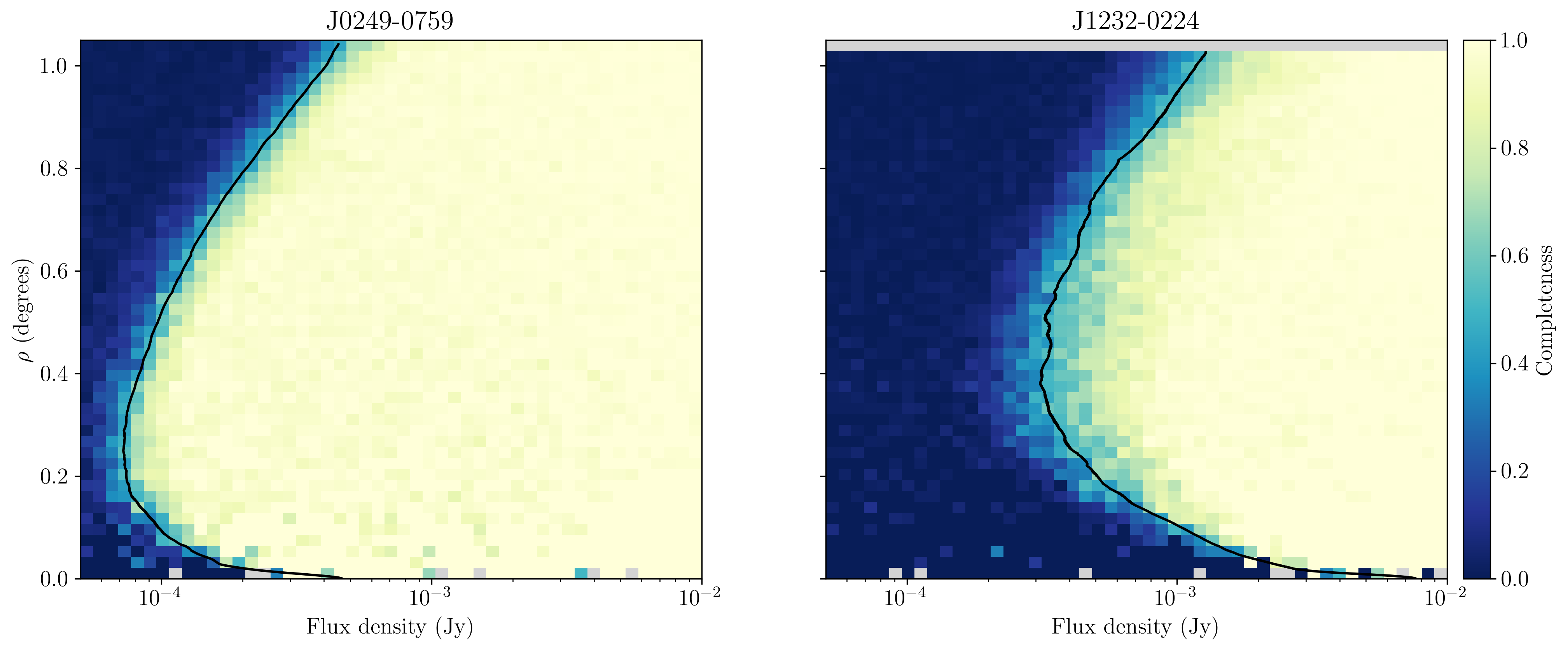}
    \caption{Completeness for unresolved sources as a function of flux density and distance from the pointing centre, $\rho$, for the fields J0249-0759 (left) and J1232-0224 (right). The radially averaged $5\sigma$ curves (black lines) for the corresponding pointings are seen to follow the zone where completeness transitions from zero to one. Due to the presence of a strong central source in J1232-0224, completeness is lower in the central region of this pointing. Pixels with no sources in them have been coloured grey.}
    \label{fig:individual_sep_flux_completeness}
\end{figure*}

With the procedure described above, we can make a statistically robust assessment of the completeness in the pointings. The (source) completeness in this case simply gives the fraction of sources that is detected, most commonly measured as a function of flux density of the source. The completeness curves for the individual pointings, for both resolved and unresolved sources, can be appreciated in Figure~\ref{fig:pointing_completeness}. Not only is there a large difference between resolved sources and unresolved sources, pointings individually have large differences between them as well. To investigate other aspects of the completeness, we look at the fields J0249-0759 and J1232-0224, which have the lowest and highest noise levels among the pointings respectively (see Table~\ref{tab:pointing_table}), which should yield the most extreme cases and allow us to probe variation between the fields.

\textbf{Unresolved sources} - As the SKADS catalogues describe the intrinsic shapes of sources, we can assess completeness for point sources by only injecting sources with a major axis of zero. The sources are defined in the image as delta functions, and convolved with the clean beam of the individual image. As the total flux density of point sources is concentrated in one peak, they are much easier to detect relative to resolved sources. Point sources allow us to assess completeness without being affected by source morphology, and so we use them to determine the completeness with respect to distance from the pointing centre. As sensitivity decreases outwards from the centre, we expect completeness to decrease as well. Figure~\ref{fig:individual_sep_flux_completeness} shows source completeness as a function of both flux density and distance from the pointing centre for the pointings J0249-0759 and J1232-0224. It is clear that indeed the completeness decreases with increased distance from the pointing centre, but is also lower near the pointing centre. This is a direct result of the strong source at the centre of each pointing pushing up the noise in its immediate vicinity. In the case of J1232-0224, which has a very strong source at the pointing centre, there is significant impact on completeness in the central portion of the image. To investigate the relation between the completeness and noise floor as a function of distance to the pointing centre we use the RMS maps created by \textsc{PyBDSF}. By radially averaging the RMS noise of the pointing, we obtain RMS noise as a function of distance to the pointing centre. As we have set the detection threshold at $5\sigma$, we plot the radially averaged $5\sigma$ detection curve in Figure~\ref{fig:individual_sep_flux_completeness}, showing that this curve almost perfectly follows the completeness `transition zone' for both pointings. In this transition zone the completeness goes up steeply from zero to one, and the flux density at which this occurs is directly related to the noise floor.

\begin{figure*}
    \centering
    \includegraphics[width=\hsize]{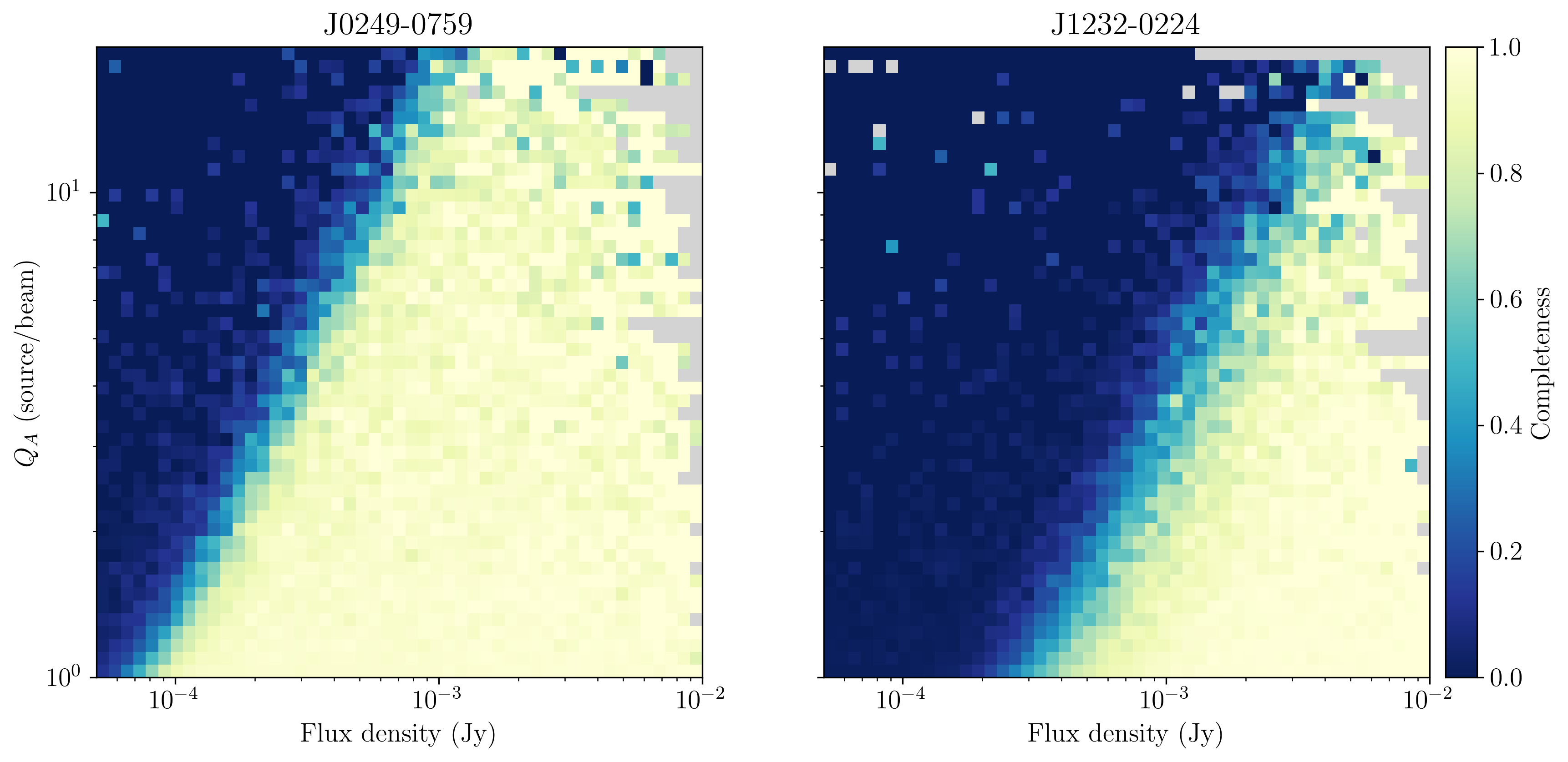}
    \caption{Completeness for resolved sources as a function of flux density and ratio between the area of the source and the beam $Q_A$, for the fields J0249-0759 (left) and J1232-0224 (right). Completeness can be seen to linearly decrease in the log-log scale as a function of $Q_A$, showing that larger sources are harder to detect. Flux densities are compensated for the local noise in order to equalise completeness for different positions in the image. Pixels with no sources in them have been coloured grey.}
    \label{fig:individual_size_flux_completeness}
\end{figure*}

\textbf{Resolved sources} - We perform the same experiment for resolved sources, where we define a resolved source as a source that has major axis and minor axis larger than 0 in the SKADS catalogues\footnote{A small subset of sources with minor axis of 0 and major axis larger than zero are not included in the simulations.}. Sources are randomly selected out of the catalogue, so the distribution of source shapes injected in the image represents the distribution of the SKADS sample. These sources are injected as Gaussians into the image, and as with point sources, convolved with the clean beam. Owing to their lower surface brightness, resolved sources are often less easily detectable compared to point sources with the same flux density. To check how the size of sources affects completeness, we define the area ratio $Q_A$ of a source as the ratio between the area of the source and the beam as defined by their Gaussian characteristics. These are the major and minor axes $\theta_{maj}$ and $\theta_{min}$ for the source and $B_{maj}$ and $B_{min}$ for the clean beam of the pointing,
\begin{equation}
    Q_A = \frac{\theta_{maj}\theta_{min}}{B_{maj}B_{min}}.
\end{equation}
We show the completeness as a function of area ratio and flux density in Figure~\ref{fig:individual_size_flux_completeness} for the pointings J0249-0759 and J1232-0224. As shown in Figure~\ref{fig:individual_sep_flux_completeness}, the completeness of unresolved sources is related to the noise floor. This same relation should be present for resolved sources, in addition to the relation between completeness and source size. In order to disentangle the two different contributions to completeness for resolved sources, we divide flux densities by the ratio of local noise to the lowest noise in the image. The result in Figure~\ref{fig:individual_size_flux_completeness} shows completeness for uniform noise, so that only the source size and flux density affect completeness.  We see a power law decrease (linear in the log-log scale) in completeness as a function of area ratio in both pointings, indicating that this is a universal feature for our source detection. This can be easily understood by considering that for larger sources the total flux density is divided over a larger area, which decreases the peak flux density that is used to detect these sources.

\subsubsection{Flux recovery}
\label{sec:flux_recovery}

\begin{figure*}
    \centering
    \includegraphics[width=\textwidth]{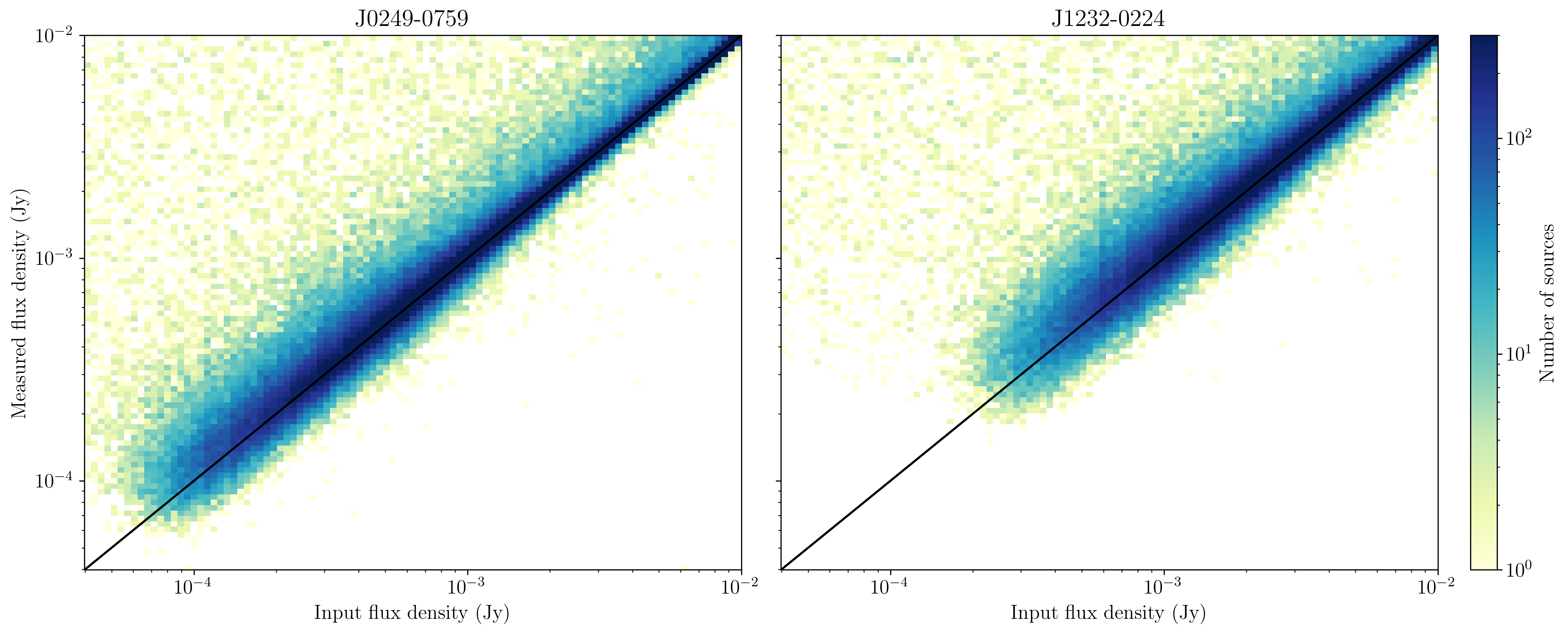}
    \caption{The input flux density plotted against the measured flux density, for the fields J0249-0759 (left) and J1232-0224 (right), based on 100 simulations. As the noise floor in J1232-0224 is relatively high ($\sigma_{20} = 80 \mathrm{
    \mu Jy /beam}$), only sources above 200~\textmu Jy are detected, while in J0249-0759 sources are detected down to 50~\textmu Jy.}
    \label{fig:individual_flux_recovery}
\end{figure*}

Using the mock catalogues, we investigate the ability of \textsc{PyBDSF} to accurately recover flux densities. This can be checked by looking at the flux densities measured by \textsc{PyBDSF} relative to the input flux densities from the mock catalogues. This is an important quality to verify as deviations from the expected 1:1 relationship are obviously undesirable. In Figure~\ref{fig:individual_flux_recovery} we show the measured flux densities against input flux densities of the pointings J0249-0759 and J1232-0224. We see that on average sources have a flux density that matches with their input value. There is however a portion of sources with lower input flux density that have a significantly higher measured flux density than their input. These sources have their flux densities boosted by noise fluctuations, which are present in various orders of strength in the images, from thermal noise to calibration artefacts. We expect these sources to land on positive as well as negative noise peaks, but only sources on positive peaks will be detected. This results in an Eddington bias \citep{Eddington1913} pushing up the distribution of flux densities. To make quantitative statements about this bias, we need to combine data from all the 10 pointings, which we will do in Section~\ref{sec:general_assessment}.

\subsubsection{Limitations of simulations}

The method we have used here for measuring completeness and flux recovery relies on injecting sources directly into the residual images and measuring their properties with \textsc{PyBDSF}. The advantage of this is a direct probe into the machinery of \textsc{PyBDSF}, as this is the only `black box' between the input sources and the measurement. However, these simulations ignore some effects that affect the flux densities and shapes of sources in radio data, such as calibration effects, clean bias, and averaging effects like time and bandwidth smearing \citep{bridle1999}. Probing these requires injecting sources directly into the visibilities and reprocessing the image, something which is not efficient for a large survey such as MALS. Finally, sources are injected into the image convolved with the clean beam as opposed to the PSF. This is well motivated for brighter sources, as these have been mostly cleaned during the imaging process. For faint sources this is not the case, especially since the masks for cleaning are generated by \textsc{PyBDSF} and are thus subject to the same selection as we have used for the final images. To make the simulations more realistic, all undetected sources should therefore be convolved with the PSF. It is not clear how this should affect source finding, but the general consequence of this is that below the detection threshold sources immediately become fainter as a consequence of being convolved with the PSF rather than the clean beam. The PSF also spreads the emission of these sources over a large area, which could affect RMS noise if source crowding is high enough. This would however only be the case if images would be close to or at the confusion limit, which is not the case for MeerKAT at L-band down to at least 0.25~\textmu Jy \citep{Mauch2020}.

\subsubsection{Purity}
\label{sec:purity}

\begin{figure}
    \centering
    \includegraphics[width=\hsize]{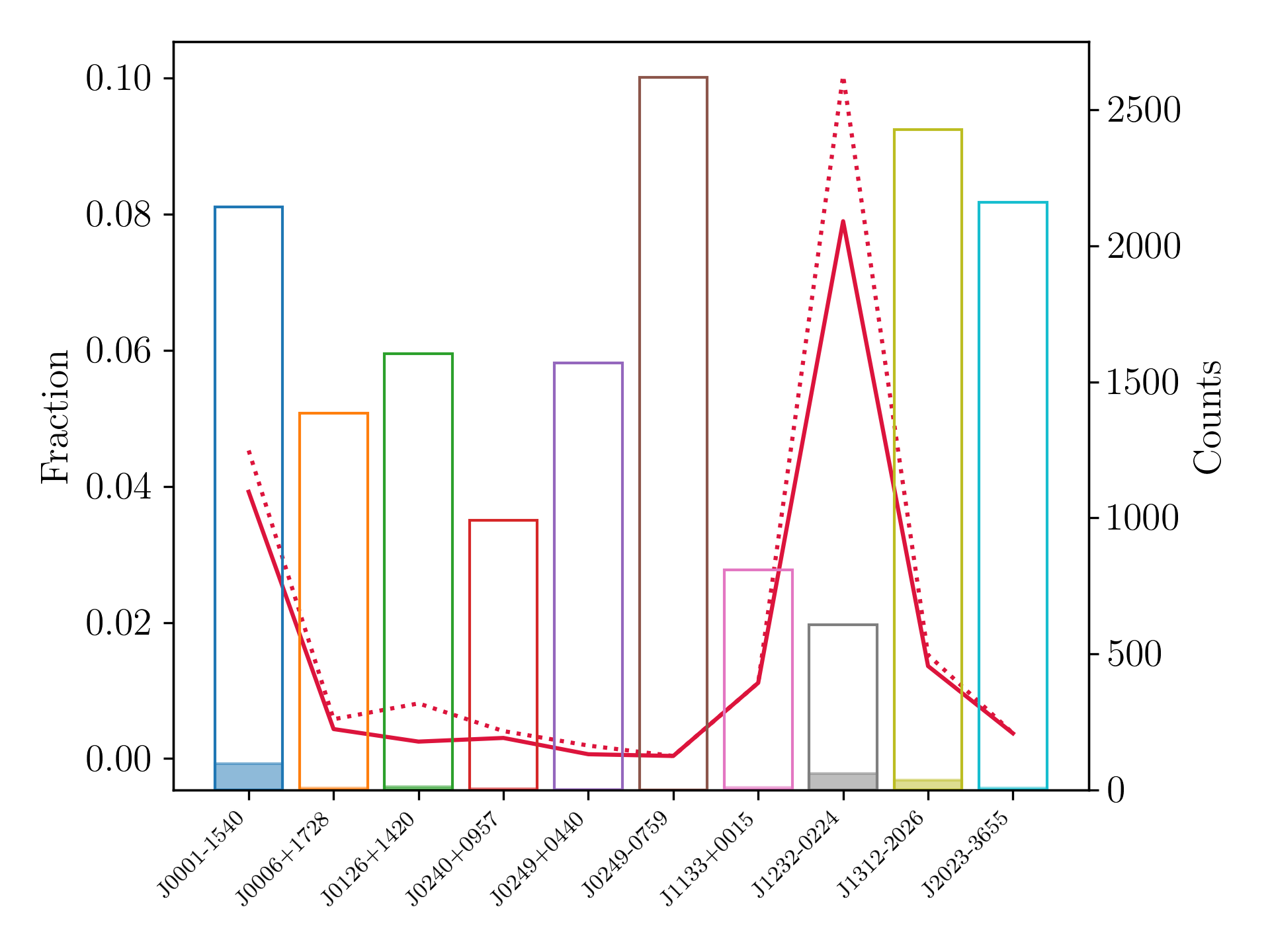}
    \caption{Purity of catalogues for different pointings. The fraction of false detections is indicated by the red line. The dashed red line indicates the fraction of false detections without flagging artefacts. The open histograms show the number of sources detected in the pointings, with the filled histograms indicating the number of false detections.}
    \label{fig:purity_pointings}
\end{figure}

The purity, or inversely the false detection rate, measures what fraction of the sources detected in the image are true detections. For a well chosen detection threshold, the amount of false detections in an image is expected to be small. It is important to have a handle on the amount of false detections, as it should be taken into account when calculating number counts. To determine the purity, we invert the pixel values of the images and run \textsc{PyBDSF} on the inverted images, using the RMS and inverted mean maps determined by \textsc{PyBDSF} from the original image. This again ensures that source finding is performed in the same way as on the original image. Since all real sources have positive flux density the only sources detected in the inverted images will be false detections. These false detections broadly fall into three categories, which we differentiate as noise peaks, (calibration) artefacts , and ghost sources. Noise peaks are statistical outliers of noise and can therefore appear at any point in the image, and are symmetric around the mean, such that these sources detected in the inverted image correspond roughly to the false detections in the normal image. Artefacts are sidelobes found around strong sources in the image, making them more easily traceable. As described in Section~\ref{sec:catalogs}, we consider a source to be an artefact if they are found within 5 times the major axis of the clean beam of the ten brightest sources in the image, and have less than 10\% of the peak flux density of the bright source. As the brightest negative sidelobe of the PSF is in general twice as bright as the brightest positive sidelobe, we would expect more artefacts to be found in the negative image. This seems consistent with the data, as using this criterion for artefacts flags 44 of the 241 sources found in the inverted images, while flagging 22 sources in the pointing catalogues. Finally there are ghost sources, which appear as negative sources too bright to be noise fluctuations, in some cases even strong enough that they have sidelobes that are detected as sources. These sources can be caused by calibration with an incomplete sky model, and only have faint positive counterparts \citep{Grobler2014}. Strong ghosts can add to the number of false detections with their sidelobes, but only a handful of such cases are seen in the images.

We plot the amount of false detections per pointing in Figure~\ref{fig:purity_pointings}, both in terms of absolute counts (coloured bars) and fraction (red line). The amount of false detections strongly depends on pointing, and we find two pointings which are most strongly affected: J0001-1540 and J1232-0224. The latter is affected by a strong central source, which leads to reduced number counts and an increased fraction of false detections, while the former would be considered a good pointing, both in terms of number counts and noise properties. There does appear however a cluster of relatively bright (10-100 mJy) sources present far from the pointing centre, which can contribute to noise. The presence of a number of strong sources far out in the field has also potentially affected self-calibration, as a high number of ghost sources are seen in the image. This result suggests that purity of any individual pointing is not always easily predictable, and should each be assessed separately. 

\section{Combined Catalogue}
\label{sec:final_catalog}

The combined source catalogue of ten pointings contains 16,313 sources, and covers 35.7 square degrees of sky. In the previous section we have mostly assessed the quality of individual pointings. Here we combine the catalogues of the individual pointings to increase statistical power, which allows us to investigate more subtle systematic effects that affect all pointings. 

\subsection{Correcting residual primary beam effects}
\label{sec:corrections}

In Section \ref{sec:pb_correction} we described primary beam corrections to both the flux densities and spectral indices in the images. Besides the main wideband primary beam corrections using holographic images, we also described corrections with simplified analytic forms. Before investigating the difference between these methods, we must make additional corrections to residual primary beam effects. In general, the simplified analytical corrections work well up to the FHWM of the primary beam, but further out results begin to diverge. This is an effect that is seen in both the spectral indices as well as the flux densities of sources, mainly caused by using the primary beam correction based on the central frequency $\nu_0 =$ 1.27~GHz for the entire bandwidth of 802.5~MHz. In order to take into account the contribution of the entire bandwidth, we recalculate the corrections by integrating over the bandwidth rather than assuming the frequency to be equal to $\nu_0$. The necessary corrections are computed for each source in the catalogue separately, depending on distance from the pointing centre and spectral shape.  

In the images corrected by the simplified analytical function of Equation~\ref{eq:pb_correction}, flux densities of sources appear higher the further they are from the pointing centre. In order to properly correct for the effect in flux density, the spectral index of the source must be known or assumed. For reasons explored in Section \ref{sec:spectral_indices}, we can not trust all spectral indices to be accurate and perform this correction assuming $\alpha=-0.75$ for all sources. We then calculate a correction factor for the flux densities as a function of distance from the pointing centre. The assumed primary beam model is as before (Equation \ref{eq:pb_correction}), and the correction is computed by integrating the primary beam over the frequency range of the band. Considering a source with some spectral index $\alpha$, the flux density of the source is described by $S(\nu,\alpha) \propto \nu^{\alpha}$. Due to the effect of the primary beam, the flux density of the source has some attenuation factor $a(\rho,\alpha)$ applied to it. This factor is described by the primary beam:
\begin{equation}
    a(\rho,\alpha) = \frac{\int_{\Delta\nu} S(\nu,\alpha) P(\rho,\nu) \mathrm{d}\nu}{\int_{\Delta\nu} S(\nu,\alpha) P(0,\nu) \mathrm{d}\nu}.
    \label{eq:flux_resid_corr}
\end{equation}
Since the flux densities have already undergone primary beam correction, we need to correct for the ratio between this term and the correction from Equation \ref{eq:flux_resid_corr},
\begin{equation}
    S_{corr} = \frac{P(\rho,\nu_0)}{a(\rho,\alpha)} S_{measured}.
\end{equation}
The effect of this correction should become visible when comparing flux densities to external catalogues. If the flux densities are properly corrected, the flux density ratio between catalogues should be constant as a function of distance to the pointing centre.

In contrast to the flux densities, both the analytical function from Equation~\ref{eq:alpha_corr} and the wideband primary beam correction from Equation~\ref{eq:wb_pb} leave a residual effect in the spectral indices of sources further from the pointing centre. Taking Equation \ref{eq:alpha_corr} to describe the spectral index induced by the primary beam variation, we correct these values taking the full bandwidth into account, recalculating the effect of the primary beam on spectral indices by integrating over the bandwidth $\Delta\nu$:
\begin{equation}
    P_{\alpha,int}(\rho) = -8\log(2)\left(\frac{\rho}{\theta_{pb}}\right)^2 \int_{\Delta\nu} \left(\frac{\nu}{\nu_0}\right)^2 \mathrm{d}\nu.
\end{equation}
To correct the already measured spectral indices present in the catalogues we subtract the difference between the integrated and original primary beam correction term,
\begin{equation}
    \alpha_{corr} = \alpha_{measured} + [P_{\alpha,int}(\rho) - P_{\alpha}(\rho, \nu_0)].
\end{equation}

\subsection{Cross matching catalogues}
\label{sec:cross_match}

To further investigate systematics that affect the pointings on a more general level, we continue the assessment from Section~\ref{sec:calibration_assessment}, now using the sources of the entire field. We cross check our sources with their counterparts from NVSS and RACS, as all our pointings here are within the sky coverage of these two surveys. Cross matching is performed by checking whether source ellipses, defined by the $3\sigma$ extent of the Gaussians describing these sources, overlap between the catalogues. We require a minimum overlap in area of 80\% to consider sources to be a match. Sources in one catalogue can be matched with any number of sources in the other, to account for different resolutions between the catalogues. Due to uncertainties in position and flux density near the NVSS detection threshold of 2.5 mJy, sources below 5 mJy in NVSS are not considered. We find that 997 sources are matched to NVSS, of which 845 are matched to a single source, and 2064 sources are matched to RACS, with 1949 matched to a single source.

\begin{figure}
    \centering
    \includegraphics[width=\hsize]{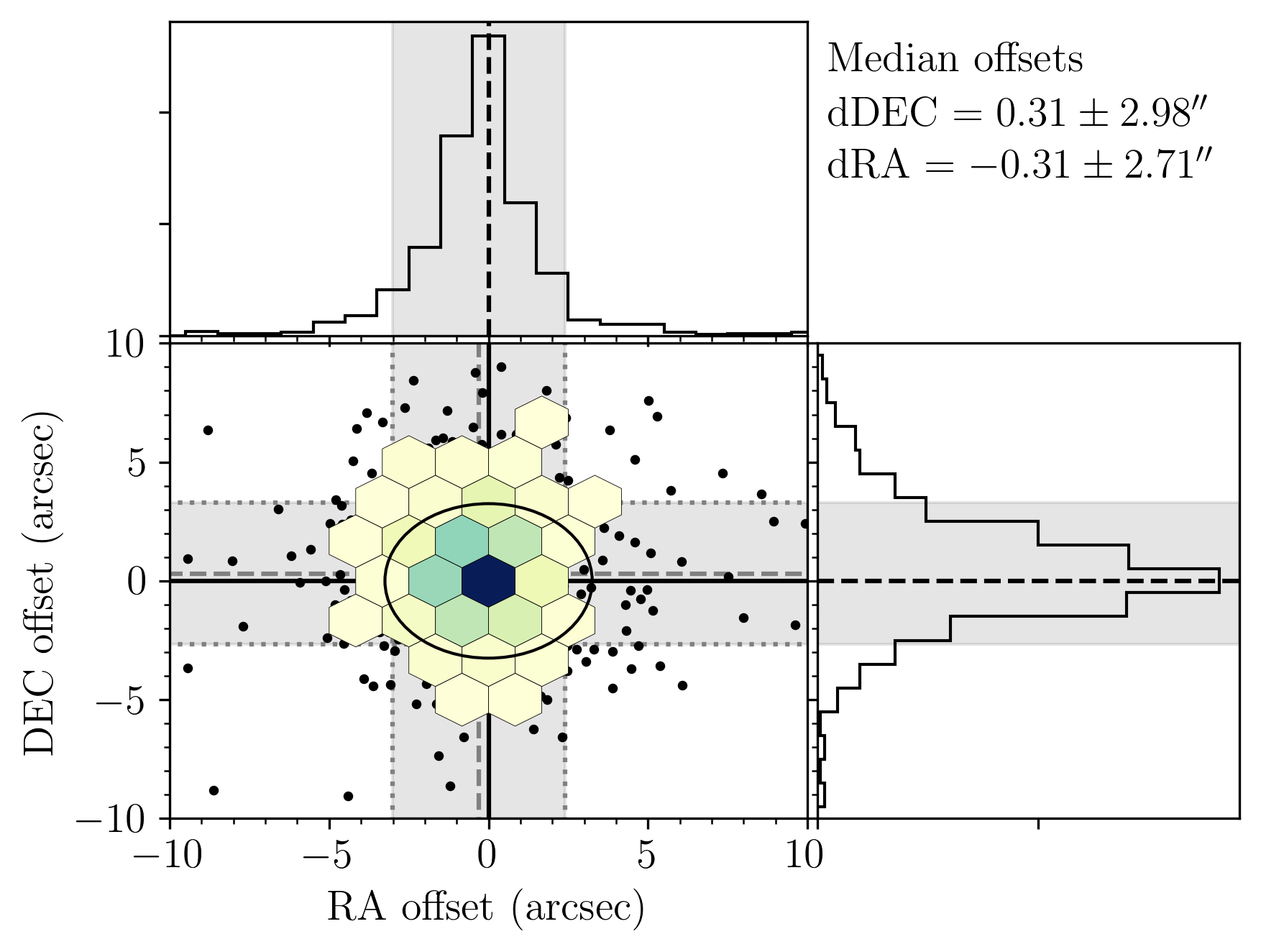}
    \caption{Astrometric offsets to NVSS for all ten pointings combined. The median offsets are given by the grey dashed lines, with the grey area indicating the uncertainty. The majority of sources lie within a FWHM of the average minor axis of the clean beam. The data is binned where five or more sources occupy the defined bin area, otherwise individual sources are shown.}
    \label{fig:full_astro}
\end{figure}

There are a number of factors that can influence astrometric precision of an observation, such as errors in the reference frequency or timestamps. Some of these errors were present in earlier MeerKAT observations \citep[e.g.][]{Mauch2020, Villiers2022}. While the errors should no longer be an issue, it is important to cross check positions in the field with an external catalogue for potential astrometric errors. The astrometric offsets of sources to their NVSS counterparts can be seen in Figure \ref{fig:full_astro}, where the offsets are shown only for single matched sources. Overall, the offsets are very small with a median offset of $\sim0.3$\arcsec, which is less than 1/6 of the image pixel size (2\arcsec) and well within the uncertainty. The scatter in both directions is smaller than 3\arcsec, which is less than the semi-minor axis of the average clean beam of 3.25\arcsec, also shown on the figure.

\begin{figure}
    \centering
    \includegraphics[width=\hsize]{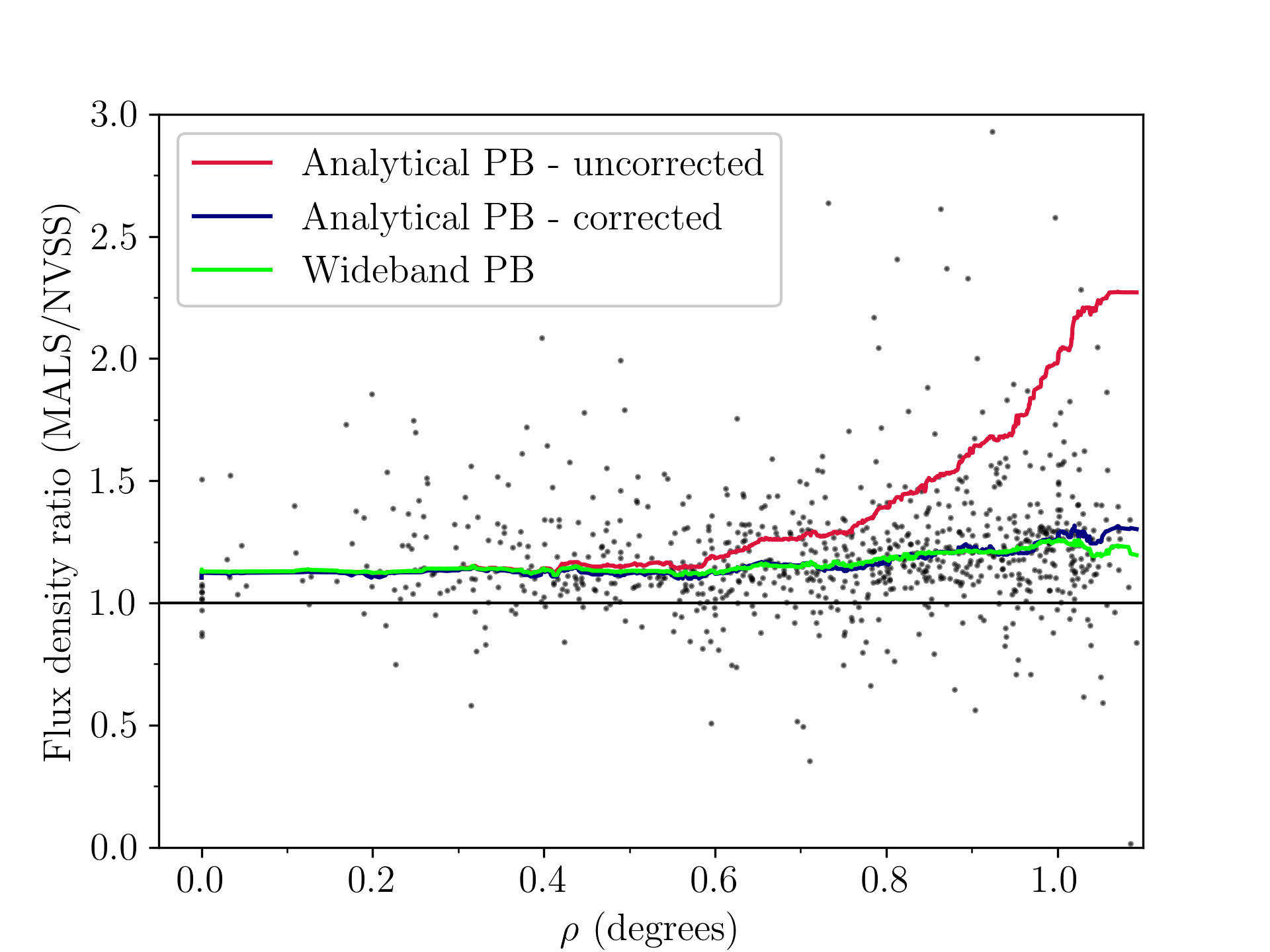}
    \caption{Ratio of flux densities of the sources in MALS compared to their NVSS counterparts as a function of distance from the pointing centre ($\rho$). The running median flux density ratio of the analytical primary beam correction both with (blue line) and without (red line) the corrections made in Section \ref{sec:corrections} are shown, as well as the running median flux density ratio of the holographic wideband primary beam correction (green line).}
    \label{fig:NVSS_flux}
\end{figure}

\begin{figure}
    \centering
    \includegraphics[width=\hsize]{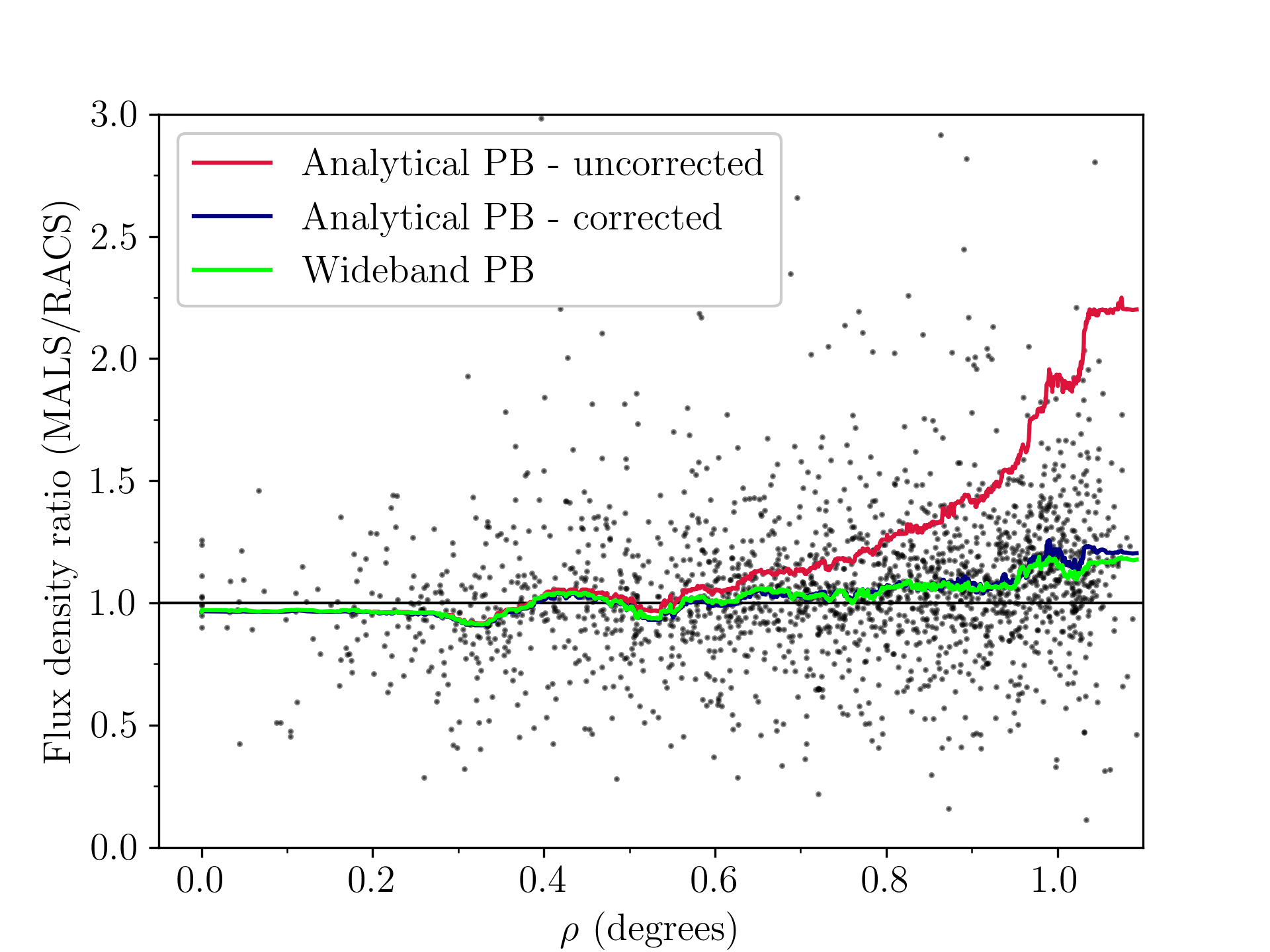}
    \caption{Ratio of flux densities of the sources in MALS compared to their RACS counterparts as a function of distance from the pointing centre ($\rho$). The running median flux density ratio of the analytical primary beam correction both with (blue line) and without (red line) the corrections made in Section \ref{sec:corrections} are shown, as well as the running median flux density ratio of the holographic wideband primary beam correction (green line).}
    \label{fig:RACS_flux}
\end{figure}

In Section~\ref{sec:corrections} we corrected spectral indices and flux densities accounting for residual effects introduced by the frequency range covered by the band. Cross matching sources with external catalogues is an important check of the correctness of their measured flux densities. Figure~\ref{fig:NVSS_flux} shows the flux density ratio of 845 MALS and NVSS sources, and Figure~\ref{fig:RACS_flux} shows the flux density ratio of 1949 MALS and RACS sources. Only sources that are matched to a single source are used, and flux densities have been converted to the MALS rest frequency (1.27 GHz) assuming $\alpha=-0.75$. In both Figures the corrections for the residual primary beam effect have properly re-scaled the flux densities, as the median flux density ratio (blue line) stays largely consistent with distance from the pointing centre, but a residual effect is left towards the edges of the image. In Figure~\ref{fig:NVSS_flux} it stands out immediately that there is a systematic offset between the MALS and NVSS flux densities, as MALS flux densities are 18\% higher on average. The overall flux density scale offset is $S_{MALS}/S_{NVSS} = 1.18\pm0.26$. In contrast to NVSS, Figure~\ref{fig:RACS_flux} shows that the flux densities between MALS and RACS agree extremely well up to $\rho\sim0.5$ degrees. The overall flux density scale offset is $S_{MALS}/S_{RACS} = 1.06\pm0.39$, with the 6\% overall offset originating mostly from the outer parts of the primary beam.

Though the result from NVSS might indicate any number of problems that could cause the offset, the additional data from RACS rules out most of these assumptions. A likely source of uncertainty would be the assumption of spectral index, however this would impact the RACS results far more significantly, with its rest frequency of 887 MHz. From RACS' flux density offset we can assume that the 6\% offset stems mostly from the residual primary beam effect, but this can only explain part of the offset seen in NVSS. If this offset is persistent, it points to a systematic effect affecting either NVSS, or both MALS and RACS. Due to the relatively low sensitivity of the surveys, only about 10\% of MALS sources are matched to a counterpart, which makes the error bars on the flux density offset measurement rather large. As such, the measured offset is within the uncertainty, preventing us from making any definitive statement on the flux density offset. Combined with the measurement from Section~\ref{sec:calibration_assessment}, the flux density scale of MALS does not currently significantly deviate from the expected value, but the offset seen here indicates that more data is needed.

\subsection{General assessment of the complete catalogue}
\label{sec:general_assessment}

\begin{figure*}
    \centering
    \includegraphics[width=\textwidth]{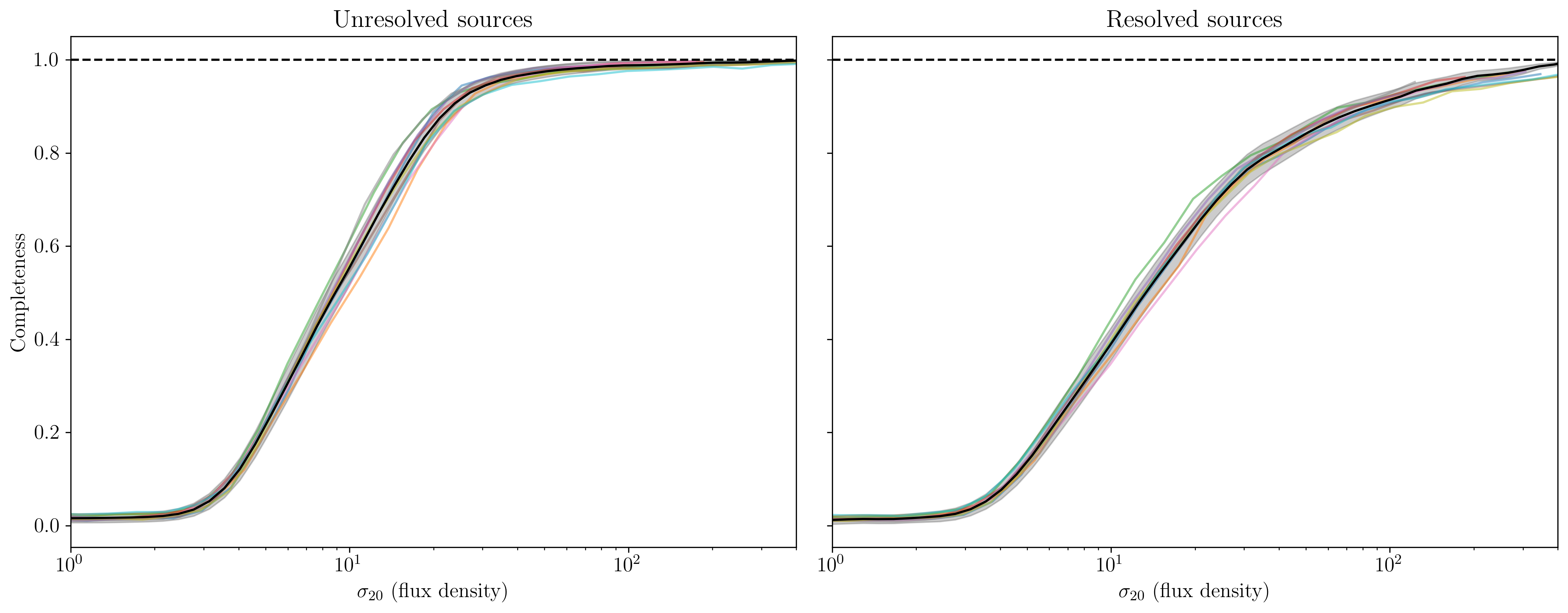}
    \caption{Completeness as a function of flux density for unresolved (left) and resolved (right) sources for the different fields, refactored with $\sigma_{20}$ and combined (black curves). Refactoring the completeness curves to $\sigma_{20}$ shows clearly that they are simply shifted with respect to each other, and we can define a unified completeness measure for the survey as a function of $\sigma_{20}$ for both resolved and unresolved sources.}
    \label{fig:completeness_refactor}
\end{figure*}

\begin{figure*}
    \centering
    \includegraphics[width=0.48\textwidth]{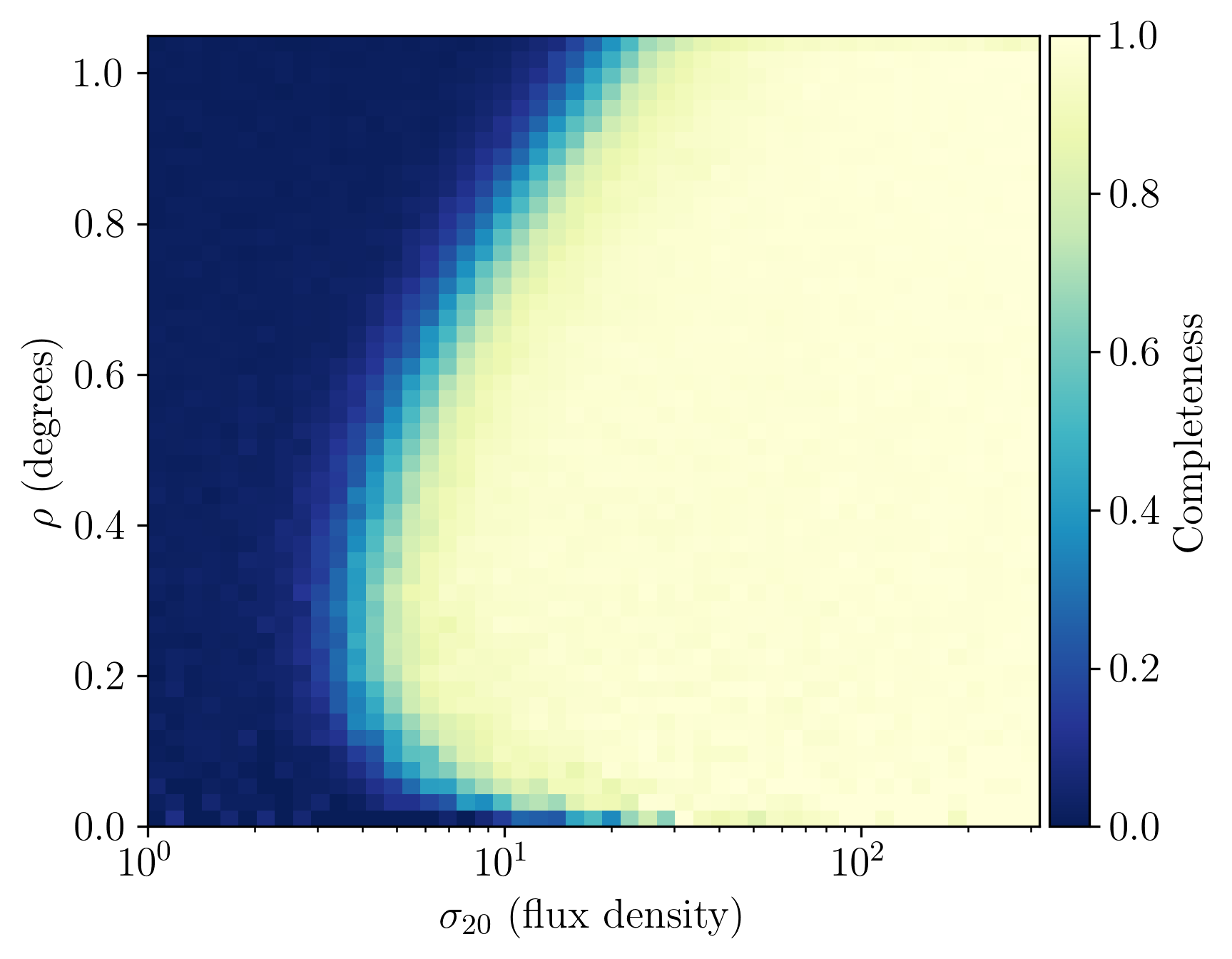}
    \includegraphics[width=0.48\textwidth]{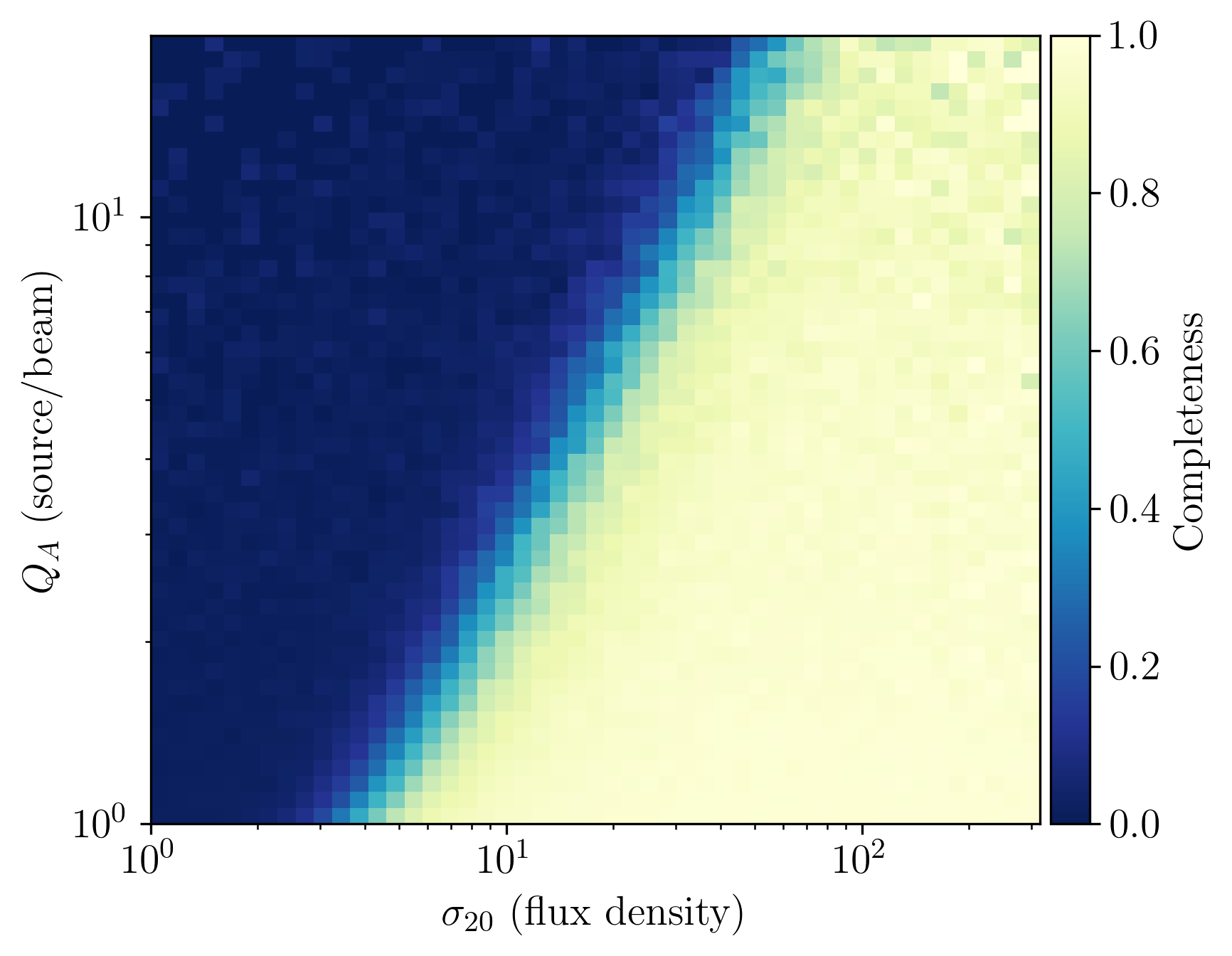}
    \caption{Combined source completeness as a function of distance from the pointing centre (unresolved sources, left) and major axis of the source (resolved sources, right). The left plot reflects the overall structure of the pointings, and shows that completeness is quite straightforwardly a radially averaged version of the noise structure as shown in Figure~\ref{fig:median_rms}. Note that the flux density is normalised by $\sigma_{20}$. The right plot indicates a clear power law relation between the size of sources and the completeness, where larger sources are on average less complete.}
    \label{fig:maj_sep_completeness}
\end{figure*}

In Section~\ref{sec:sourcefinding} the individual pointings have been evaluated with respect to completeness, purity, and flux density recovery. Here these properties are assessed on the entire catalogue in order to understand the impact of these characteristics on the final data product. 

\subsubsection{Completeness}
\label{sec:combined_completeness}

To statistically determine completeness of the data, we refactor the completeness such that it is consistent between different pointings. In order to achieve this, instead of expressing completeness as a function of flux density in units of Jansky, we show flux density in units of $\sigma_{20}$ as defined in Section~\ref{sec:noise}. As we showed in Section~\ref{sec:completeness}, completeness for point sources scales linearly with local noise, and thus should be 0.2 at $5\sigma_{20}$ for all pointings. Though $\sigma_{20}$ has units of Jy/beam, for point sources integrated flux density and peak flux density are in principle equal, which means in this definition $\sigma_{20}$ has the same value in Jy. Figure~\ref{fig:completeness_refactor} shows that in terms of $\sigma_{20}$, pointings have very similar completeness curves, which allows us to combine the individual pointings and evaluate completeness for the whole survey, as indicated by the black combined completeness curve. 

Combined completeness is also assessed as a function of separation from the pointing centre using only point sources, and as a function of major axis of the source using resolved sources. Both are shown in Figure~\ref{fig:maj_sep_completeness}. Combining the completeness from all the pointings gives enough statistical power to paint a clear picture of how the completeness is dependent on these variables. A clear relation is shown between completeness and distance from the pointing centre. The major difference between individual pointings seems to be the influence of the central source on the completeness. These differences are however extremely well modelled by the radially averaged RMS noise (see Figure~\ref{fig:individual_sep_flux_completeness}), indicating that completeness is related to the local noise. The right plot in Figure~\ref{fig:maj_sep_completeness} shows that there is a power law decrease in completeness for larger sources. This was already suggested in Figure~\ref{fig:individual_size_flux_completeness}, but with the combined catalogues we have enough number counts to fully cover the space.

\begin{figure*}
    \centering
    \includegraphics[width=\textwidth]{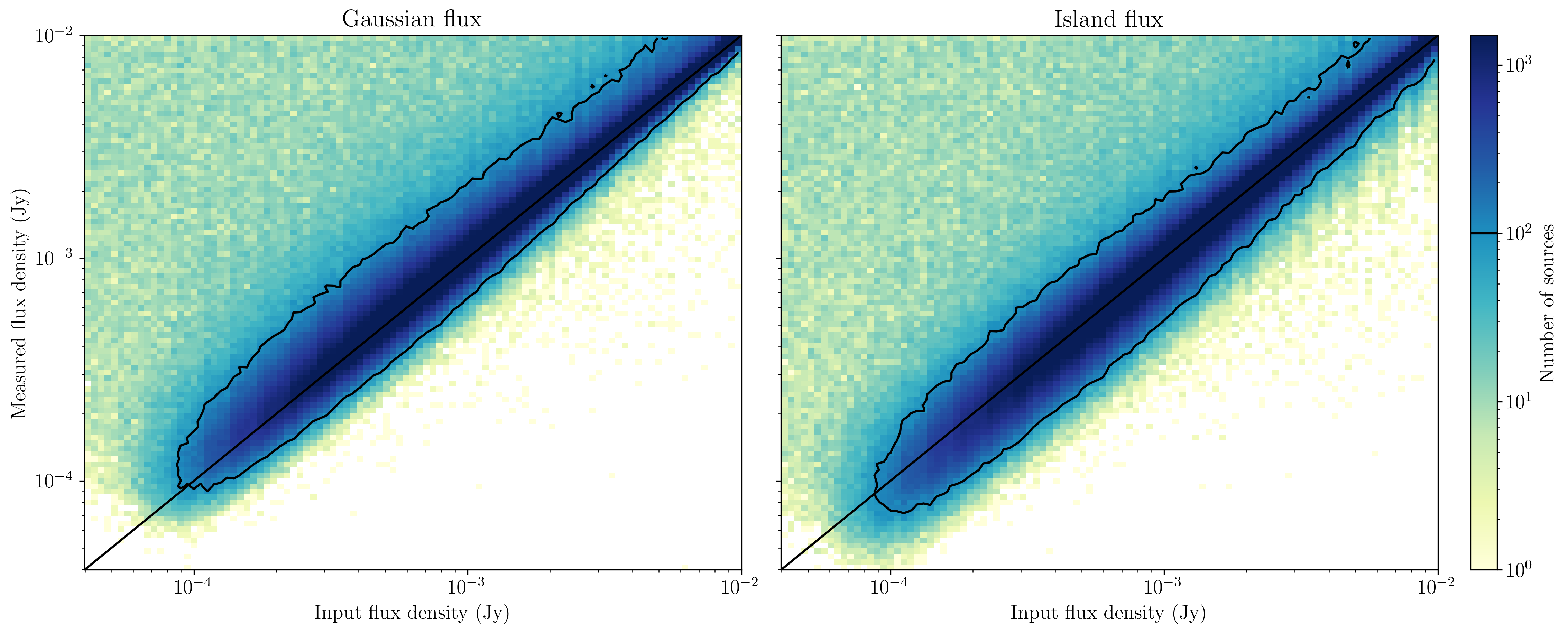}
    \caption{Input flux density plotted against the measured flux density for both Gaussian flux densities (left) and Island flux densities (right). The threshold of 100 sources per bin (black contour) shows quite clearly the bias present in Gaussian flux densities compared to island flux densities.}
    \label{fig:flux_recovery}
\end{figure*}

\begin{figure*}
    \centering
    \includegraphics[width=0.48\textwidth]{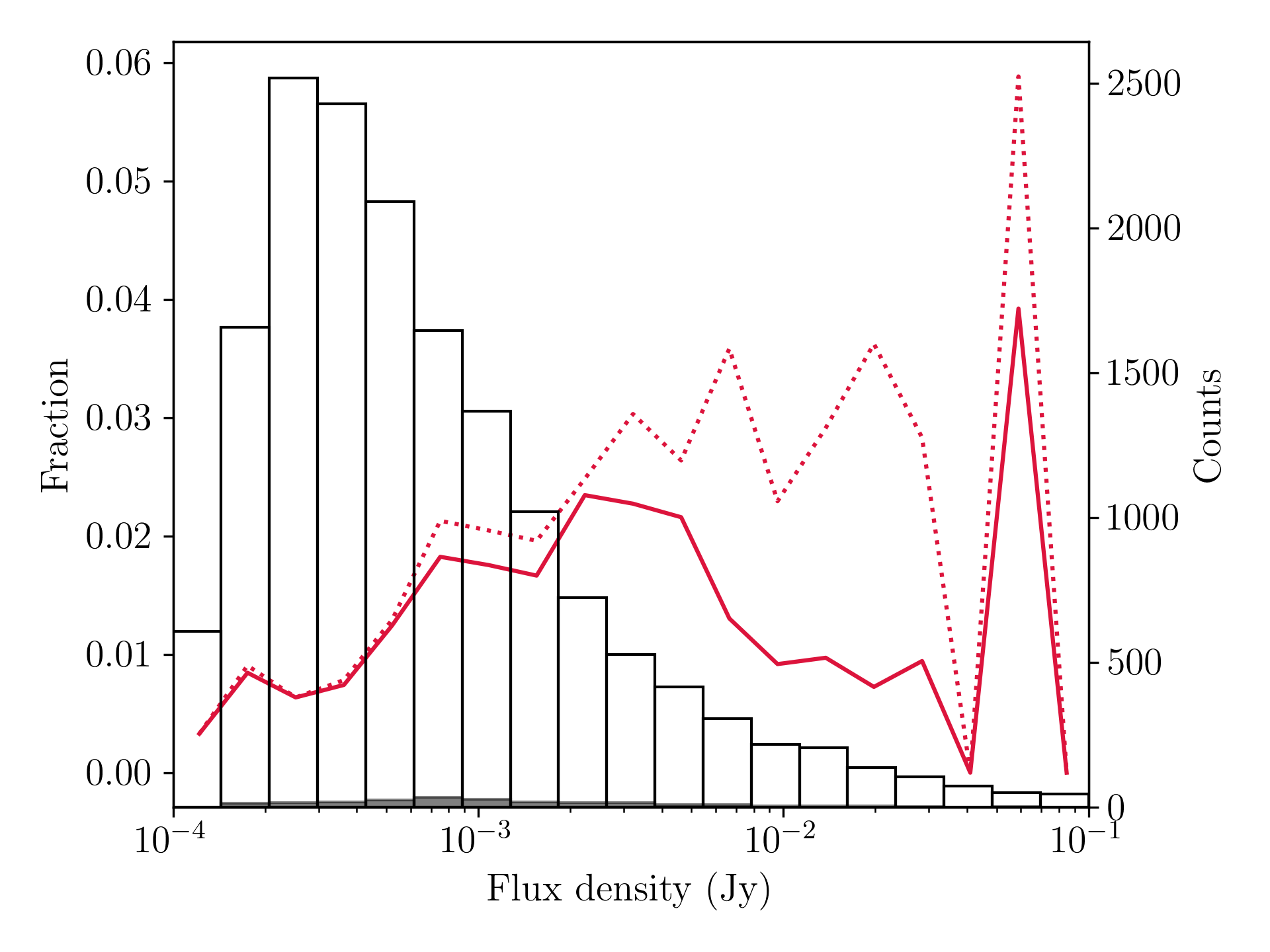}\hfill
    \includegraphics[width=0.48\textwidth]{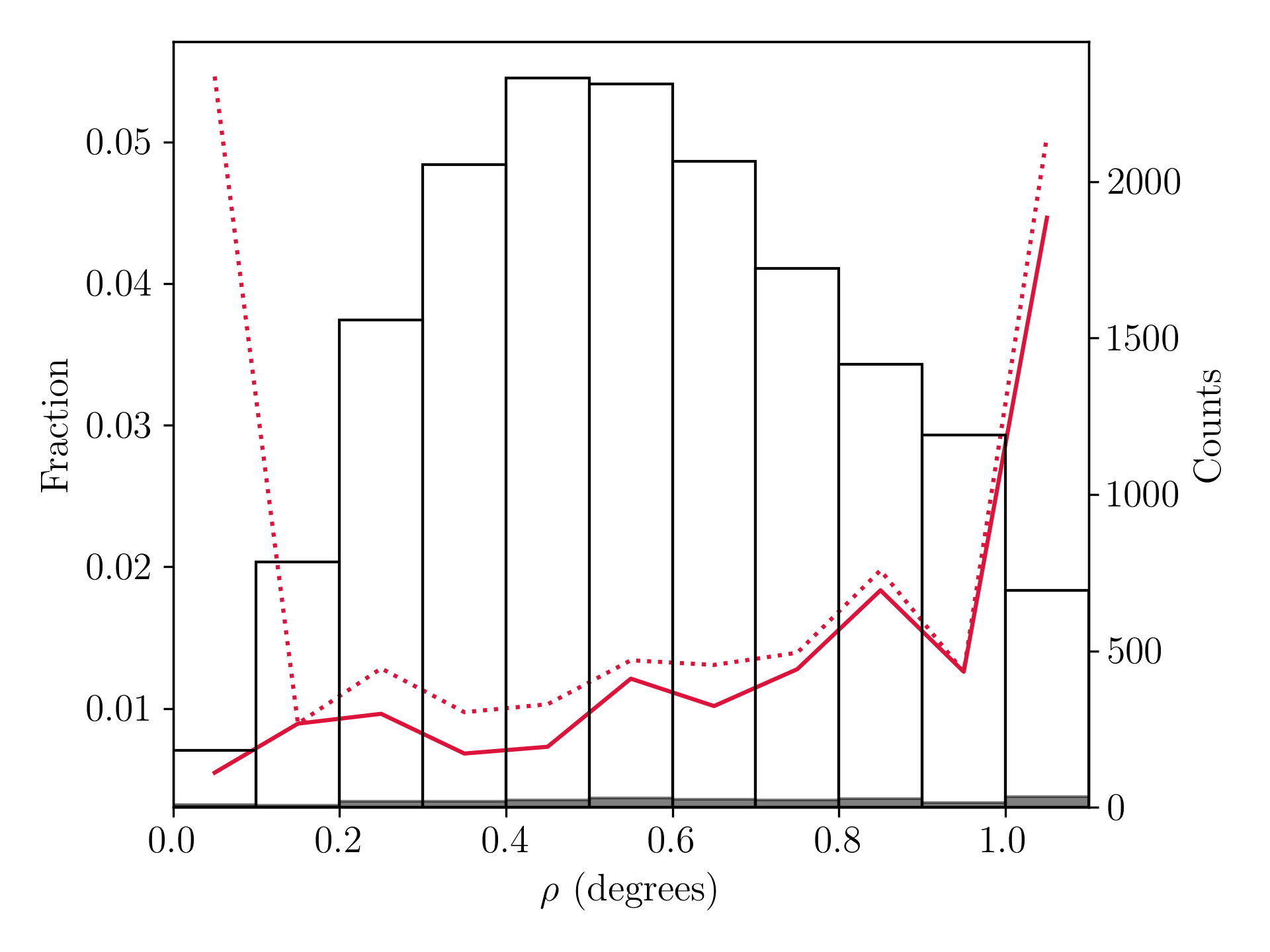}
    \caption{Purity of catalogs as a function flux density (left) and separation from the pointing center (right). The fraction of false detections is indicated by the red line, both the total fraction (dotted line) and with removal of sidelobes (solid line). The open histograms show the number of sources detected in the pointings, with the filled histograms indicating the number of false detections. Though there seems to be no strong relation between flux density and purity, the number of false detections is strongly dependent on distance from the pointing centre, increasing both towards the centre and towards the edges of the pointing. Our criterion for identifying artefacts flags most of the false detections around the central source.}
    \label{fig:purity}
\end{figure*}
 
\subsubsection{Flux recovery}
\label{sec:combined_flux_recovery}

In evaluating the individual pointings in Section \ref{sec:individual_assessment}, faint sources on average had higher measured flux densities compared to input flux densities. To fully assess this effect, we combine the flux recovery statistics from all pointings in Figure~\ref{fig:flux_recovery}. In the combined statistics the effect is more clear, with bins further away from the flux density ratio of unity being occupied with on average 10 sources per bin. There is no visible dependence on flux density or distance from the flux density ratio of unity. Assuming a Poisson distribution of these bins with mean and variance $\lambda=10$, we take all bins with less than 25 sources ($5\sigma$) to be part of this distribution. These bins combined contain 1.7\% of all sources, indicating that this effect is rather small in terms of induced bias. Up to this point we have assumed that the flux density measured from the Gaussian fitting (the \verb|Total_flux| column in the catalogues) best represents the flux density of the sources. In Figure~\ref{fig:flux_recovery} we compare the flux recovery between the Gaussian flux density and the integrated flux density from the island which the source occupies, where the contour indicates the threshold of 100 sources per bin. We see that across the board Gaussian flux densities are skewed towards higher values, where island flux densities remain symmetric around the input flux density. This is an effect that can significantly affect our catalogues, especially considering the increased number counts at lower flux densities. Consequently, in further analysis we will assume that flux densities of sources are more accurately represented by the island flux density.

\subsubsection{Purity}
\label{sec:combined_purity}

Combining false detections from all ten pointings, 241 sources are detected in the inverted images, making up 1.5\% of the combined catalogue. As described in Section~\ref{sec:purity}, our artefact identification criterion flags 44 of these, leaving 197 sources, or 1.2\% of the combined catalogue. With the combined catalogue of false detections, we can investigate how purity is affected by other variables such as flux density and distance from the pointing centre. This will allow us to properly account for the purity of the catalogue, in order to not overestimate number counts. Given the variety of `source types' seen in negative images and what counterparts we expect to see in the positive, the overall amount of false detections should be lower than the amount of sources seen in the negative image. In this sense, the purity is more appropriately an upper limit rather than a direct measure of false detections.

Figure \ref{fig:purity} shows the combined purity as a function of flux density and distance from the pointing centre. The left plot shows the purity as a function of flux density and shows that the fraction of false detections increases with higher flux density. This is largely a result of the overall number of sources decreasing at higher flux density, but does show that the flux density distribution of false detections is not the same as that of real sources. The lack of false detections at low flux densities shows that our $5\sigma$ detection threshold does not lead to a lot of spurious detections. It is noteworthy that more sources are flagged as artefacts at higher flux densities, indicating that artefacts around bright sources have higher flux densities on average. The right plot of Figure \ref{fig:purity} shows a strong dependence of purity on the separation from the pointing centre, similar to the completeness. False detections increase near the central source because of strong artefacts, and there is a steady increase towards the edges of the primary beam. We see that our artefact selection criterion correctly picked out most of the artefacts originating from sidelobes of the central source, which dramatically increases the purity in the central portion of the image. Although the number of false detections restricts the statistical power of these results, the relations already show clear trends for the purity as a function of flux density and distance from the pointing centre that can be used when assessing number counts. 

\subsection{Resolved and unresolved sources}
\label{sec:resolved}

\begin{figure}
    \centering
    \includegraphics[width=\hsize]{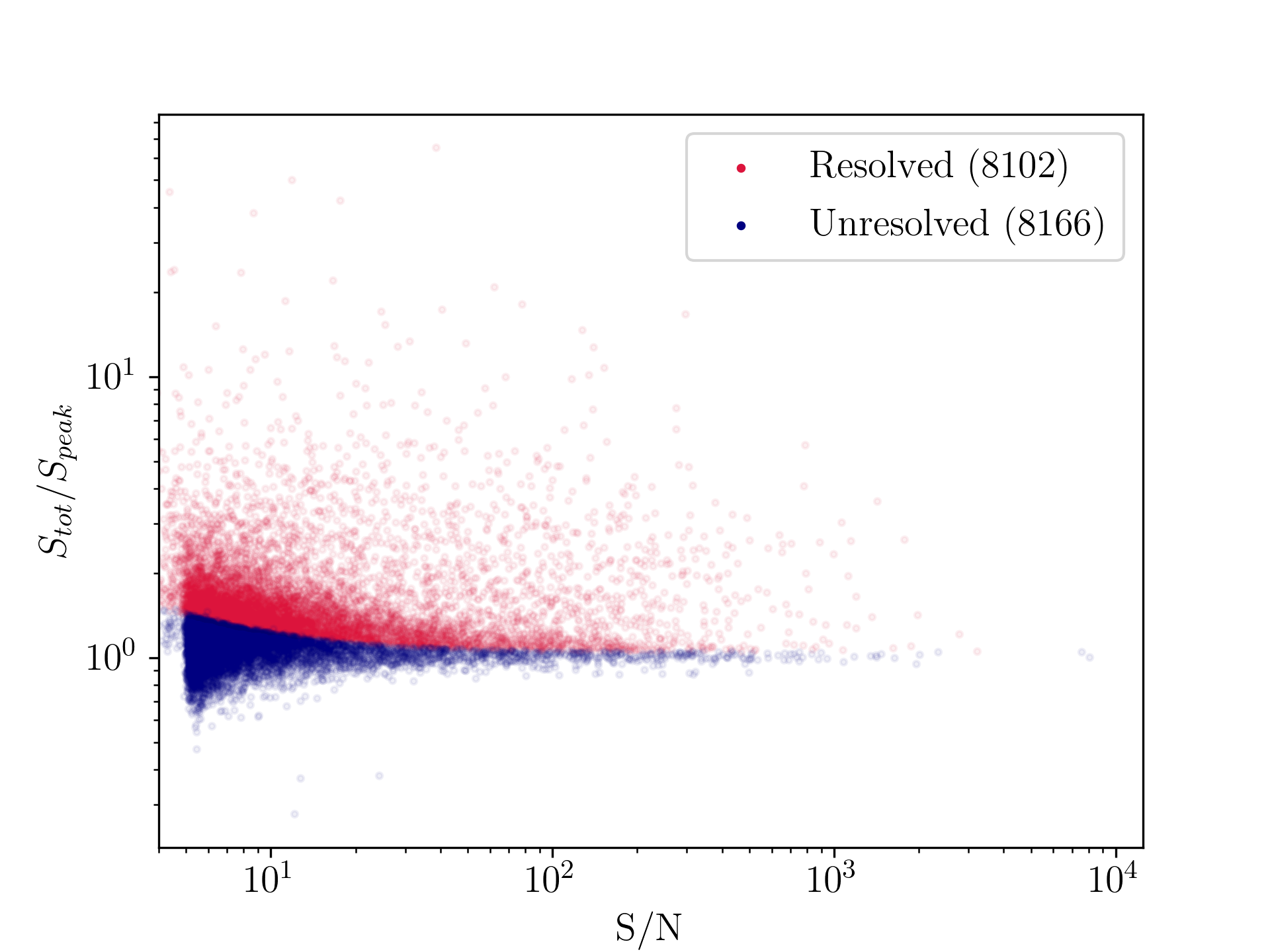}
    \caption{The ratio of total to peak flux as a function of signal to noise of both unresolved (blue) and resolved (red) sources in the combined catalogue.}
    \label{fig:resolved_fraction}
\end{figure}

For the analysis of completeness in Sections~\ref{sec:completeness} and \ref{sec:combined_completeness}, we assumed that our catalogues are populated with both unresolved or resolved sources, and that these should be assessed separately. Figure \ref{fig:pointing_completeness} shows this distinction is warranted, as these sources types have very different completeness relations. If we want to apply this knowledge to real sources in the catalogue, we must have a reliable way of determining whether a source is resolved. We expect sources are resolved when their size exceeds the size of synthesised beam of the image, however we must take into account the uncertainties introduced by noise in the image and fitting errors. 

We determine source size by measuring the ratio between integrated flux density $S$ and peak flux density $S_{peak}$ of the source, which should be equal to one for an unresolved source. Figure~\ref{fig:resolved_fraction} shows $S/S_{peak}$ as a function of S/N, for both resolved and unresolved sources in our combined catalogue. Here and in subsequent usage of S/N we define it as the ratio between the peak flux density of the source and the local RMS. Due to a combination of uncertainties, unresolved sources follow a log-normal distribution in $S/S_{peak}$ \citep{Franzen2015}, and thus a normal distribution in $R = \ln (S/S_{peak})$ with mean 0 and standard deviation $\sigma_R$,
\begin{equation}
    \sigma_R = \sqrt{\left(\frac{\sigma_S}{S}\right)^2 + \left(\frac{\sigma_{S_{peak}}}{S_{peak}}\right)^2}.
\end{equation}
We take both the uncertainties $\sigma_S$ and $\sigma_{S_{peak}}$ as the sum in quadrature of their errors as determined by \textsc{PyBDSF} and a calibration error of 3\%. The magnitude of the error in calibration is motivated by assessing Gaussian fits of bright unresolved sources and the flux density offset determined in Section~\ref{sec:calibration_assessment}. Using these quantities, the compactness criterion is then
\begin{equation}
\ln \left(\frac{S}{S_{peak}}\right) > 1.25\sigma_R.
\end{equation}
The factor of $1.25\sigma_R$ encloses 95\% of sources below $S/S_{peak} = 1$, so with the symmetry of the Gaussian distribution, 95\% of all unresolved sources should be correctly identified with this criterion. As can be appreciated in Figure~\ref{fig:resolved_fraction}, according to this metric, 50\% of all sources in the combined catalogue are resolved.

\subsection{Catalogue columns}
\label{sec:columns}

In the final catalogue, the majority of the columns are preserved from the \textsc{PyBDSF} source catalogues. Additional columns are however added in subsequent steps where required. Our aim is to only add information, and not remove any. This means that, for example, sources can be flagged as artefacts, but will still be present in the catalogue. When performing additional corrections on source flux densities and spectral indices, the correction factors are inserted into the catalogues so that the original values can be easily reproduced. The catalogue has 49 columns in total. Several lines of the catalogue are shown in Table \ref{tab:catalog_lines} as an example.

\begin{itemize}
    \item \verb|Pointing_id| - The ID of the pointing where the source has been found formatted as PT-JHHMM$\pm$HHMM.
    \item \verb|Source_name| - Name of the source, following IAU convention, formatted as JHHMMSS.S$\pm$HHMMSS.S with prefix MALS.
    \item \verb|Source_id| - Source ID as assigned by \textsc{PyBDSF}.
    \item \verb|Isl_id| - Island ID as assigned by \textsc{PyBDSF}.
    \item \verb|RA| and \verb|DEC| (and errors) - The J2000 position of the source, defined as the centre of the composite Gaussian of the source, and associated errors.
    \item \verb|Sep_PC| - Distance of the source from the pointing centre.
    \item \verb|Total_flux| (and error)- Total flux density of the source and associated error.
    \item \verb|Flux_correction| - The correction factor for residual primary beam effects on the flux density of the source.
    \item \verb|Peak_flux| (and error) - Measured peak flux of the source and associated error.
    \item \verb|Spectral_index| (and error) - Spectral index of the source, measured from the spectral index image, and associated error.
    \item \verb|Spectral_index_correction| - The correction factor for residual primary beam effects on the spectral index of the source.
    \item \verb|RA_max| and \verb|DEC_max| (and errors) - Position of maximum intensity of the source and associated errors.
    \item \verb|Maj|, \verb|Min|, and \verb|PA| (and errors) - FWHM of the major axis, minor axis and position angle of the source fit by \textsc{PyBDSF} and associated errors.
    \item \verb|DC_Maj|, \verb|DC_Min|, \verb|DC_PA| (and errors) - FWHM of deconvolved major axis, minor axis, and position angle, and associated errors.
    \item \verb|Isl_Total_flux| (and error) - Total integrated flux of the island in which the source is located, and associated error.
    \item \verb|Isl_rms| - Average background RMS noise of the island in which the source is located.
    \item \verb|Isl_mean| - Average background mean value of the island in which the source is located.
    \item \verb|Resid_Isl_rms| - Average residual background RMS noise of the island in which the source is located.
    \item \verb|Resid_Isl_mean| - Average residual background mean value of the island in which the source is located.
    \item \verb|S_Code| - Value generated by \textsc{PyBDSF} indicating whether a source is: fit by a single Gaussian (`S'), fit by multiple Gaussians (`M'), or one of multiple sources on the same island (`C').
    \item \verb|Resolved| - Boolean indicating whether the source is resolved according to the metric defined in Section~\ref{sec:resolved}.
    \item \verb|Flag_Artifact| - Boolean indicating whether the source is a likely artefact according to the criterion described in Section~\ref{sec:catalogs}.
    \item \verb|N_Gaus| - Number of Gaussian components fit to the source.
    \item \verb|RA_mean| and \verb|DEC_mean| - Mean intensity weighted position of all pixels of the island in which the source is located, measured if a source is fit with multiple Gaussians.
    \item \verb|Cutout_Spectral_index| - Intensity weighted average spectral index of all pixels of the island in which the source is located, measured if a source is fit with multiple Gaussian components.
    \item \verb|Cutout_Total_flux| - Total flux density of all pixels above the island threshold, measured if source fit with multiple Gaussian components.
    \item \verb|Cutout_flag| - Flag assigned to cutout in certain conditions: the mean position falls outside the island (`M'), the position of the brightest pixel does not correspond to the maximum position measured by \textsc{PyBDSF} (`C'), the difference between \verb|Cutout_total_flux| and \verb|Isl_Total_flux| is more than 20\% (`F').
    \item \verb|Cutout_class| - Classification assigned at visual inspection as described in Section~\ref{sec:catalogs}, indicating whether a source is well described by the Gaussian model (`G'), is better described by the island characteristics (`I'), is better described by a single Gaussian component (`P'), or an artefact (`A').
\end{itemize}

\section{Source characteristics}
\label{sec:science}

When considering source counts in the radio regime, extra care must be taken in understanding the population of sources that is being probed. Depending on observing frequency and flux density, different source populations may appear in the sample. Our reference for source counts are the SKADS simulations, as the simulated sample is built up by different source populations. Given the theoretical noise limit of 10~\textmu Jy/beam, we can expect to detect sources down to 50~\textmu Jy. As shown in \citet{Wilman2008}, the radio population is dominated by AGN above flux densities of 1~mJy, while below that star-forming galaxies start to make up a significant fraction of the source counts. There are several important distinctions between these source types that can influence source counts and a dipole measurement. Among them is source morphology, as multi-component sources can easily be mistaken for multiple separate sources, biasing number counts. Star-forming galaxies are primarily found at lower flux densities and can be morphologically described by a single component, such that we expect them to appear as faint point sources in our fields. Consequently, these sources can be easily counted as they are unambiguously unique sources and are thus statistically independent. At higher flux densities however, some sub-classes of AGN, such as Fanaroff-Riley type I (FRI, core-dominated) and type II (FRII, lobe-dominated) sources \citep{Fanaroff1974}, can boast extended structures that can complicate automated source finding methods. 

\subsection{Extended sources in \textsc{PyBDSF}}
\label{sec:ext_sources}

\begin{figure}
    \centering
    \includegraphics[width=\hsize]{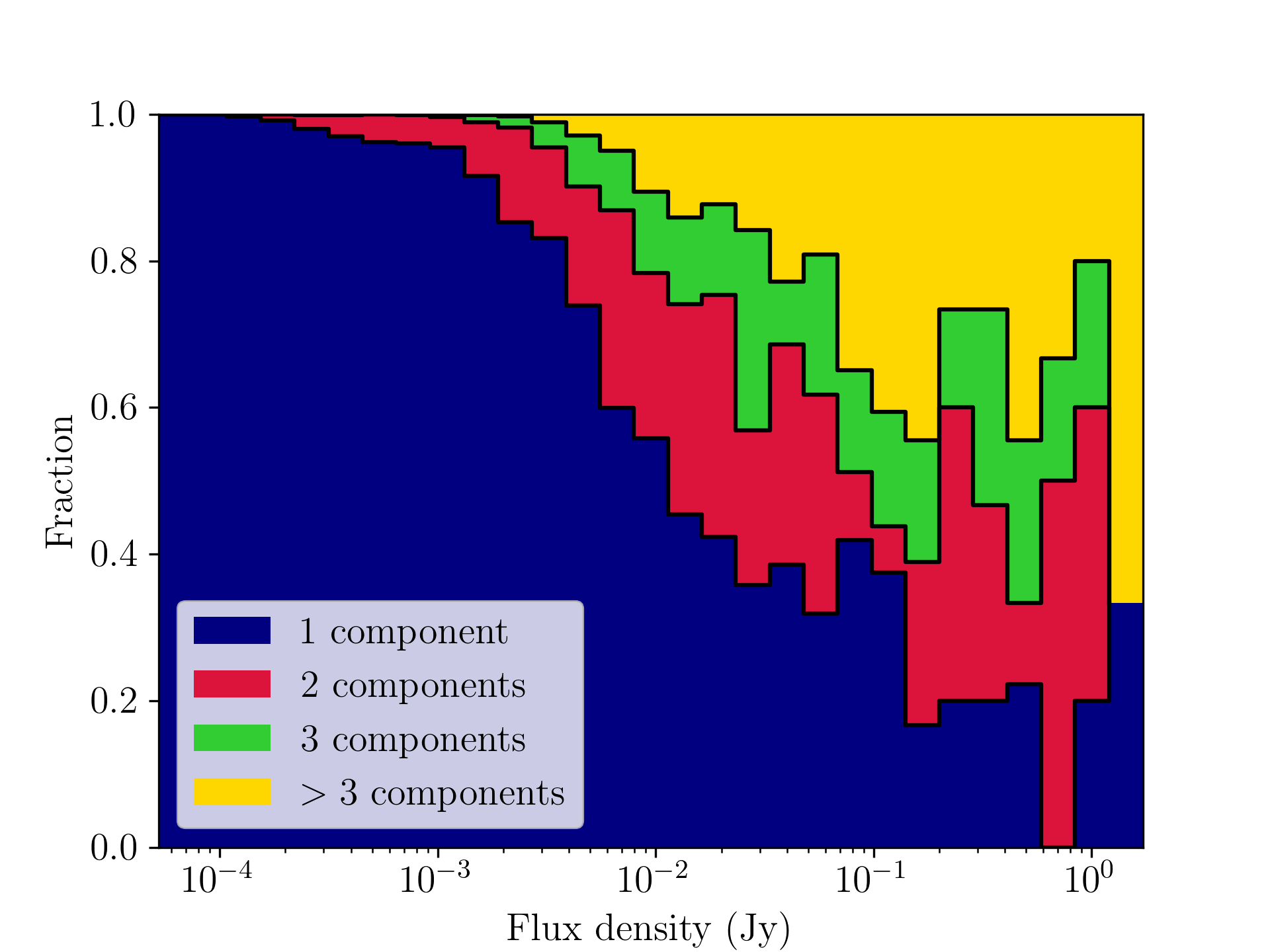}
    \caption{The fractions of sources fit with varying numbers of Gaussians as a function of flux density. The number of components fit to sources increases steadily towards high flux density, but flattens out around 100 mJy, which can be caused by sources splitting up at these flux densities.}
    \label{fig:ngaus}
\end{figure}

\textsc{PyBDSF} operates by fitting Gaussians to sources, which is an effective method for most radio sources, but breaks down in sources with more complex structure. \textsc{PyBDSF} offers multiple ways of improving the fit to extended sources, as specified in Section~\ref{sec:sourcefinding}, but this chiefly improves detection of extended and more diffuse sources. To ensure that complex sources are accurately fit by a combination of Gaussians, we employ a special recipe for these types of sources, which make up 8\% of the all sources found in the fields. These sources are flagged in our workflow for visual inspection upon which the Gaussian fit is assessed as described in Section~\ref{sec:catalogs}. 

However, automated source finding algorithms will only recognise objects as a single source if they are closely connected, and will therefore fail on a subset of sources. This effect is strongest for FRII sources, as increased luminosity in the lobes makes them appear as separate radio sources. Source association is one of the outstanding problems in radio astronomy, and beyond the scope of this paper. Instead, we try to characterise this effect and how it might influence source counts. As the components of these sources are not statistically independent, they will naturally bias source counts. To get an estimate of how complex sources are fit by \textsc{PyBDSF}, we look at the number of components fit to sources as a function of flux density in Figure~\ref{fig:ngaus}. We see a steady increase in the number of Gaussian components at higher flux densities, however at $\gtrsim100$ mJy the increase in components flattens out. This may indicate that at these flux densities we are seeing all the emission from these sources, and this is the `true' distribution of components. However, an alternative explanation is that at these flux densities some extended sources no longer have connecting emission and are not recognised as single sources anymore. In this case the number of components keep increasing, but individual components will split off and be detected as separate sources, effectively keeping the number of components per source the same. Another indication of this happening might be seen in the differential source counts in Figure \ref{fig:diff_counts}, where there is an excess in source counts above 100 mJy, relative to the expected values. 

To get an alternative measure of this, we look at FRII galaxies in SKADS and the separation of their components. For an upper limit estimate on how many sources we expect to split up, we count components as separated when the distance between them exceeds 6.5\arcsec, which is the minor axis of the average clean beam. With this, 11\% of sources in the range 10 mJy - 1 Jy have two separated components. Furthermore, 6.3\% and 24\% of sources in the ranges 10-100 mJy and 100 mJy-10 Jy respectively have three separated components. Outside of these ranges the fractions are negligible. Doubling the distance required for separation mostly exchanges the amount of triple component sources for double component sources in the range 100 mJy - 1 Jy. The amount of triple component sources does not change in the range 1-10 Jy, indicating that the brightest sources are also the largest and most likely to separate. If we define the excess fraction of sources detected as $f_n = \frac{\tilde{n}-n}{n}$, where $\tilde{n}$ is the amount of sources detected counting separate components and $n$ the actual amount of sources, $f_n=0.4$ at 10-100 mJy, $f_n=0.9$ at 100 mJy-1 Jy, and $f_n=0.7$ at 1-10 Jy. The values of $f_n$ are given for each of the flux density bins used to determine number counts in Table~\ref{tab:diff_counts}. The maximum value $f_n$ can take is 2, when a bin is entirely occupied by sources with three separate components. From this we can conclude that this effect is more important at higher flux densities, and is most significant at flux densities $\gtrsim 100$ mJy, which contains only a very tiny subset (102 sources, 0.6\%) of the full catalogue.

\subsection{Spectral indices}
\label{sec:spectral_indices}

\begin{figure*}
    \centering
    \includegraphics[width=\textwidth]{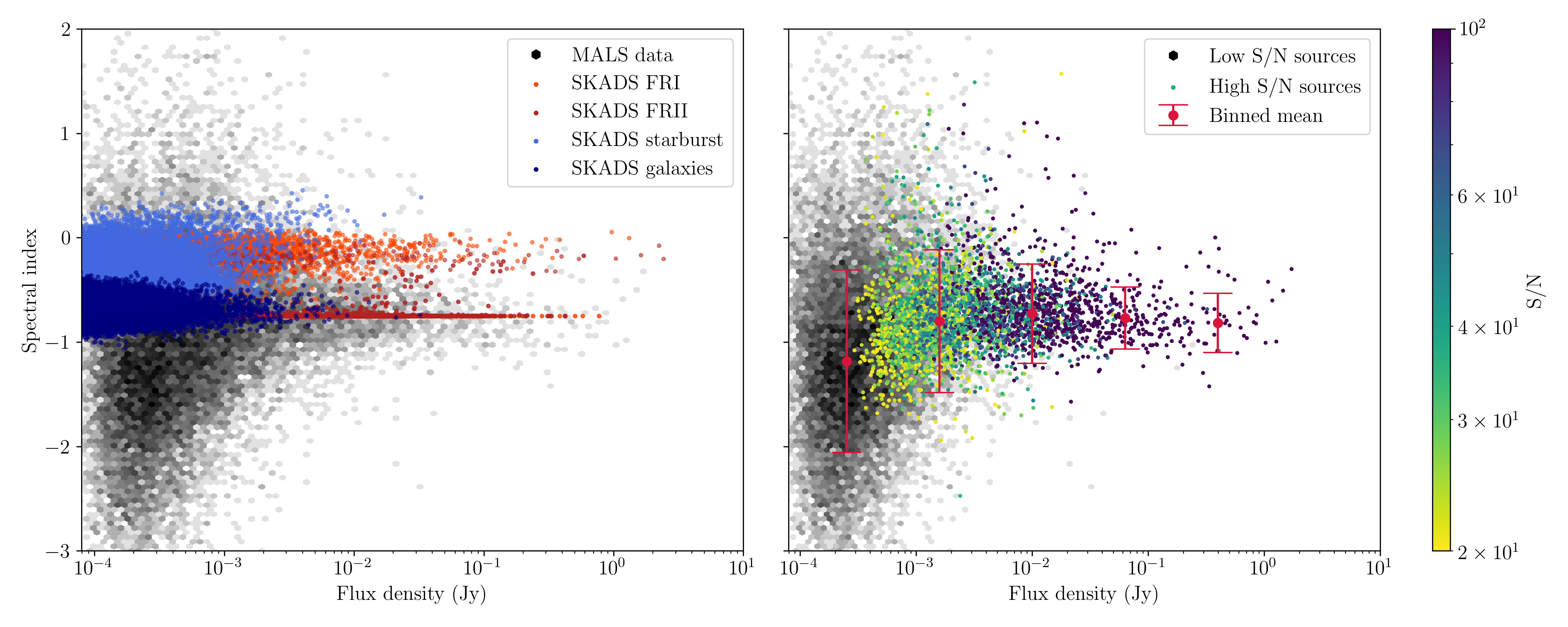}
    \caption{Distribution of spectral indices of MALS sources. Left: MALS spectral indices (black) compared to AGN (red) and star-forming galaxies (blue) from SKADS as a function of flux density. Right: MALS spectral indices as a function of flux density, sources with S/N above 20 are coloured by S/N. The median value and error of spectral indices of different flux bins are indicated by red error bars, indicating that at lower flux densities spectral indices tend towards lower values.}
    \label{fig:alpha_flux}
\end{figure*}

\begin{figure}
    \centering
    \includegraphics[width=\hsize]{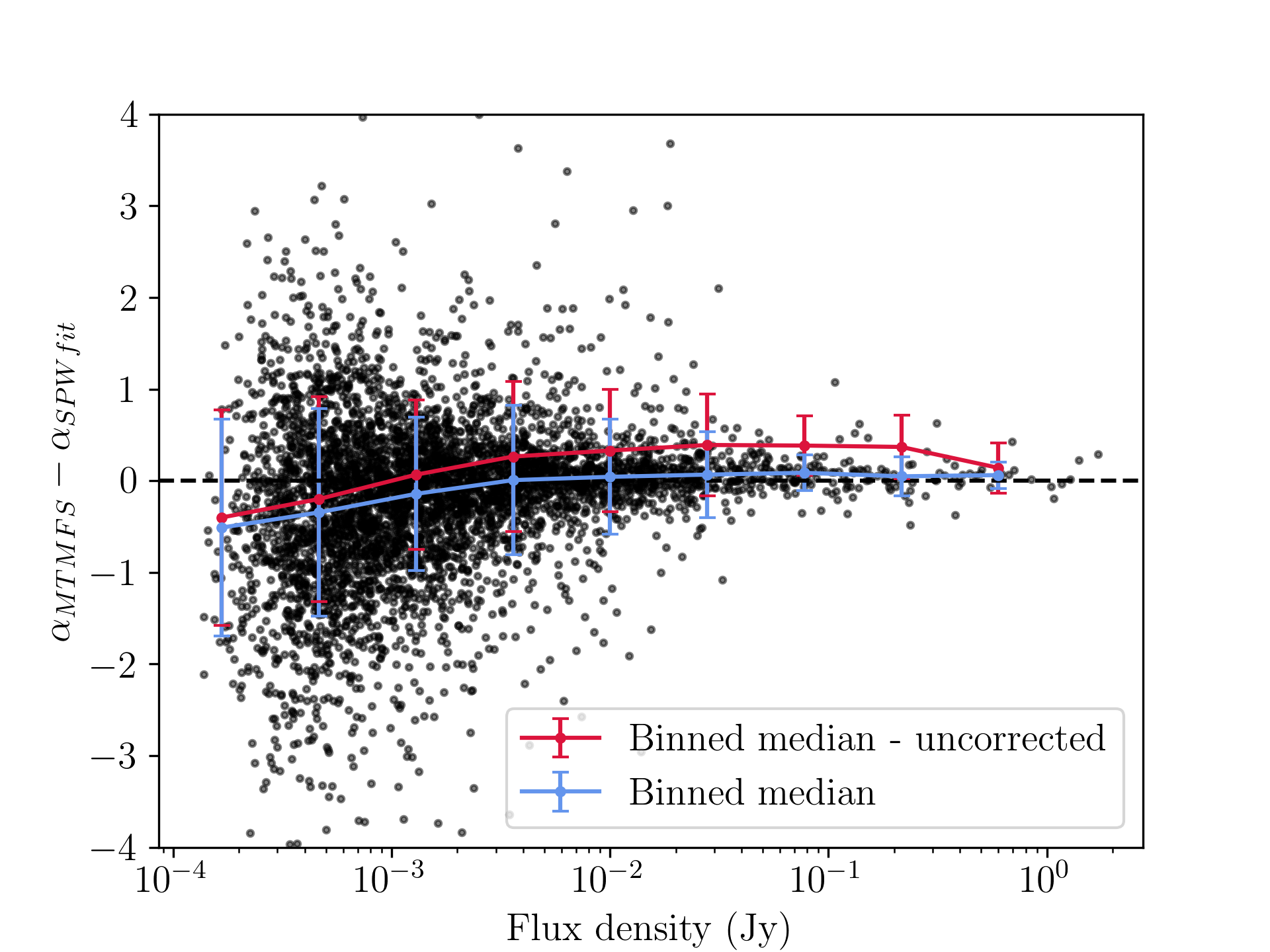}
    \caption{Offsets of spectral indices measured from the wideband MTMFS images with respect to spectral indices derived from processing the full bandwidth using 15 individual SPW images from \citet{Deka2023} as a function of flux density. The spectral indices are calculated by using the SPW2 and SPW9 images. Binned median offsets are shown (blue), along with the offsets without the correction applied in Section~\ref{sec:corrections} (red), showing that the spectral indices are properly corrected.}
    \label{fig:alpha_comparison}
\end{figure}

Due to different emission mechanisms and sources of emission, there can be differences in spectral index distribution between starburst/star-forming galaxies and AGN. Additionally, the Doppler shift observed as a consequence of the motion of the observer induces a change in observed flux density which depends on the spectral index of the source. Thus, the spectral indices of sources influences the magnitude of the radio dipole. In general, dipole studies assume a single value for spectral index based on the physics of synchrotron emission, $\alpha \approx -0.75$ near 1 GHz \citep[e.g.][]{Rubart2013,Tiwari2013,Siewert2021}. Measuring the spectral index of sources generally requires either large bandwidth, or measurements at different frequencies, which in turn requires high S/N to ensure that sources are detected at both ends of the frequency range.  With the large bandwidth (802.5~MHz) of MALS we are able to create spectral index images, as described in Section \ref{sec:alpha_images}, and measure spectral indices of sources in the catalogue. 

To compare the source types, we look at the spectral indices of all sources with respect to flux density. While we cannot completely separate these source types based on flux density or spectral index, these source populations are labelled accordingly in the SKADS simulated sample \citep{Wilman2008}. Figure~\ref{fig:alpha_flux} shows the distribution of spectral indices measured compared to the populations in the SKADS simulated sample. Here, the AGN are separated into FRI (orange) and FRII (red) sub-populations, as are star-forming galaxies separated into `starburst' (light blue) and `normal' (dark blue) galaxies. It is noteworthy that the spectral index distribution of AGN boasts two peaks, corresponding to lobes at $\alpha \approx -0.75$, and cores at $\alpha \approx -0.25$. These peaks are dominated by FRII and FRI galaxies respectively, but there is cross-contamination present as we have plotted source components rather than combined source characteristics. 

The right plot of Figure \ref{fig:alpha_flux} highlights the MALS sources with high S/N, which are more likely to be detected across the full band and thus have more reliable spectral index measurements. Though less tight, the peak at $\alpha \approx -0.75$ is present in the MALS sample. The distribution is broad enough to have mixed with the peak $\alpha \approx -0.25$, so this second peak is possibly lost in the data. To see if we can retrieve these two populations, we take all sources with S/N > 20 and assert that the spectral index of the sources above and below $\alpha = -0.5$ represent FRI and FRII sources respectively. With this definition, there are 626 FRI sources, of which 54\% are resolved and 16\% are fit with multiple Gaussian components. The brightest of these can be appreciated in Figure~\ref{fig:bright_FRI}, showing mostly point sources or sources where core emission dominates. FRII sources are more numerous, with 2581 present in the catalogue, of which 75\% resolved and 30\% fit with multiple Gaussian components. A set of the brightest FRII are also shown in Figure~\ref{fig:bright_FRII}, showing more sources with two or more components representing radio lobes. These results show that the expected dichotomy in morphology between these sources is indeed present, with FRII sources being more likely identified as extended and/or resolved.

At lower flux densities there appears to be a discrepancy between measured and theoretical spectral indices. We see in Figure~\ref{fig:alpha_flux} that not only is there a wider distribution in the MALS data, spectral indices are steeper at lower flux densities compared to the SKADS sources. The median spectral index for sources with $S>1$ mJy is $\alpha = -0.76$, while for sources with $S<1$ mJy it is $\alpha = -1.17$. Because of the high sensitivity that is needed, spectral indices are not commonly measured at lower flux densities. Looking at deep field surveys however, we see that this result is inconsistent with the spectral indices found in the XMM-LSS/VIDEO deep field \citep{Heywood2020}, where it is found that at lower flux densities spectra flatten out. The S/N > 20 sources shown in the right plot of Figure~\ref{fig:alpha_flux}, though increasing in spread at lower flux densities, are not affected by the same bias. 

We further investigate the bias seen in low S/N sources, and verify the corrections made in Section~\ref{sec:corrections}. We compare our spectral indices to those generated by comparing flux densities in SPW2 (1.0 GHz) and SPW9 (1.38 GHz) of the same MALS data by \citet{Deka2023}. \citet{Smolcic2017} find a discrepancy between the spectral indices generated by MTMFS deconvolution and those generated by comparing flux densities at different frequencies, so we make the same comparison in Figure~\ref{fig:alpha_comparison}, showing the offset between the MTMFS and SPW derived spectral indices. The median offsets for both corrected (blue) and uncorrected (red) are shown, indicating that spectral indices have been properly corrected for residual primary beam effects. Though the offset trends negatively at lower flux densities, it is well within the uncertainties. Overall, the offset between spectral indices is $-0.06\pm0.92$ ($0.16\pm0.94$ without corrections) for all sources, and $-0.01\pm0.78$ for S/N > 20 sources, agreeing well between the catalogues. There is no systematic effect seen of the corrected spectral indices with respect to the distance to the pointing centre, indicating no residual primary beam contribution. \citet{Deka2023} observe an overall flattening at low S/N compared to our overall steepening, creating a discrepancy which is clearly showing at lower flux densities. Overall spectral indices appear to be reliable down to mJy flux densities, or S/N of 20. Considering the flux densities of these sources, it is unlikely that many star-forming galaxies are included in the high S/N sample, precluding an analysis of these sources.

\subsection{Number counts}
\label{sec:number_counts}
\begin{figure}
    \centering
    \includegraphics[width=\hsize]{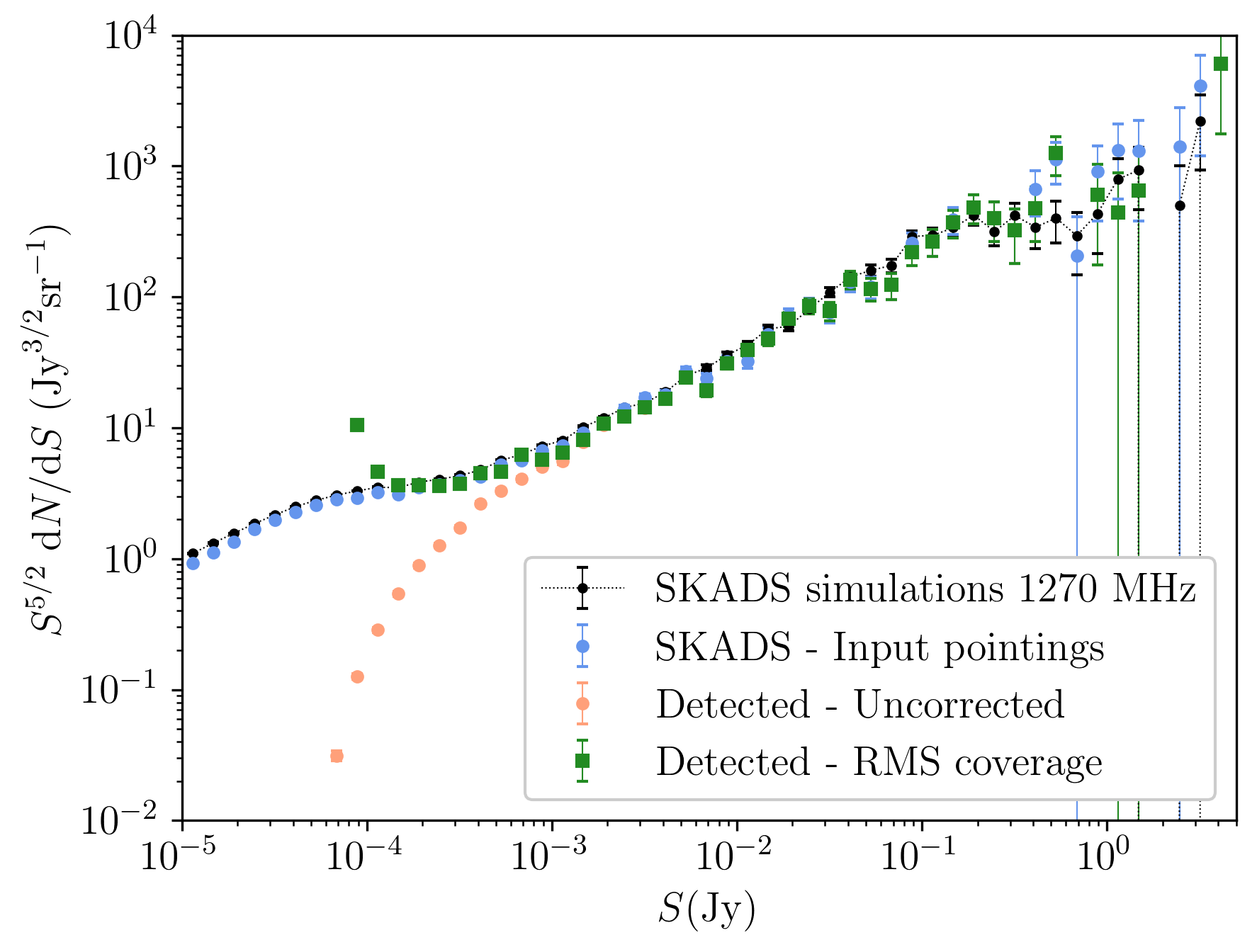}
    \caption{Differential source counts from the SKADS simulations. The the complete SKADS sample is shown (black), as well as the sample extracted from SKADS and injected into the images (blue). The uncorrected number counts (beige) indicate the sources detected by the source finding routine, which are then corrected with the RMS noise coverage (green).}
    \label{fig:diff_counts_sim}
\end{figure}

\begin{figure*}
    \centering
    \includegraphics[width=\textwidth]{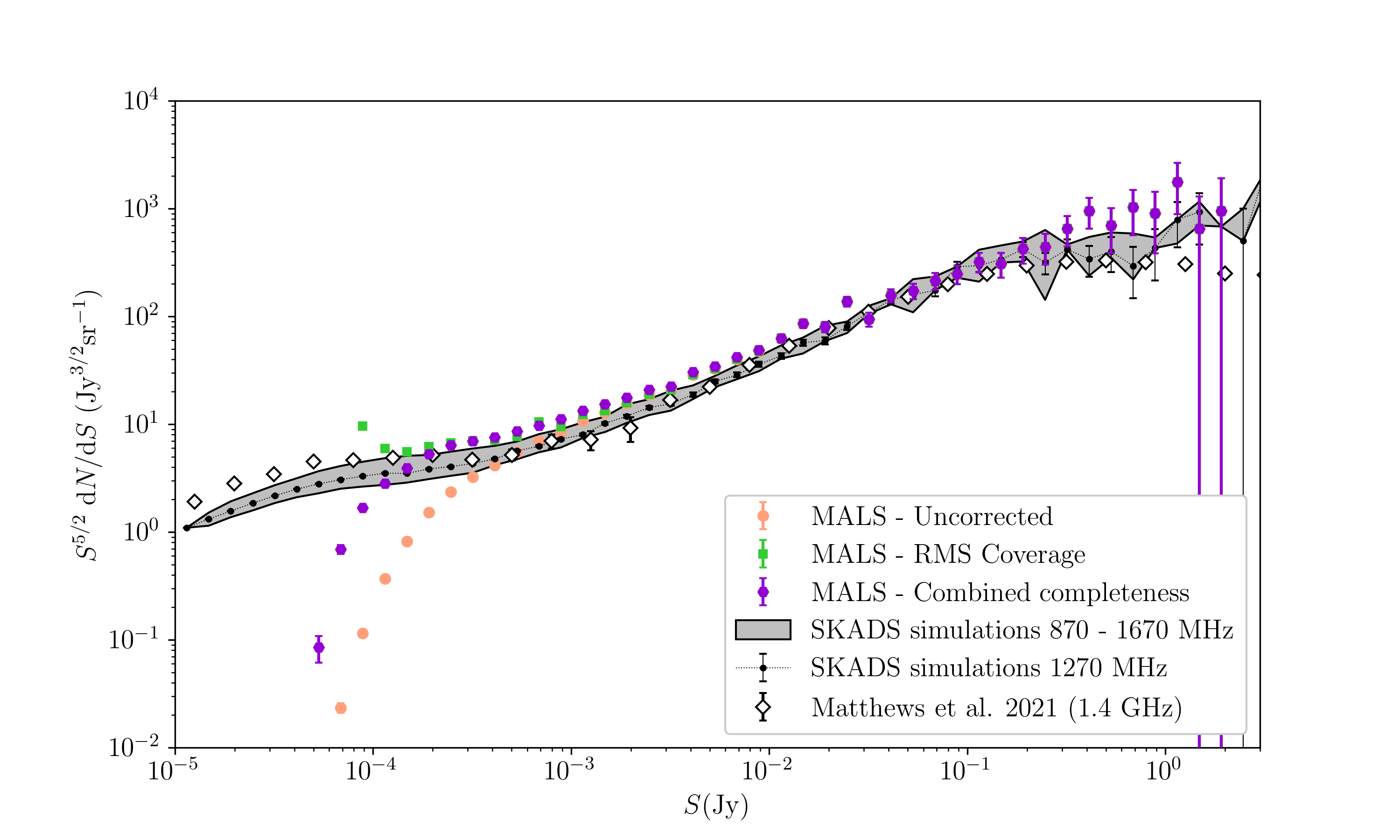}
    \caption{Differential source counts from MALS, uncorrected (beige circles) and corrected with completeness (purple hexagons) taking into account whether a source is resolved or unresolved. Lastly, source counts are also corrected with RMS noise coverage (green squares), which for the lowest flux bins goes to zero, causing solutions to diverge. This is all compared to the SKADS source counts, both for the central frequency of 1270~MHz (black), and for the full frequency range (grey area), and to the source counts derived from the MeerKAT DEEP2 image and NVSS from \citet{Matthews2021} (white diamonds). The number counts are tabulated in Table~\ref{tab:diff_counts}.}
    \label{fig:diff_counts}
\end{figure*}

Now it is left for us to assert that we will have the necessary number counts for a dipole measurement. Extrapolating from the combined catalogue of the ten pointings, a catalogue of the first 391 pointings is expected to carry ${\sim}650,000$ sources, enough to produce a dipole estimate if most sources can be used. However, because our pointings are inhomogeneous, both in terms of internal structure as well as with respect to other pointings, we have to assess to which extent this affects number counts and whether corrections can be made to homogenise the catalogues. The most common method of comparing number counts to other surveys or simulations is to compute differential source counts, which describes the number of sources $\mathrm{d}N$ within a given flux density bin $S + \mathrm{d}S$ per steradian on the sky. This is usually multiplied by $S^{5/2}$ which would yield a flat curve in a static Euclidean Universe \citep{Condon2016}.

To verify that we would be able to retrieve the correct number counts, we repeat the experiment carried out in Section~\ref{sec:individual_assessment}, injecting and retrieving sources in the residual images. To simulate a realistic physical distribution of sources we cut out an area equal to the size of the pointing from the SKADS simulated catalogue. We repeat this experiment for every pointing, each time choosing a random position in the SKADS sample as pointing centre. Figure~\ref{fig:diff_counts_sim} shows the differential number counts for all the stages of the experiment. The reference sample from the SKADS simulations (black) represents the full $10\times10$ degree area simulated in \citet{Wilman2008}. Out of the full sample we cut out 10 pointings with the same sky area as the MALS pointings that are injected into the residual images (blue). On these images we perform our source finding routine and see what number counts we can retrieve. Figure~\ref{fig:diff_counts_sim} shows that below mJy flux densities detected source counts begin to fall off (pink), indicating the limit of 100\% completeness for the full catalogue. These number counts are normalised by the area coverage of the pointings, however we can make a simple correction based on the fact that the area coverage is not constant between different flux bins due to varying RMS noise. The actual area covered in a certain flux bin $S + \mathrm{d}S$ can be obtained by taking the RMS noise coverage (as shown in Figure~\ref{fig:rms_coverage}) assuming a detection limit of $5\sigma$. This basic correction (green squares) produces correct number counts down to 100-200 \textmu Jy, showing that we can account for completeness of the catalogue down to this flux density. Below this, we reach the absolute sensitivity limit of the pointings, as the RMS noise coverage is so low that it produces diverging results. Above 100 mJy results are more scattered, mainly because of low number counts at these flux densities and the smaller sky coverage of our pointings (35.7 sq. deg.) compared to SKADS (100 sq. deg.).

\begin{table*}[]
    \centering
    \caption{Differential source counts of MALS, including corrected counts using RMS coverage, completeness of unresolved sources and completeness of resolved sources. Counts are normalised for the sky area of 35.7 degrees. Raw counts ($N$) and number of false detections in ($N_{false}$) per bin are also given. The excess fraction of sources due to separated components $f_n$ is also given, based on a separation distance of 6.5\arcsec of FRII sources in SKADS.}
    \begin{tabular}{c c c c c c c c | c}
    $S$ & $S_{mean}$ & $N$ & $N_{false}$ & Sky coverage & $S^{5/2}\frac{\mathrm{d}N}{\mathrm{d}S}$ & Corrected $S^{5/2}\frac{\mathrm{d}N}{\mathrm{d}S}$ & Corrected $S^{5/2}\frac{\mathrm{d}N}{\mathrm{d}S}$ & $f_n$ \\
    & & & & & & RMS coverage & Completeness & SKADS \\
    (mJy) & (mJy) & & & (sq. deg.) & (Jy$^{3/2}$ sr$^{-1}$) & (Jy$^{3/2}$ sr$^{-1}$) & (Jy$^{3/2}$ sr$^{-1}$) \\
    \hline \hline
    0.1 - 0.13 & 0.11 & $850\pm33$ & 4 & 2.2 & $0.377\pm0.015$ & $6.06\pm0.24$ & $2.90\pm0.11$ & 0.0\\ 
    0.13 - 0.17 & 0.15 & $1254\pm43$ & 11 & 5.3 & $0.816\pm0.028$ & $5.51\pm0.19$ & $3.93\pm0.13$ & 0.0\\ 
    0.17 - 0.22 & 0.19 & $1575\pm51$ & 13 & 8.7 & $1.50\pm0.05$ & $6.16\pm0.20$ & $5.23\pm0.15$ & 0.0\\ 
    0.22 - 0.28 & 0.25 & $1708\pm54$ & 12 & 12.5 & $2.39\pm0.076$ & $6.82\pm0.22$ & $6.47\pm0.18$ & 0.0\\ 
    0.28 - 0.36 & 0.32 & $1558\pm50$ & 20 & 16.6 & $3.20\pm0.10$ & $6.89\pm0.22$ & $6.91\pm0.19$ & 0.0\\ 
    0.36 - 0.46 & 0.41 & $1393\pm46$ & 22 & 20.9 & $4.21\pm0.14$ & $7.18\pm0.24$ & $7.58\pm0.21$ & 0.0\\ 
    0.46 - 0.6 & 0.53 & $1257\pm43$ & 19 & 25.4 & $5.57\pm0.19$ & $7.83\pm0.27$ & $8.76\pm0.25$ & 0.0\\ 
    0.6 - 0.77 & 0.69 & $1053\pm38$ & 15 & 23.3 & $6.85\pm0.25$ & $10.5\pm0.4$ & $9.67\pm0.30$ & 0.0\\ 
    0.77 - 1 & 0.89 & $881\pm34$ & 16 & 31.6 & $8.41\pm0.33$ & $9.49\pm0.37$ & $11.1\pm0.4$ & 0.0\\ 
    1 - 1.3 & 1.1 & $769\pm31$ & 10 & 30.9 & $10.8\pm0.5$ & $12.5\pm0.5$ & $13.4\pm0.5$ & 0.0\\ 
    1.3 - 1.7 & 1.5 & $635\pm28$ & 7 & 34.4 & $13.1\pm0.6$ & $13.6\pm0.6$ & $15.5\pm0.6$ & 0.0\\ 
    1.7 - 2.2 & 1.9 & $503\pm24$ & 4 & 35.0 & $15.2\pm0.7$ & $15.5\pm0.8$ & $17.3\pm0.8$ & 0.0\\ 
    2.2 - 2.8 & 2.5 & $411\pm21$ & 7 & 35.3 & $18.2\pm1.0$ & $18.4\pm1.0$ & $20.3\pm1.0$ & 0.0\\ 
    2.8 - 3.6 & 3.2 & $318\pm18$ & 2 & 35.5 & $20.7\pm1.2$ & $20.8\pm1.2$ & $22.5\pm1.3$ & 0.0\\ 
    3.6 - 4.6 & 4.1 & $307\pm18$ & 3 & 35.6 & $29.3\pm1.8$ & $29.4\pm1.8$ & $31.3\pm1.8$ & 0.01\\ 
    4.6 - 6 & 5.3 & $232\pm15$ & 6 & 35.7 & $32.5\pm2.2$ & $32.5\pm2.2$ & $34.3\pm2.3$ & 0.02\\ 
    6 - 7.7 & 6.9 & $199\pm14$ & 1 & 35.7 & $40.9\pm3.0$ & $40.9\pm3.0$ & $42.7\pm3.0$ & 0.07\\ 
    7.7 - 10 & 8.9 & $160\pm13$ & 1 & 35.7 & $48.3\pm3.9$ & $48.3\pm4.0$ & $49.6\pm4.0$ & 0.11\\ 
    10 - 13 & 11 & $137\pm12$ & 1 & 35.7 & $60.7\pm5.3$ & $60.7\pm5.3$ & $60.7\pm5.3$ & 0.16\\ 
    13 - 17 & 15 & $136\pm11$ & 0 & 35.7 & $88.5\pm7.8$ & $88.5\pm7.8$ & $88.5\pm7.8$ & 0.22\\ 
    17 - 22 & 19 & $84\pm9$ & 1 & 35.7 & $80.2\pm8.9$ & $80.2\pm8.9$ & $80.2\pm8.9$ & 0.32\\ 
    22 - 28 & 25 & $100\pm10$ & 0 & 35.7 & $140\pm14$ & $140\pm14$ & $140\pm14$ & 0.5\\ 
    28 - 36 & 32 & $48\pm6$ & 0 & 35.7 & $98.7\pm14.4$ & $98.7\pm14.4$ & $98.7\pm14.4$ & 0.5\\ 
    36 - 46 & 41 & $50\pm7$ & 0 & 35.7 & $151\pm22$ & $151\pm23$ & $151\pm22$ & 0.47\\ 
    46 - 60 & 53 & $36\pm6$ & 0 & 35.7 & $160\pm27$ & $160\pm27$ & $160\pm27$ & 0.63\\ 
    60 - 77 & 69 & $34\pm5$ & 0 & 35.7 & $221\pm38$ & $221\pm38$ & $221\pm38$ & 0.73\\ 
    77 - 100 & 89 & $24\pm4$ & 0 & 35.7 & $229\pm47$ & $229\pm47$ & $229\pm47$ & 0.82\\ 
    100 - 130 & 110 & $21\pm4$ & 0 & 35.7 & $294\pm65$ & $294\pm65$ & $294\pm65$ & 0.93\\ 
    130 - 170 & 150 & $19\pm4$ & 0 & 35.7 & $391\pm90$ & $391\pm90$ & $391\pm90$ & 1.0\\ 
    170 - 220 & 190 & $13\pm3$ & 0 & 35.7 & $392\pm109$ & $392\pm109$ & $392\pm109$ & 0.67\\ 
    220 - 280 & 250 & $10\pm3$ & 0 & 35.7 & $443\pm140$ & $443\pm140$ & $443\pm140$ & 0.75\\ 
    280 - 360 & 320 & $10\pm3$ & 0 & 35.7 & $650\pm206$ & $650\pm206$ & $650\pm206$ & 0.89\\ 
    360 - 460 & 410 & $9\pm3$ & 0 & 35.7 & $859\pm287$ & $859\pm287$ & $859\pm287$ & 1.9\\ 
    460 - 600 & 530 & $6\pm2$ & 0 & 35.7 & $841\pm344$ & $841\pm344$ & $841\pm344$ & 0.75\\ 
    600 - 770 & 690 & $5\pm2$ & 0 & 35.7 & $1030\pm460$ & $1030\pm460$ & $1030\pm460$ & 0.5\\ 
    770 - 1000 & 890 & $3\pm1$ & 0 & 35.7 & $906\pm523$ & $906\pm523$ & $906\pm523$ & 1.5\\ 
    1000 - 1300 & 1100 & $4\pm2$ & 0 & 35.7 & $1770\pm890$ & $1770\pm890$ & $1770\pm890$ & 1.2\\ 
    1300 - 1700 & 1500 & $1\pm1$ & 0 & 35.7 & $650\pm650$ & $650\pm650$ & $650\pm650$ & 0.75\\ 
    1700 - 2200 & 1900 & $1\pm1$ & 0 & 35.7 & $955\pm955$ & $955\pm955$ & $955\pm955$ & 0.0\\ 
    \hline
    \end{tabular}
    \label{tab:diff_counts}
\end{table*}

Having shown that we can reproduce number counts for a large range of flux densities, we measure differential number counts for the combined MALS catalogue. Both corrected and uncorrected differential source counts are tabulated in Table~\ref{tab:diff_counts} and shown in Figure~\ref{fig:diff_counts}, where they are compared to the SKADS simulated sample. We apply a completeness correction (purple hexagons) using either unresolved or resolved completeness based on whether the source is classified as such using the criterion from Section~\ref{sec:resolved}. Once again we also correct number counts using the RMS noise coverage (green squares). For comparison, number counts from the full SKADS simulated sample are also shown, both for the central frequency of 1.27 GHz and for the full frequency band, accounting for the fact that for most sources flux density is not equal across the band. The number counts derived from the MeerKAT DEEP2 image and NVSS by \citet{Matthews2021} are also shown (white diamonds). Error bars are computed taking into account Poisson uncertainties as well as source clustering following \citet{heywood2013}. The same data from Figure~\ref{fig:diff_counts} is tabulated in Table~\ref{tab:diff_counts}, showing the numerical values of flux bins, raw number counts, and uncorrected and corrected differential number counts. Though it is not taken into consideration in Figure~\ref{fig:diff_counts}, the number of false detection per bin is included in Table~\ref{tab:diff_counts}. In all cases the number of false detections is smaller than the uncertainty on the number counts, from which we infer that the purity is a small factor compared to completeness of the catalogue.

As for the simulated sample, the corrections appear to hold down to 100-200 \textmu Jy, after which we reach the sensitivity limit and solutions diverge. Until solutions diverge, the completeness and RMS noise coverage corrections produce similar results, indicating that a simple sky coverage correction performs well given the ease with which it can be generated. Given the agreement between the completeness and RMS noise coverage corrections, we can safely say that the major contributor to completeness is the inhomogeneous sky coverage of the pointings. This results in the dependency of completeness on the local noise seen for point sources in Figure~\ref{fig:individual_sep_flux_completeness}. With these corrections applied, Figure~\ref{fig:diff_counts} shows a wide range of flux densities where the differential number counts deviate from the expected values. At high flux densities (>100~mJy), we see an increase in sources that can for some part be attributed to the central sources in the images, as well as single sources being classified as multiple sources, as described in Section~\ref{sec:ext_sources}. However we see that number counts, except for the range 20 mJy - 200 mJy, are higher across the board than what we might expect given theoretical predictions. There is some evidence that the SKADS simulations underestimated the number of star-forming galaxies, causing lower counts at low flux densities compared to what is seen in nature \citep[e.g.][]{Hale2022}. Though this can explain an offset at the lowest flux densities, this effect would only be significant up to mJy flux densities, whereas our number counts are higher up to an order of magnitude above that. An alternative explanation is that with its selection of high flux density sources as pointing targets, MALS is probing overdensities which naturally boosts the number of sources in the pointings. Lastly, such an effect can also be produced by a systematic overestimation of flux densities, which is an option that cannot be ruled out at this stage. We expect that this offset might also be caused by low number statistics, and may disappear once more data is added.

\section{Towards the cosmic radio dipole}
\label{sec:dipole}

\begin{figure}
    \centering
    \includegraphics[width=\hsize]{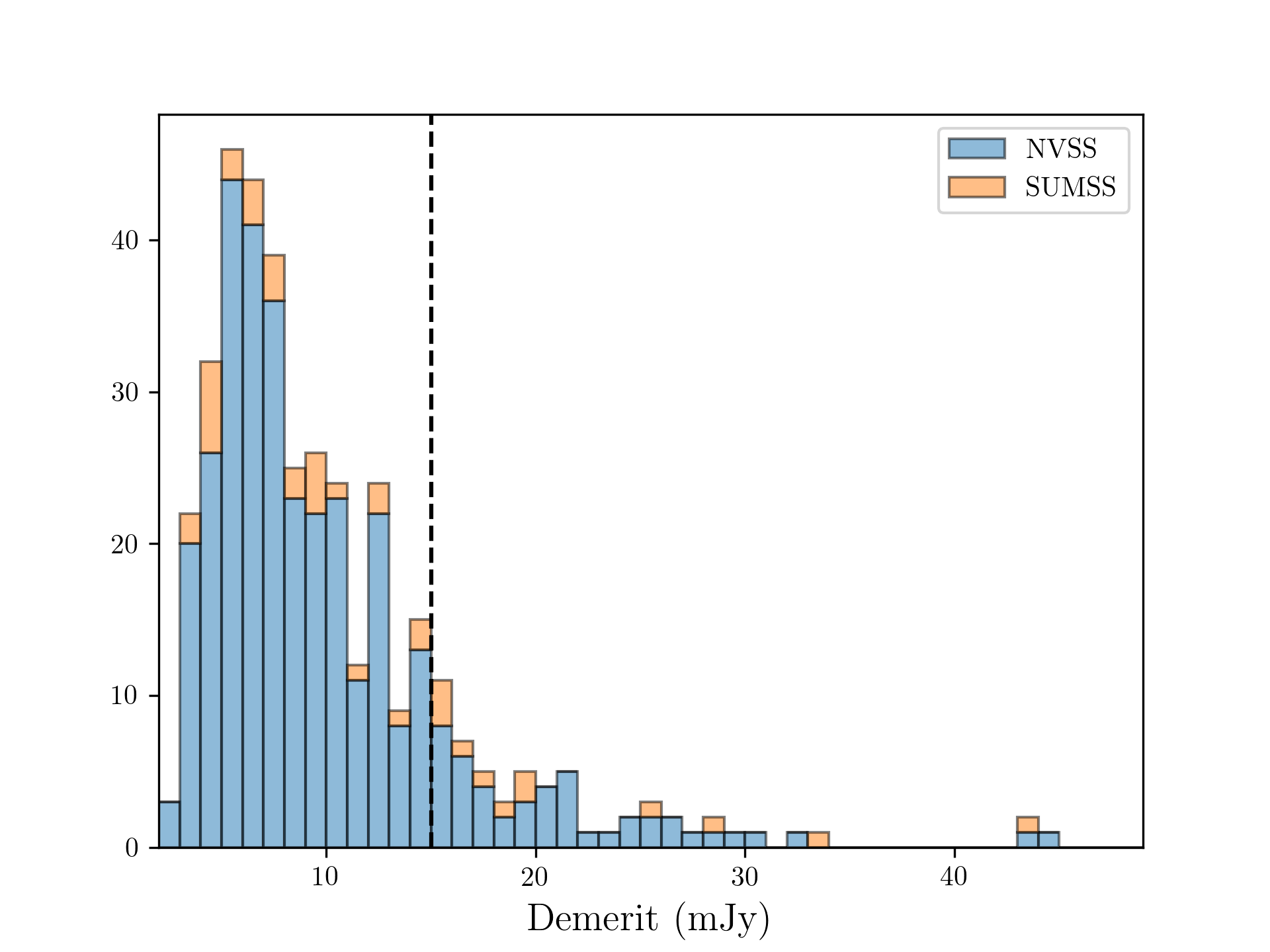}
    \caption{Demerit scores calculated for the 391 currently observed MALS pointings, using bright ($>100$ mJy) sources retrieved from their corresponding surveys (NVSS and SUMSS). The dotted line indicates a quality threshold of $d < 15$ mJy based on the quality of the pointings inspected in this paper.}
    \label{fig:mals_fluxes}
\end{figure}

With a thorough assessment of the quality of the pointings described in this work we have the opportunity to extrapolate our findings to the larger survey of 391 pointings, both in estimating how many pointings will be needed for a dipole estimate, as well as how to effectively homogenise the catalogues. Given these results, there are however some questions and limitations that remain, and these will have to be addressed in later works.   

We first estimate the statistical power we can reach with the observed MALS pointings. As we determined in Section~\ref{sec:noise}, the demerit score is a strong indicator for pointings with high noise and thus low number counts. Consequently, the demerit score allows us to make predictions about the quality of other pointings in the survey. Of the ten pointings we have investigated we consider seven of them to be of good quality based on their noise and source count values. Since all MALS pointings are in the footprint of either NVSS or SUMSS, we use these surveys to match sources and calculate the demerit score to predict quality of the images. Based on this principle Figure \ref{fig:mals_fluxes} shows the distribution of the demerit scores of the first 391 observed MALS pointings. Defining a quality threshold of $d < 15$ mJy based on the ten pointings we have investigated here, we see that 322 of 391 pointings are below this threshold. The seven good quality pointings average ${\sim}2000$ sources per pointing, such that 100 such pointings will approximately result in $2\times10^5$ sources. If we choose our pointings to properly cover the sky along the axis of the dipole, this is the minimum number of sources required for a $3\sigma$ measurement of the cosmic radio dipole, assuming an amplitude equal to that of the CMB dipole. 

Based on the demerit scores of the ten pointings and those of the first 391 pointings, we also see that our ten pointings well represent the average pointing, and we can use the differential number counts to extrapolate them to the rest of the survey. This allows us to take into account the completeness of the survey so far. As shown in Figures~\ref{fig:diff_counts_sim} and \ref{fig:diff_counts}, we get correct source counts down to 100-200~\textmu Jy. Naturally, this means that sources below that flux density can not be included in a dipole estimate, which leaves 13,663 sources in the combined catalogue. However, the corrections are essentially compensating for the missing sources, yielding effectively ${\sim}28,000$ sources down to 200~\textmu Jy, meaning even less than 100 pointings would again suffice for a $3\sigma$ dipole measurement. 

In both cases we may expect that the first 391 observed pointings will yield around a million sources, which should yield a dipole measurement at a significance level of $6.5\sigma$, assuming an adequate coverage of the dipole axis and a dipole amplitude equal to that of the CMB \citep{Ellis1984}. Though it is clear that MALS will deliver the number counts needed for a significant measurement of the cosmic radio dipole, the larger challenge is making a measurement while accounting for the systematics present in the survey. In order to thus successfully measure the dipole with MALS, the way forward is to build on the corrections introduced in Section~\ref{sec:number_counts}.

\subsection{Compiling a homogeneous catalogue}

We have effectively shown that we can make a unified description of the properties of the average MALS pointing, which should now allow us to homogenise the catalogue. Though the structure of the survey, with deep coverage over distinct patches of sky, appears to not lend itself especially well to large scale cosmology, the fact that these pointings are all equal area by design allows for straightforward discretisation. In the simplest use case, each pointing of MALS can thus be treated as a single unit simply containing $N$ number of sources. When measuring number counts over a full sky, inhomogeneities between the pointings induce higher order multipoles in the data that will spill over into a dipole measurement. To homogenise the data and get an unbiased estimate of the dipole, we must account for the individual differences between the pointings and calculate the corrected \textit{effective} number counts $N_{eff}$.

To get to effective number counts, the starting point is the corrections to number counts as shown in Figure~\ref{fig:diff_counts} that are seen to largely compensate for incomplete catalogues. We can extend this treatment by assigning a `completeness factor' to each individual source. As discussed in Section~\ref{sec:number_counts}, the largest contributor is the incomplete sky coverage, which is well modelled with the RMS noise coverage of the pointings. Immediately we can disentangle the completeness into a detection probability $P(det)$ and sky coverage $\Omega$,
\begin{equation}
    Comp(S, Q_A, \sigma) = P(det|S,Q_A)\Omega(\sigma).
\end{equation}
Here the detection probability has a power law dependence on source size $Q_A$ (linear in log-log space, right plot of Figure~\ref{fig:maj_sep_completeness}), and assuming Gaussian errors on the flux density, detection probability should follow a Gaussian cumulative distribution function or similar sigmoid function\footnote{Sigmoid is the collective name of functions following `S'-shaped curves, which are well suited to describe the detection probability of data near the detection boundary.} as a function of flux density. Sky coverage is exclusively determined by the local RMS $\sigma$ of the source, encoding the noise structure of the pointing. This completeness factor can largely correct for the imhogeneities present in the catalogues, but we can make it even more robust by including information on other investigated quantities. Using information on purity we can define a `purity factor' 
\begin{equation}
    Purity(S,\rho, pointing) = Purity(S,\rho)Purity_{pointing},
\end{equation}
which indicates how likely a source is to be a true positive. This depends on distance from the pointing centre $\rho$, flux density $S$ (Figure~\ref{fig:purity_pointings}) and has a multiplicative factor that indicates a purity level that is different per pointing (Figure~\ref{fig:purity}). Finally, individual sources have associated uncertainties that can be used to weigh each source accordingly. An obvious choice is a weight based flux density, as the uncertainty in flux density $\sigma_S$ is dependent on flux density $S$ (Figure~\ref{fig:flux_recovery}), which can be combined with a potential flux density scale error $\Delta S$. We are not limited to one uncertainty factor, and a second choice that is relevant for a dipole measurement is the uncertainty in position $\sigma_{\vec{\phi}}$, which in absence of any systematics (as we see in Figure~\ref{fig:full_astro}) is simply equal to the measured uncertainty $\Delta\vec{\phi}$. Combining all these measures, we can assign a weight to sources based on the quantities laid out,
\begin{align}
    w_S &= \sigma_S^{-1}(S,\Delta S), \\
    w_{\vec{\phi}} &= \sigma_{\vec{\phi}}^{-1}(\Delta \vec{\phi}), \\
    w_{eff} &= w_{\vec{\phi}}w_S\frac{Purity(S,\rho, pointing)}{Comp(S, Q_A, \sigma)}. \\
            &= w_{\vec{\phi}}w_S\frac{Purity(S,\rho)Purity_{pointing}}{P(det|S,Q_A)\Omega(\sigma)}
\end{align}
The effective weight factor $w_{eff}$ fulfils a dual purpose in estimating the number counts. The completeness and purity factors correct the number counts, while the weights from the flux density and position errors then serve as a quality measure for each source, allowing us to measure the effective number density of the individual pointings,
\begin{equation}
    N_{eff} = \frac{\sum_i^n w_{eff,i}}{\sum_i^n w_{\vec{\phi},i}w_{S,i}}.
\end{equation}

\subsection{Limitations and future prospects}

With this prescription, the systematic effects that we have characterised can be accounted for when computing the number counts and estimating the dipole. There remain however some effects that have not been explicitly characterised that could influence a dipole measurement. By checking the corrections to number counts as we did in Section~\ref{sec:number_counts} on simulated data, we essentially calibrated the corrections on the SKADS sample, which as a simulation might not perfectly represent the number counts found in nature (this can be plainly seen in Figure~\ref{fig:diff_counts}, where SKADS number counts do not always agree with the counts from \citet{Matthews2021}). While this can introduce an unknown error into the process, the error is expected to be in overall number counts, and therefore not directionally dependent. Similarly, an important aspect of MALS is the selection of the pointings, as every pointing has a bright radio source at the centre. Although pointings are distributed isotropically, this is not equivalent to a random selection as bright central sources are more likely to be embedded in overdensities. This effect seems to be very pronounced in our measured source counts already, which are larger than expected. Again, this effect is expected to be directionally independent, but whether this is truly the case remains to be determined. Finally, the depth of MALS might be to its detriment when measuring a dipole, as the reached depth of 200~\textmu Jy probes into the population of starburst and normal galaxies. The brightest sub-population of these fainter sources occurs at lowest redshifts, which exhibits stronger clustering than AGN. To what extent this affects a dipole measurement is explored in \citet{Bengaly2019}, who perform several redshift cuts and a significant improvement is made with $z_{cut} = 0.1$ compared to including all sources. This is a more stringent cut than \citet{Blake2002}, who claim to eliminate local clustering effects with $z_{cut} = 0.03$. Although no direct redshift information is available from the MALS data, we will be able to investigate the effect of clustering due to nearby star-forming galaxies using photometric redshifts from current surveys such as the Dark Energy Camera Legacy Survey \citep[DeCALS,][]{Blum2016}, and the Rubin Observatory Legacy Survey of Space and Time \citep[LSST,][]{LSSTScienceCollaboration2009} in the near future.

It is worth noting that an increase in number counts as seen in Figure~\ref{fig:diff_counts} can also be (partially) caused by a systematic flux density offset. The results seen in Figure~\ref{fig:NVSS_flux} hint to a systematic flux density offset with respect to NVSS, which is worth investigating with the larger MALS catalogue. If this offset turns out to be indeed significant, there are many possibly explanations given that such a systematic effect could potentially be introduced at many points in the data processing pipeline. We have already shown in Section~\ref{sec:combined_flux_recovery} that the island flux density from \textsc{PyBDSF} properly recovers the flux density from simulated sources, which verifies that the source finding step is not inducing a systematic flux density offset. To further narrow down the options, we cross-check our results with those from \citet{Deka2023}, which use the same data, calibration pipeline and source finding strategy but show an overall agreement with NVSS. The most notable difference between these catalogues is the fact that we utilise the full band while \citet{Deka2023} only use individual SPWs with 50~MHz bandwidth. Such a different between results from the full bandwidth and individual SPWs could point to systematic effects introduced in the imaging stage, as we model the emission with two Taylor terms in frequency as opposed to a single Taylor term in case of individual SPW images. Though the ten pointings explored here have already provided a wealth of statistics and insight into systematic effects, these will be further investigated with the full suit of MALS pointings. 

With regards to a dipole measurement, there are still some questions which remain to be answered. Throughout this work we have used the estimate by \citet{Ellis1984} of $2\times10^5$ sources properly distributed along the axis of the dipole. Though the MALS pointings properly cover the axis of the CMB dipole, if the direction of the radio dipole deviates from this, for example towards the northern hemisphere, the coverage of MALS pointings might not be adequate. Furthermore, the exact amount of sources needed for a dipole estimate can vary depending on this coverage and has been differently estimated in different works. \citet{Crawford2009} claims $2\times10^6$ sources are necessary for a $3\sigma$ dipole estimate, which is an order of magnitude more than the number from \citet{Ellis1984}. Dipole studies using NVSS have generally reached $3\sigma$ significance with $3\times10^5$ sources \citep[e.g.][]{Singal2011,Rubart2013,Secrest2022}. This is in closer agreement to the \citet{Ellis1984} numbers, however this significance is only reached because of the anomalously high dipole amplitude. Therefore, the significance with which the dipole can be measured ultimately depends on many factors, including sky coverage, number counts, dipole amplitude, frequency, and the employed estimator. Consequently, another important step towards measuring the cosmic radio dipole with MALS is defining an appropriate dipole estimator \citep[see e.g.][]{Siewert2021} which beyond the scope of this paper but will be explored in a future work. 

\section{Summary and conclusion}
\label{sec:conclusion}

In this work we have presented a thorough analysis of the first ten deep continuum pointings of the MeerKAT Absorption Line Survey \citep[MALS,][]{Gupta2016}, and have compiled a catalogue with 16,313 sources covering 35.7 square degrees of deep radio sky. We have set out to extensively analyse the properties of the first ten pointings of MALS, with the ultimate goal of measuring the cosmic radio dipole. To achieve a measurement of number counts unbiased by the inhomogeneities present between the MALS pointings, we have characterised systematic effects that can influence such a measurement. This assessment of systematic effects in the ten pointings as presented in this work shows that these effects are for the most part predictable and can be properly accounted for. This will eventually not only benefit a dipole measurement, but all continuum science carried out with MALS. In the current literature on the cosmic dipole there are many examples of systematic effects that limit the sensitivity of these estimates, that could not be further pushed due to a lack of information on the inner workings of the surveys that were used. For MALS, we have a complete assessment of the inner workings of the survey, with insight and access into the processing pipeline. Looking forward, we determine that 100 MALS pointings suffice for a dipole measurement. Paired with the analysis on source characteristics and counts in the pointings, we are poised to perform the most complete dipole estimate of a radio survey thus far.

Calibration and imaging of all the data is carried out through ARTIP. After imaging we separately create spectral index images, and perform primary beam correction averaged over the frequency range on these and the continuum images with a primary beam model derived from holographic measurements. We make an initial assessment of calibration by checking the flux density scale of the calibrators and central sources, and find that flux densities are consistent with those reported in the literature. We investigate the quality of the images by looking at the RMS noise maps created by \textsc{PyBDSF}. Measuring RMS noise coverage shows that all pointings have similar noise structure, but overall noise levels are offset between the pointings. We quantify this offset with $\sigma_{20}$, which gives the noise level at 20\% RMS noise coverage for a pointing. To try to explain the difference in pointing quality, we calculate demerit scores for each pointing to estimate the contributions of bright sources to the noise. Though there is a correlation between the noise in a pointing and demerit score, other factors play a part that introduce a scatter in this relation, especially for pointings with lower demerit scores. As $\sigma_{20}$ directly describes the overall noise in the image, it is the quality measure of choice for the pointings.  

We perform source extraction on all the images using \textsc{PyBDSF}, and convert \textsc{PyBDSF} catalogues to full Stokes I catalogues, extending them to include spectral indices and flagging artefacts. We consider sources that have been fit with multiple Gaussian components to be potentially complex and visually inspect them to investigate how well the Gaussian fit describes these sources. We further assess the quality of the individual pointings and how it affects source finding by measuring completeness, flux recovery, and purity. For completeness and flux recovery we need to know the intrinsic properties of the sources in the images, so we create mock catalogues of sources from the SKADS simulated sample and inject these into the residual images of the pointings. We assess completeness separately for unresolved and resolved sources. Using unresolved sources, we assess completeness as a function of distance from the pointing centre, and for resolved sources we investigate completeness as a function of source size.

Combining the catalogues of the individual pointings, we correct for residual primary beam effects in the flux densities and spectral indices sources originating from the frequency dependence of the primary beam. To check these corrections and other potential systematic effects, we cross-match the catalogues with NVSS to check if positions and flux densities are consistent. There is no appreciable astrometric offset, however there is a 18\% offset in flux density compared to NVSS, though still within the uncertainties. Using $\sigma_{20}$ as a normalisation factor, we combine completeness measures from the individual pointings and find unified completeness relations that hold for all pointings. Combining flux recovery statistics from all pointings, we find that a systematic bias is present in the integrated flux densities from the fitted Gaussians in the catalogue. This bias is not present in the integrated flux densities of the islands which the sources occupy, making this quantity the logical choice of flux density for the here presented analysis. Combining purity from all pointings, we find that we can account for 20\% of false detections with a suitable artefact identification scheme. The remaining false detections are shown to make up an increasing fraction of sources further out from the pointing centre.  

As the full catalogues are expected to be populated by various sources types, we assess how these can influence source counts. Looking at the number of Gaussian components fit to sources, we see that the number of components needed to describe sources increases as a function of flux density until it stagnates at around 100 mJy. This implies that around this flux density sources separate and can be counted multiple times, however as only a tiny percentage of sources is present at these flux densities, it is unlikely to bias the source counts. To further differentiate between source populations, we look at spectral indices of sources, and find that we can reasonably separate core-dominated FRI and lobe-dominated FRII sources by making a cut at $\alpha = -0.5$. At low flux densities we find that spectral indices are much steeper than expected which is likely caused by low S/N and inadequate modelling by the MTMFS deconvolution scheme.

Finally, to show that we can account for the systematic effects present in the catalogues, we calculate and correct differential number counts for a set of simulated catalogues using the SKADS simulated sample. We find that a simple correction using only the RMS noise coverage produces correct number counts down to 100-200 \textmu Jy. We then compute differential number counts for the full catalogue and correct number counts using combined completeness measures found for unresolved and resolved sources, as well as the RMS noise coverage. Once again, corrections seem to hold down to 100-200 \textmu Jy. Comparing these number counts to the expected number counts from the SKADS sample, we see that our number counts are higher in the full range of probed flux densities. This is independent of corrections, so a likely explanation is that MALS is probing overdense regions, as it targets bright sources. The same effect can however be (partially) produced by a systematic flux density offset, which should be taken into consideration given the results on the flux density scale.  

Using both the demerit score to predict the quality of other MALS pointings and the corrected number counts of this sample, we show that we will require 100 MALS pointings to reach the necessary number counts for a $3\sigma$ measurement of the dipole. Going further, we assert that we can assess the dipole on the level of individual sources using the information on flux density scale, completeness, purity, flux errors, and position errors. The precise implementation is left to later works, along with a exploration of viable dipole estimators.

\begin{acknowledgements}
We thank the anonymous referee for their useful comments and feedback.
We thank Charles Walker for his insightful comments and feedback on the text.
JDW acknowledges the support from the International Max Planck Research School (IMPRS) for Astronomy and Astrophysics at the Universities of Bonn and Cologne.
EB acknowledges support by NASA under award number 80GSFC21M0002.
PK is partially supported by the BMBF project 05A17PC2 for D-MeerKAT II.
The MeerKAT telescope is operated by the South African Radio Astronomy Observatory, which is a facility of the National Research Foundation, an agency of the Department of Science and Innovation.
The MeerKAT data were processed using the MALS computing facility at IUCAA (https://mals.iucaa.in/releases). 
The Common Astronomy Software Applications (CASA) package is developed by an international consortium of scientists based at the National Radio Astronomical Observatory (NRAO), the European Southern Observatory (ESO), the National Astronomical Observatory of Japan (NAOJ), the Academia Sinica Institute of Astronomy and Astrophysics (ASIAA), the CSIRO division for Astronomy and Space Science (CASS), and the Netherlands Institute for Radio Astronomy (ASTRON) under the guidance of NRAO.
The National Radio Astronomy Observatory is a facility of the National Science Foundation operated under cooperative agreement by Associated Universities, Inc.
\end{acknowledgements}

\bibliographystyle{aa}
\bibliography{main_paper}

\newpage
\begin{appendix}

\section{Cutouts of bright FRI and FRII sources}
\noindent\begin{minipage}{\textwidth}
\begin{minipage}[c]{0.33\textwidth}
  \vspace*{\fill}
  \centering
  \includegraphics[width=\textwidth]{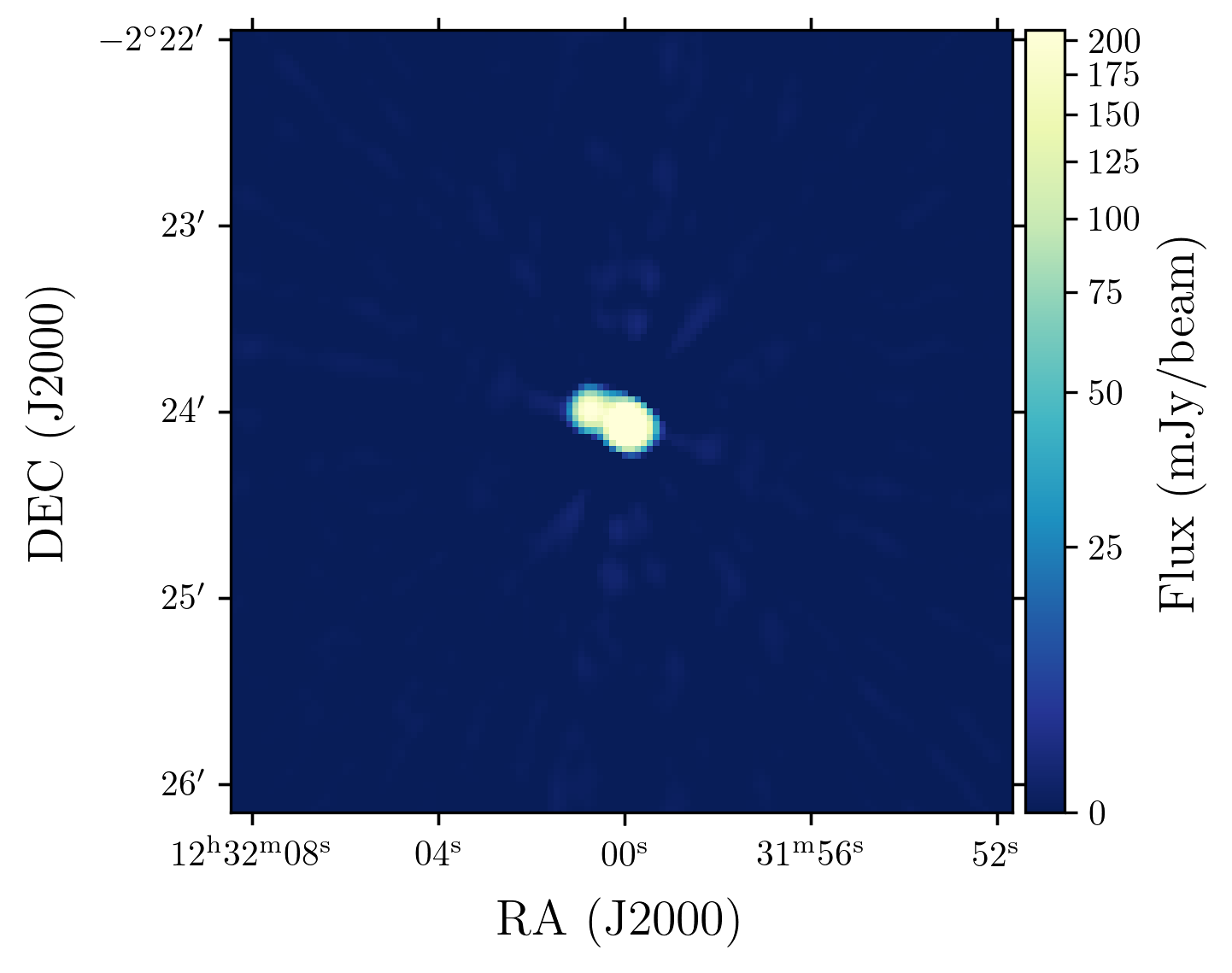}
  \label{subfig:frI_1}
\end{minipage}
\begin{minipage}[c]{0.33\textwidth}
  \vspace*{\fill}
  \centering
  \includegraphics[width=\textwidth]{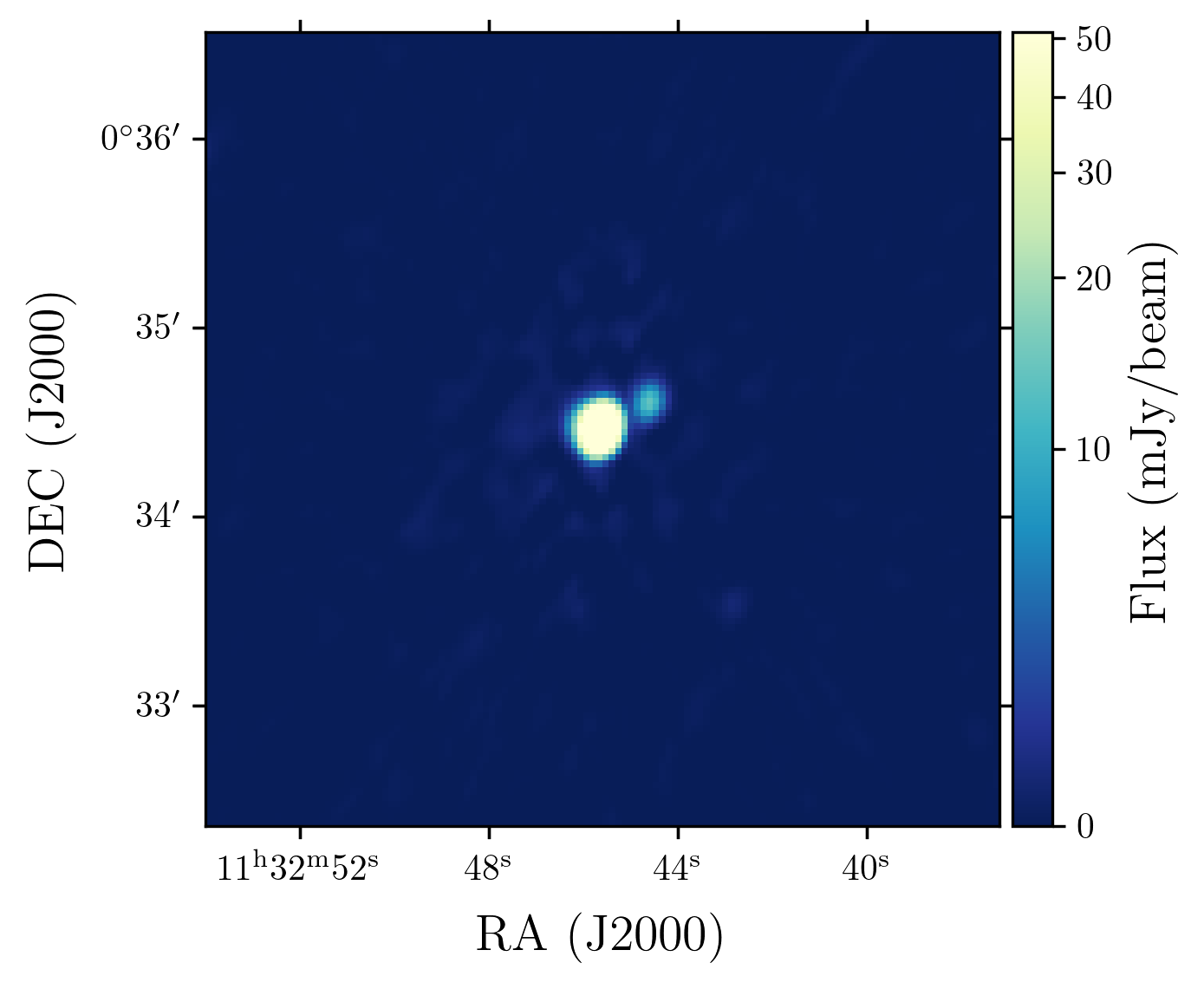}
  \label{subfig:frI_2}
\end{minipage}
\begin{minipage}[c]{0.33\textwidth}
  \vspace*{\fill}
  \centering
  \includegraphics[width=\textwidth]{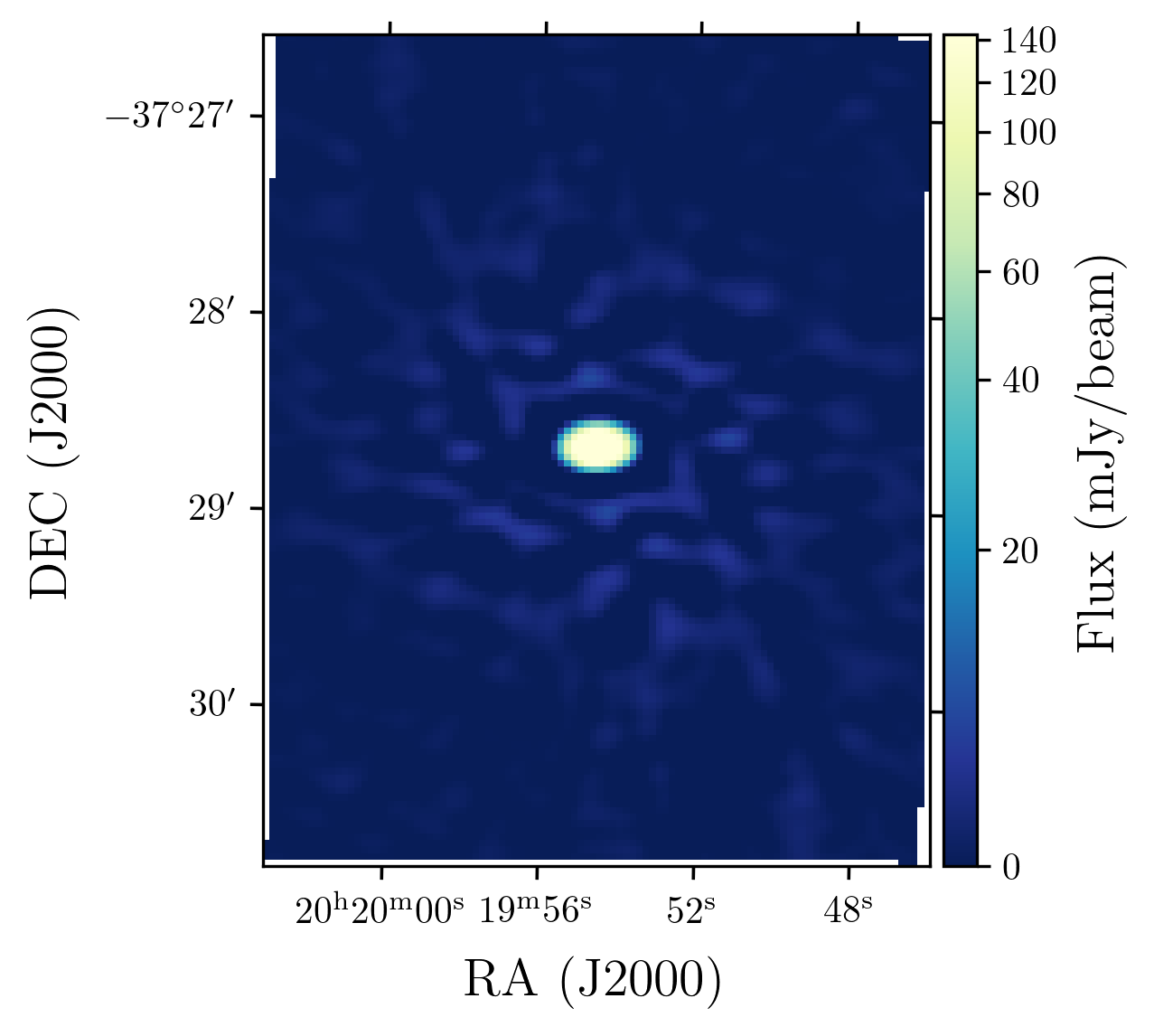}
  \label{subfig:frI_3}
\end{minipage}

\begin{minipage}[c]{0.33\textwidth}
  \vspace*{\fill}
  \centering
  \includegraphics[width=\textwidth]{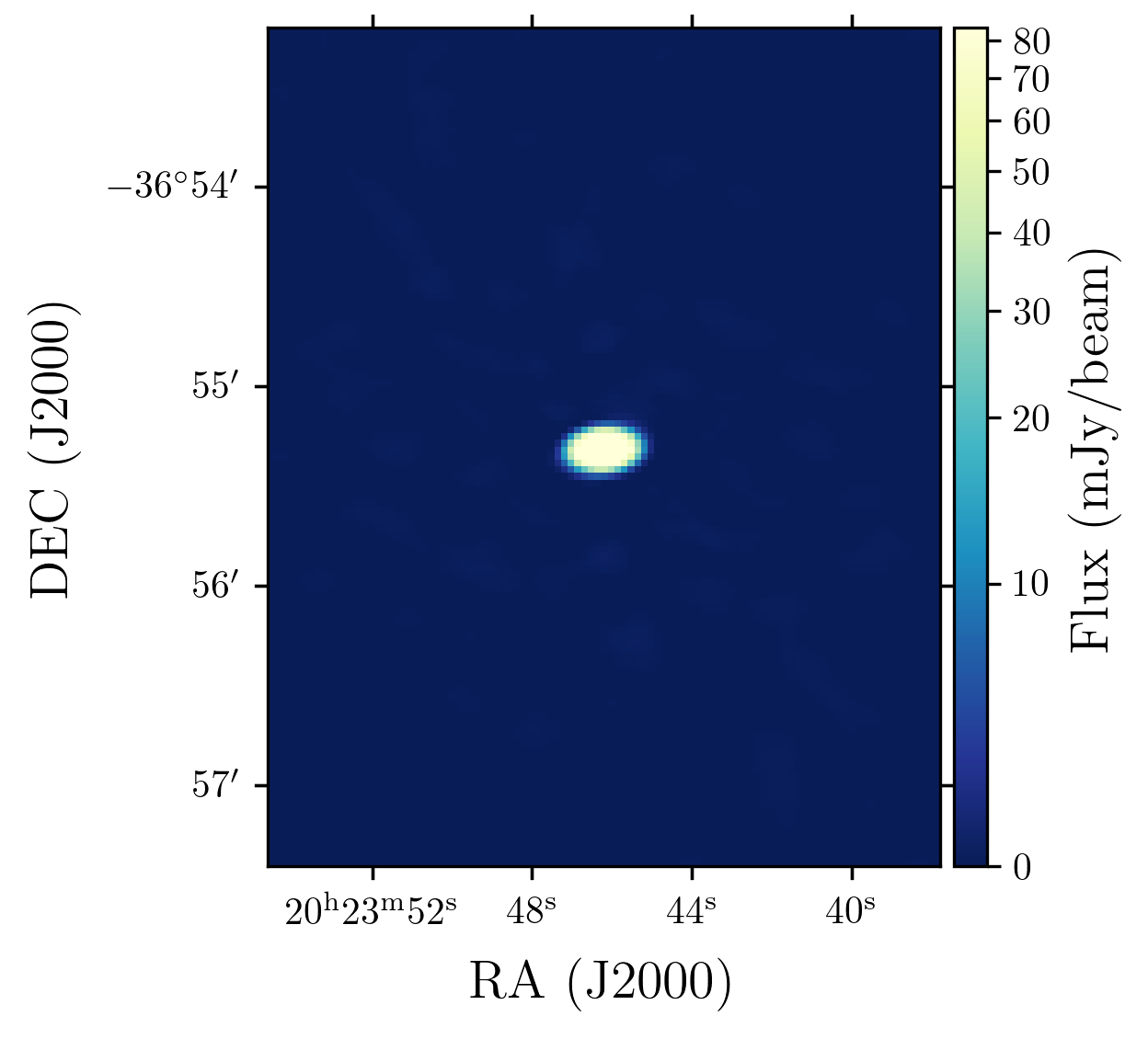}
  \label{subfig:frI_4}
\end{minipage}
\begin{minipage}[c]{0.33\textwidth}
  \vspace*{\fill}
  \centering
  \includegraphics[width=\textwidth]{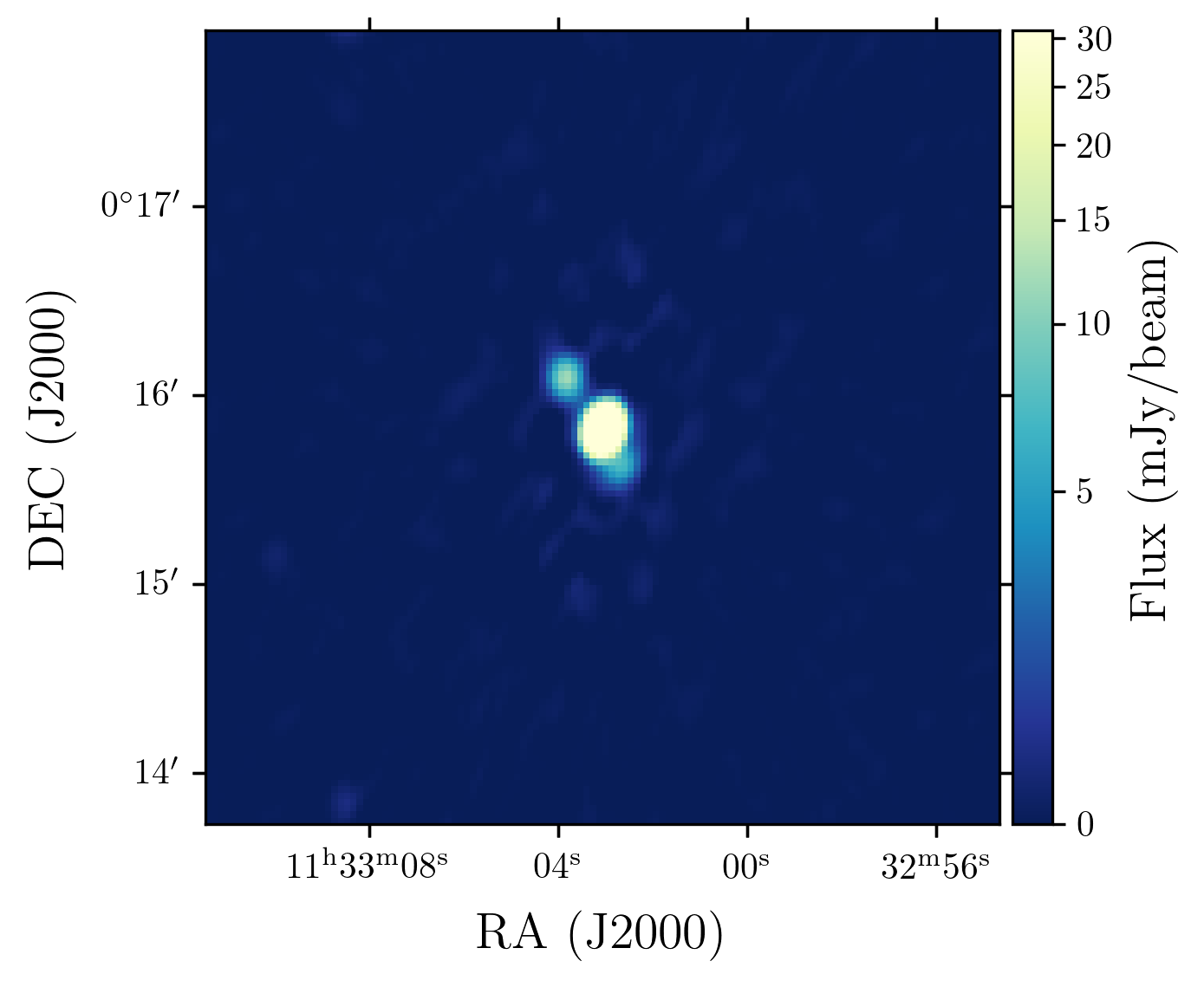}
  \label{subfig:frI_5}
\end{minipage}
\begin{minipage}[c]{0.33\textwidth}
  \vspace*{\fill}
  \centering
  \includegraphics[width=\textwidth]{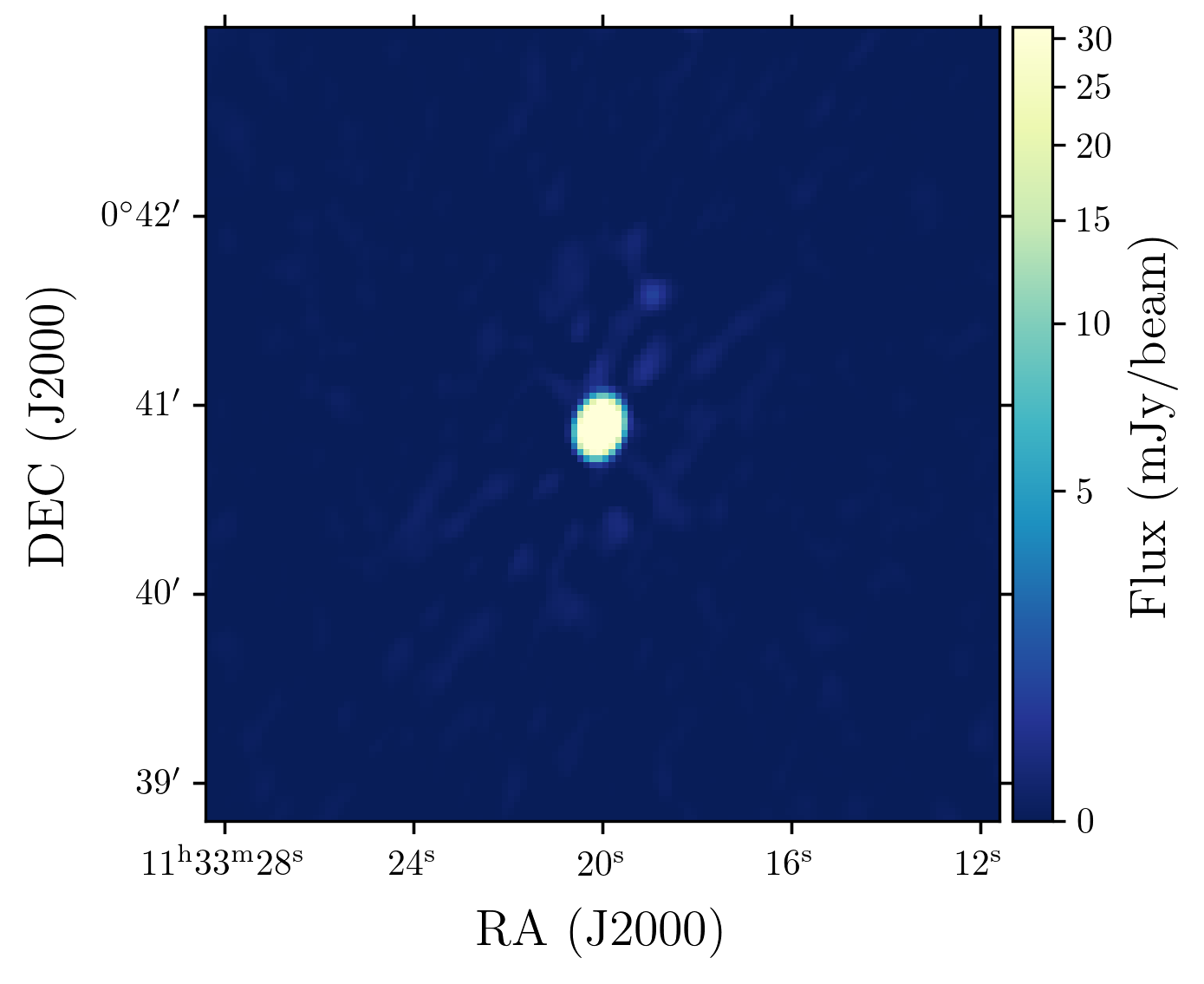}
  \label{subfig:frI_6}
\end{minipage}

\begin{minipage}[c]{0.33\textwidth}
  \vspace*{\fill}
  \centering
  \includegraphics[width=\textwidth]{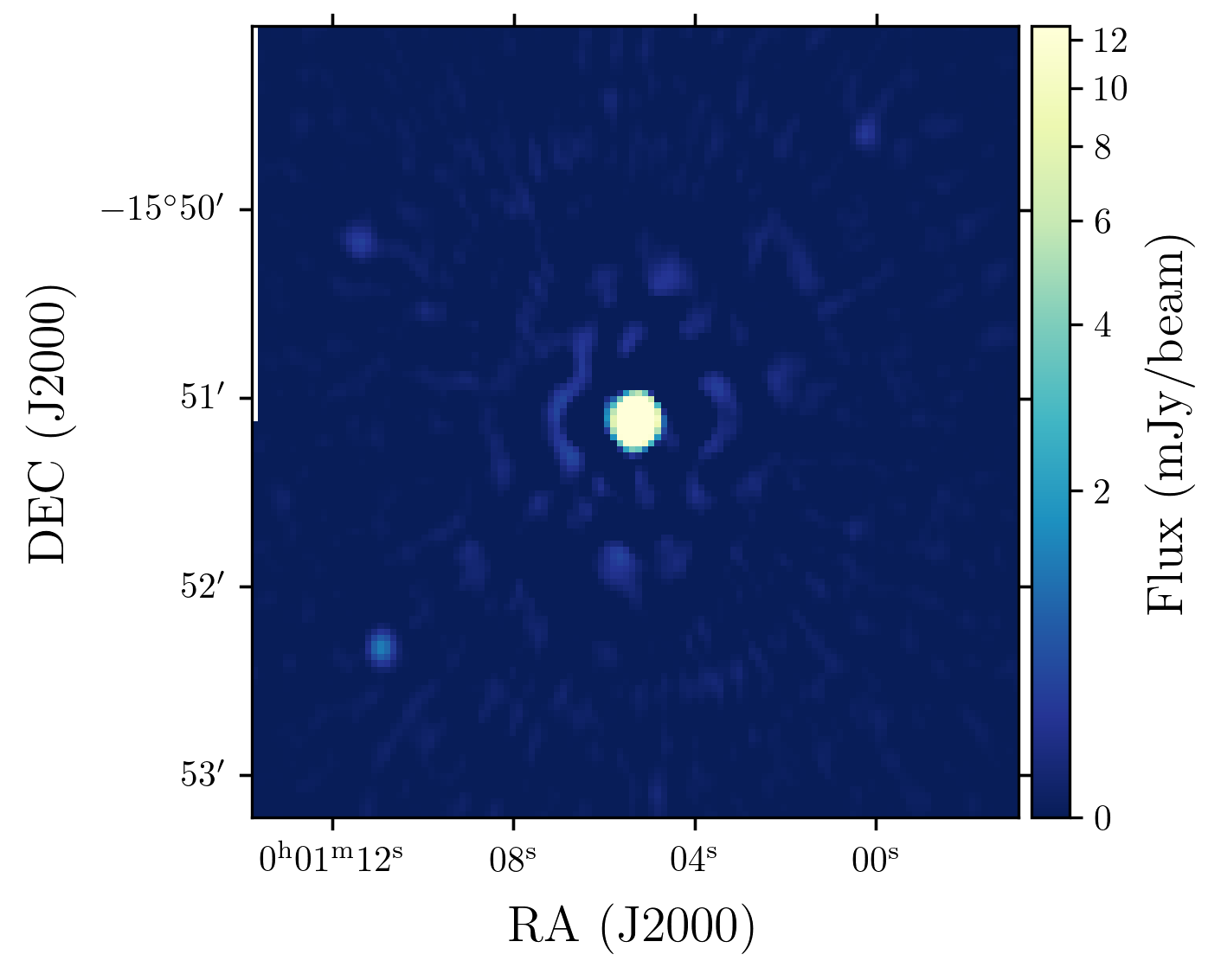}
  \label{subfig:frI_7}
\end{minipage}
\begin{minipage}[c]{0.33\textwidth}
  \vspace*{\fill}
  \centering
  \includegraphics[width=\textwidth]{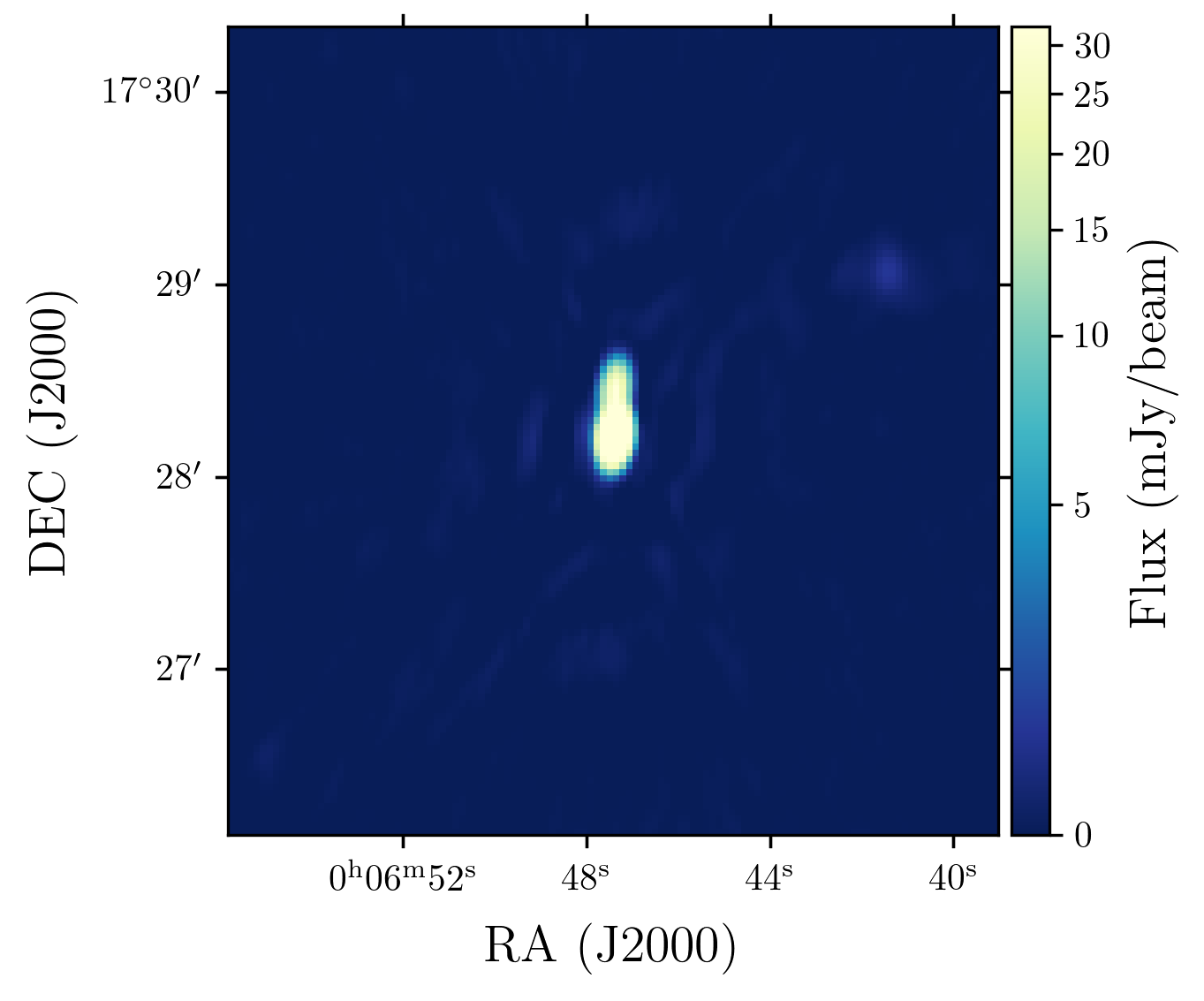}
  \label{subfig:frI_8}
\end{minipage}
\begin{minipage}[c]{0.33\textwidth}
  \vspace*{\fill}
  \centering
  \includegraphics[width=\textwidth]{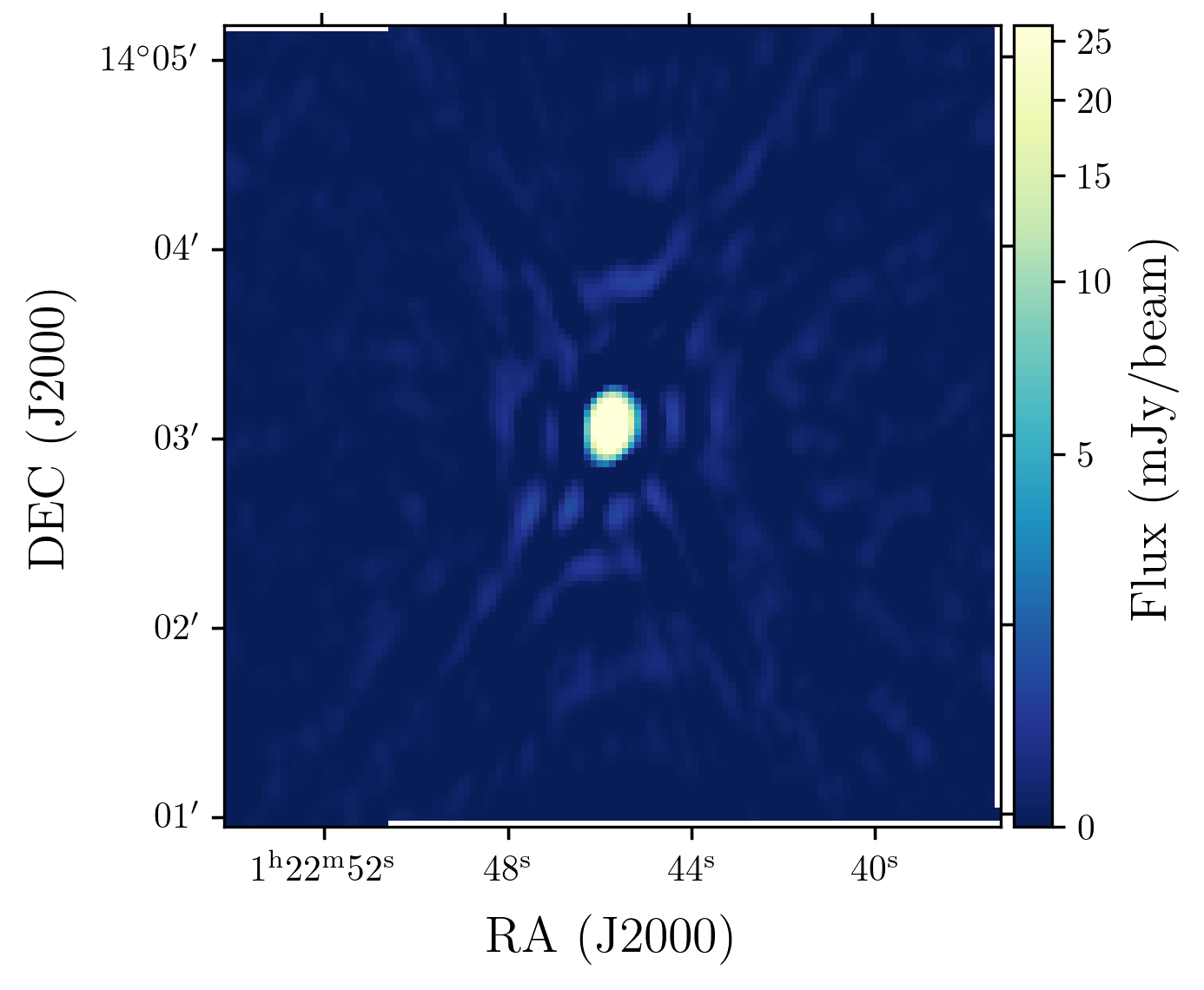}
  \label{subfig:frI_9}
\end{minipage}

\begin{minipage}[c]{0.33\textwidth}
  \vspace*{\fill}
  \centering
  \includegraphics[width=\textwidth]{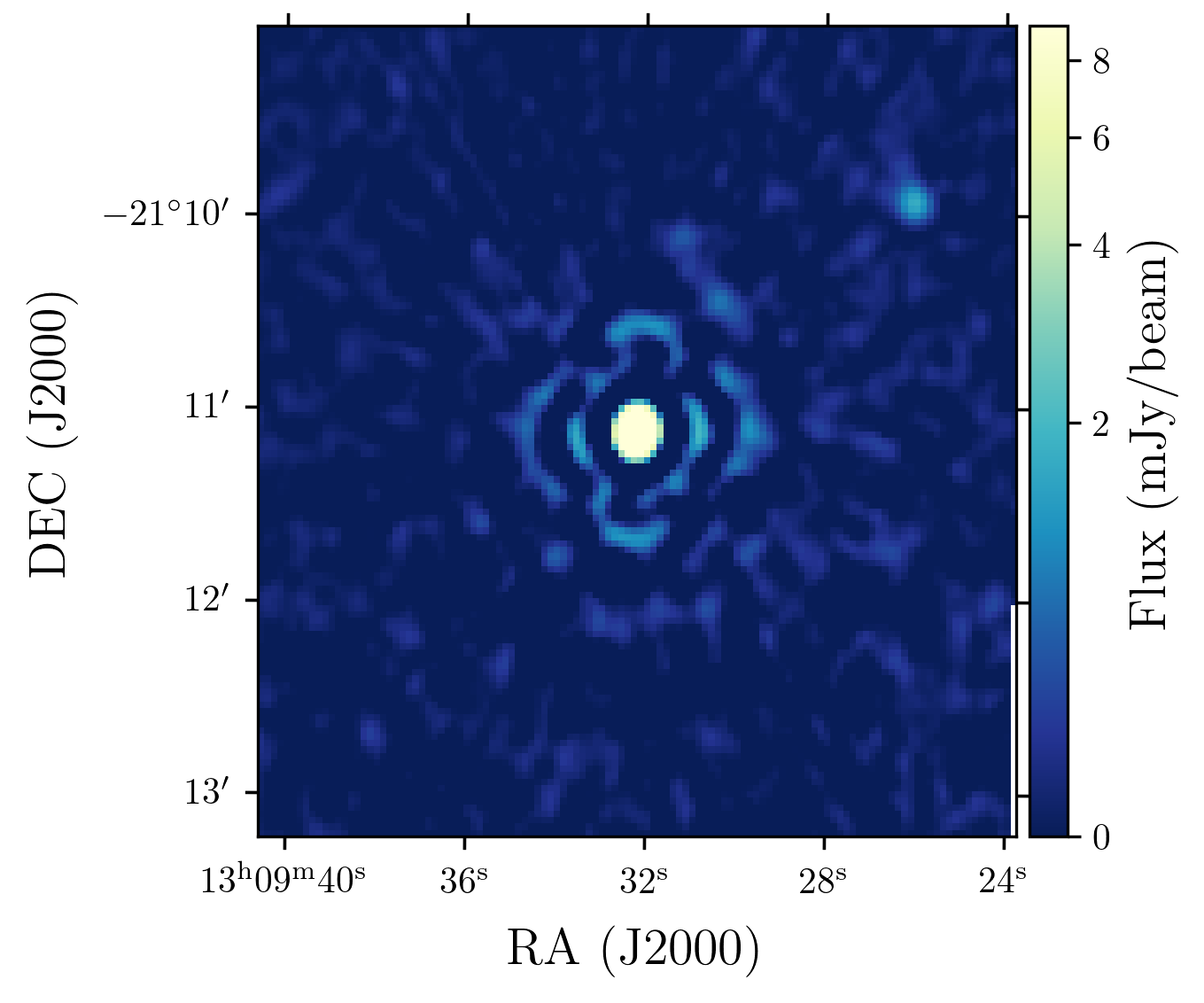}
  \label{subfig:frI_10}
\end{minipage}
\begin{minipage}[c]{0.33\textwidth}
  \vspace*{\fill}
  \centering
  \includegraphics[width=\textwidth]{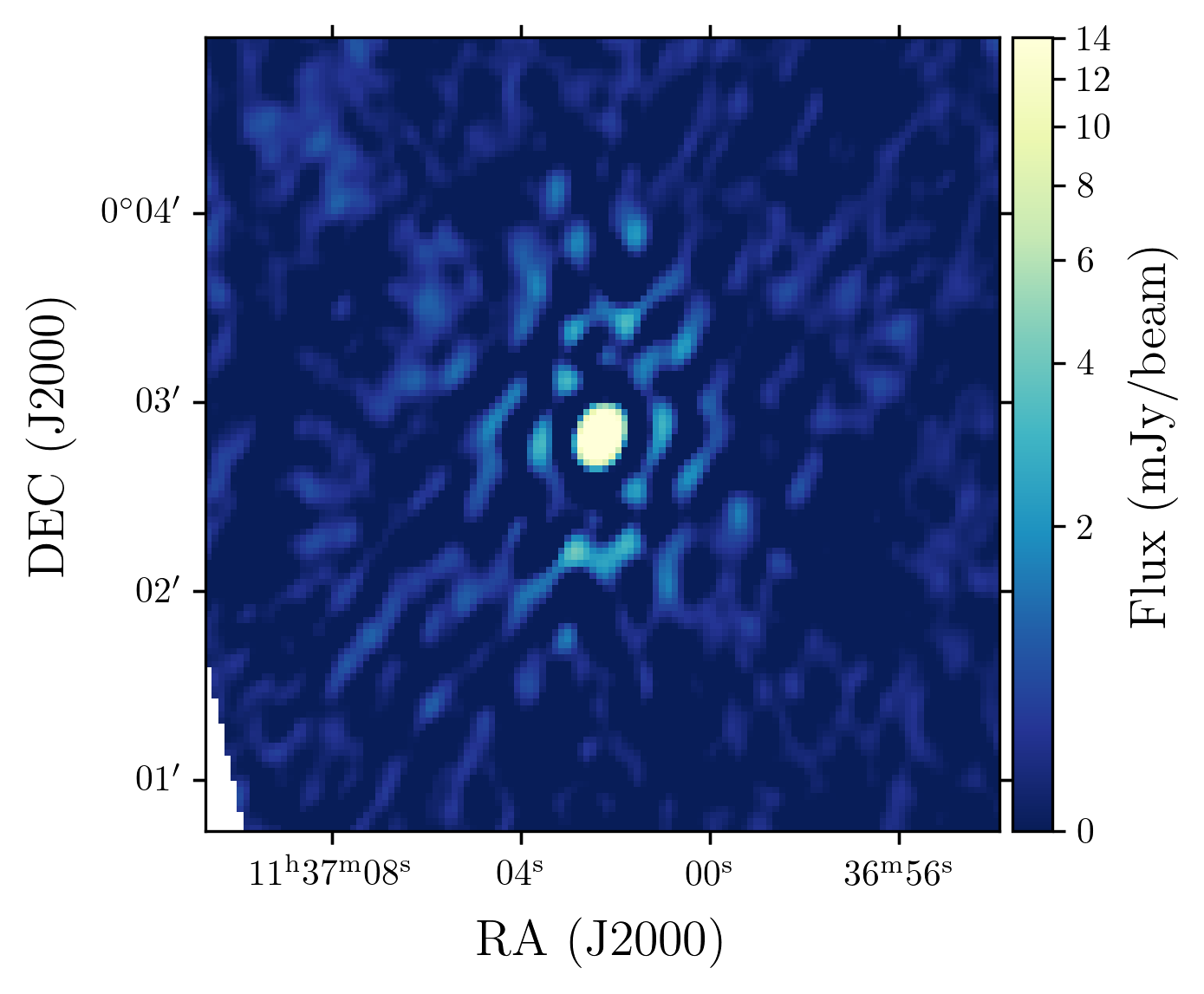}
  \label{subfig:frI_11}
\end{minipage}
\begin{minipage}[c]{0.33\textwidth}
  \vspace*{\fill}
  \centering
  \includegraphics[width=\textwidth]{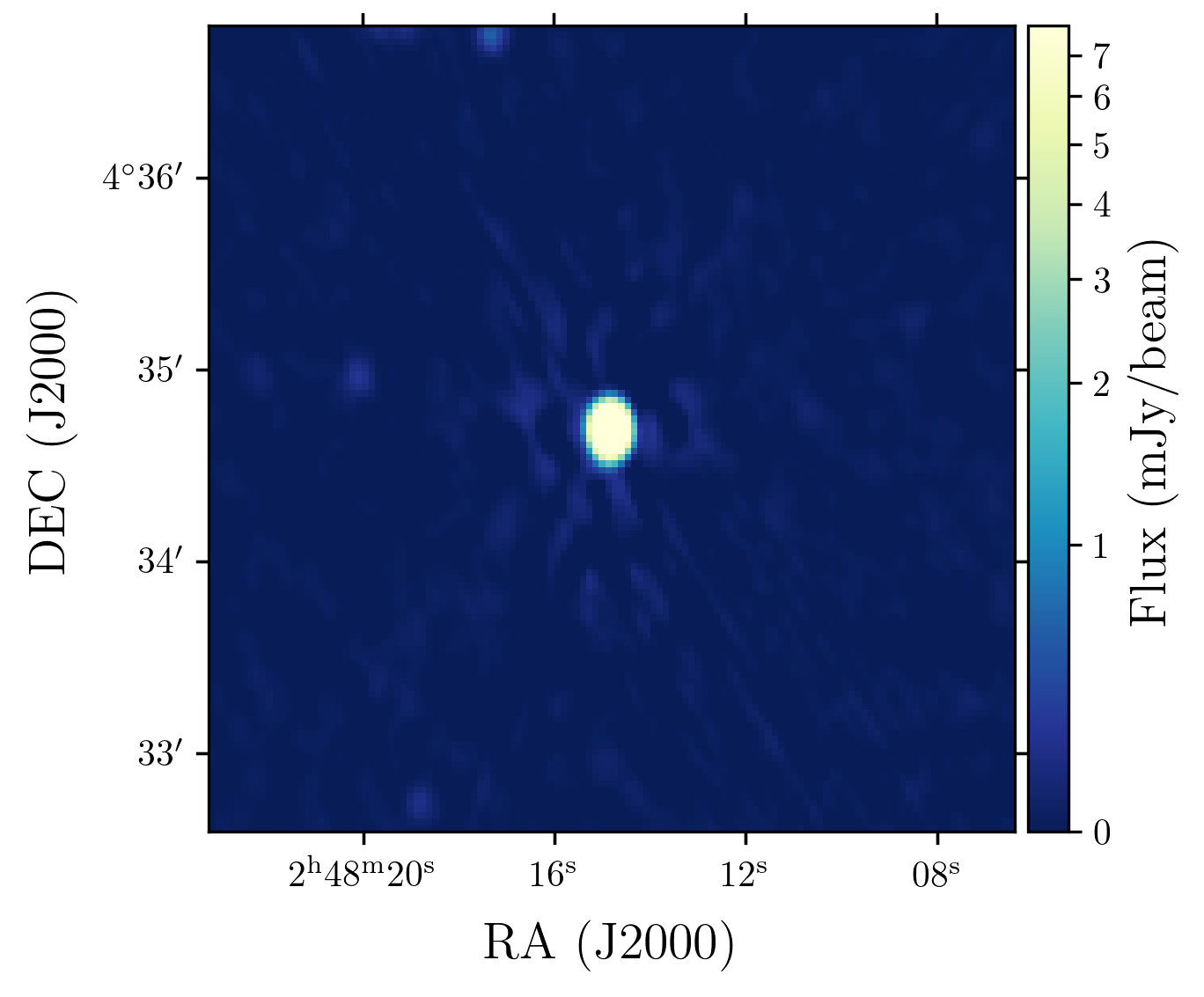}
  \label{subfig:frI_12}
\end{minipage}

\captionof{figure}{Selection of brightest sources in the combined catalogue with $\alpha > =-0.5$. As discussed in Section~\ref{sec:spectral_indices}, this range of spectral indices is expected to come from emission originating in AGN cores, which are generally point sources. The majority of this sample indeed is unresolved or can be seen to dominate the emission from their associated lobes.}
\label{fig:bright_FRI}
\end{minipage}

\begin{figure*}
\begin{minipage}[c]{0.66\textwidth}
  \vspace*{\fill}
  \centering
  \includegraphics[width=\textwidth]{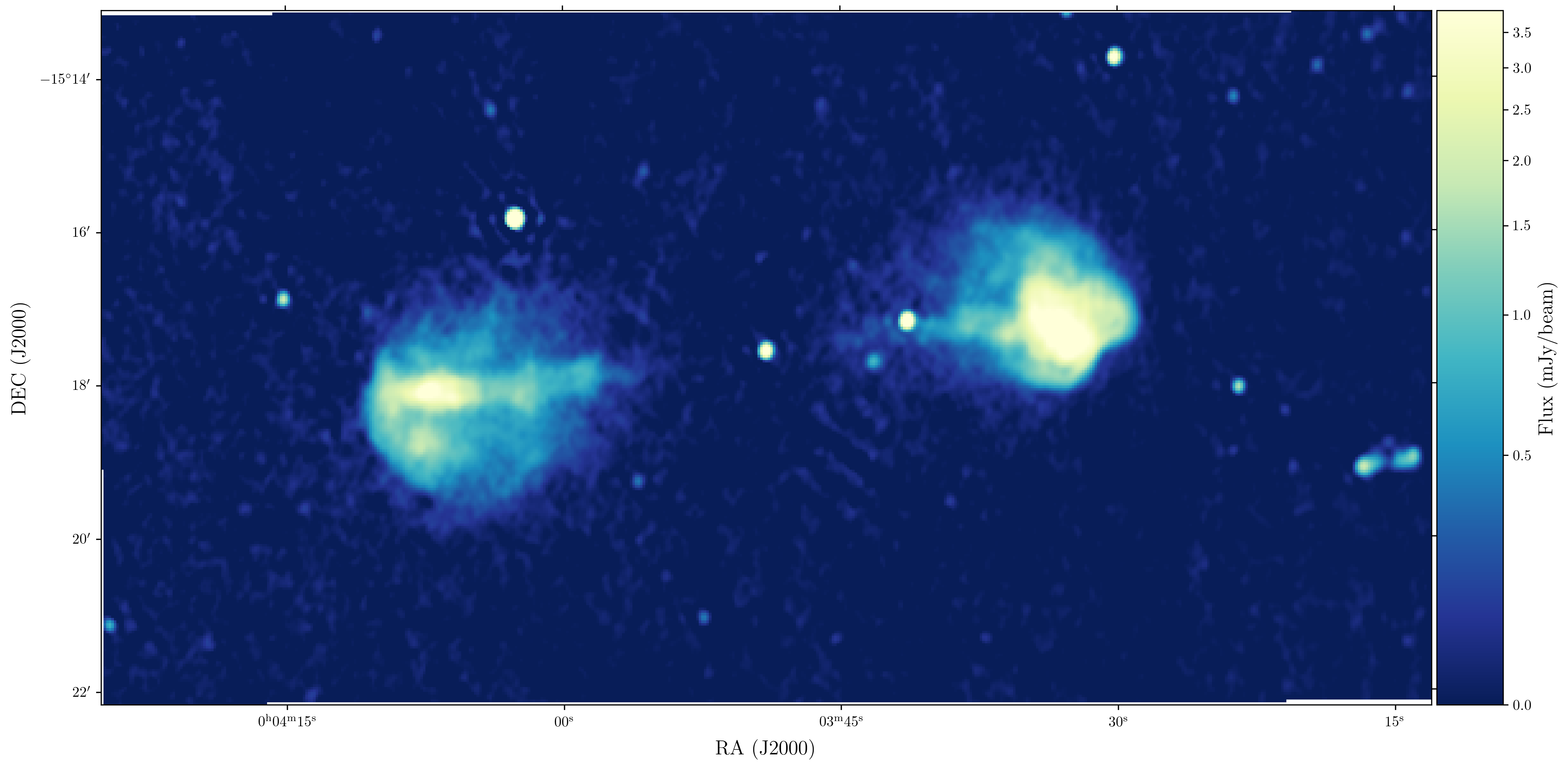}
  \label{subfig:frII_1}
\end{minipage}
\begin{minipage}[c]{0.33\textwidth}
  \vspace*{\fill}
  \centering
  \includegraphics[width=\textwidth]{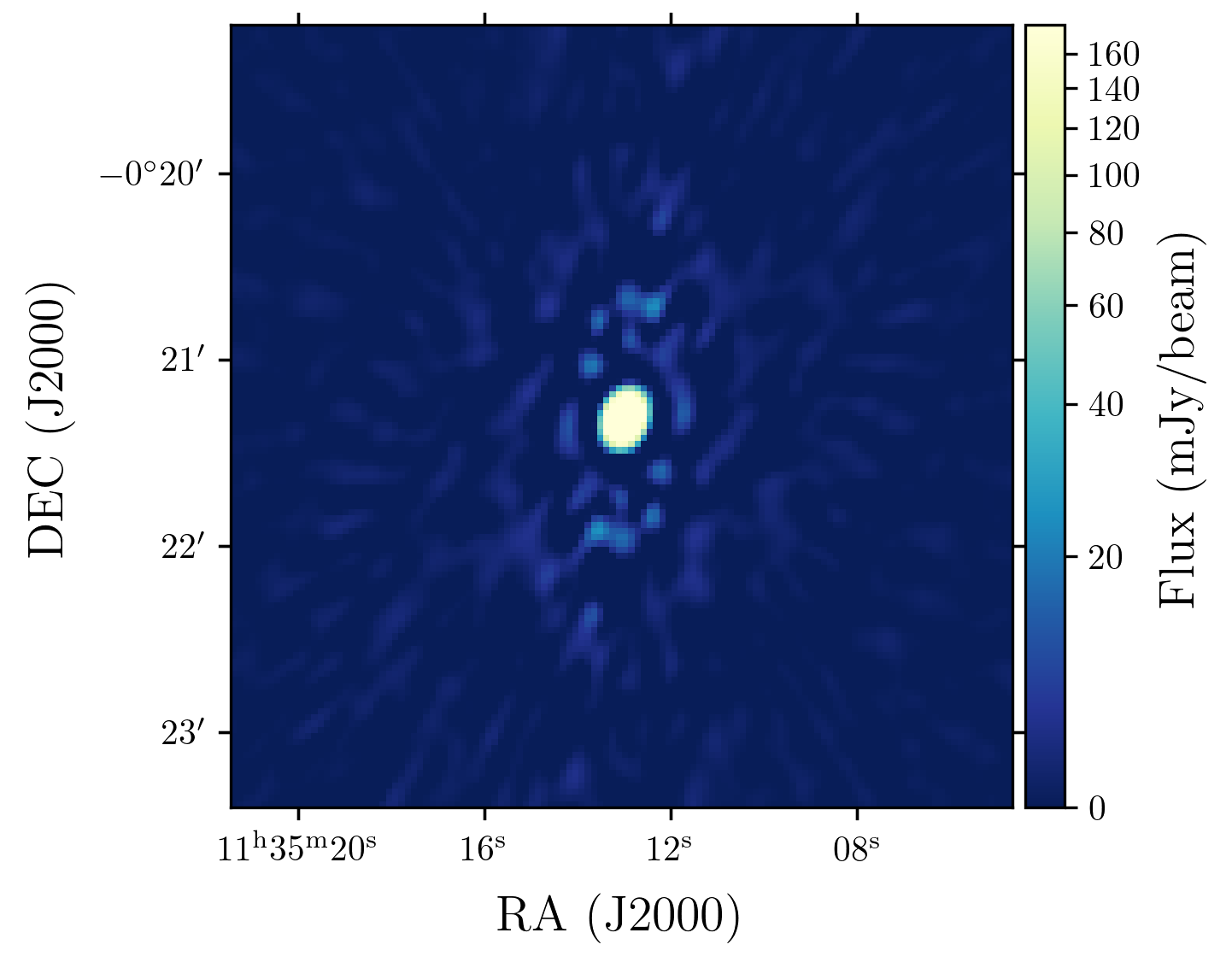}
  \label{subfig:frII_2}
\end{minipage}

\begin{minipage}[c]{0.33\textwidth}
  \vspace*{\fill}
  \centering
  \includegraphics[width=\textwidth]{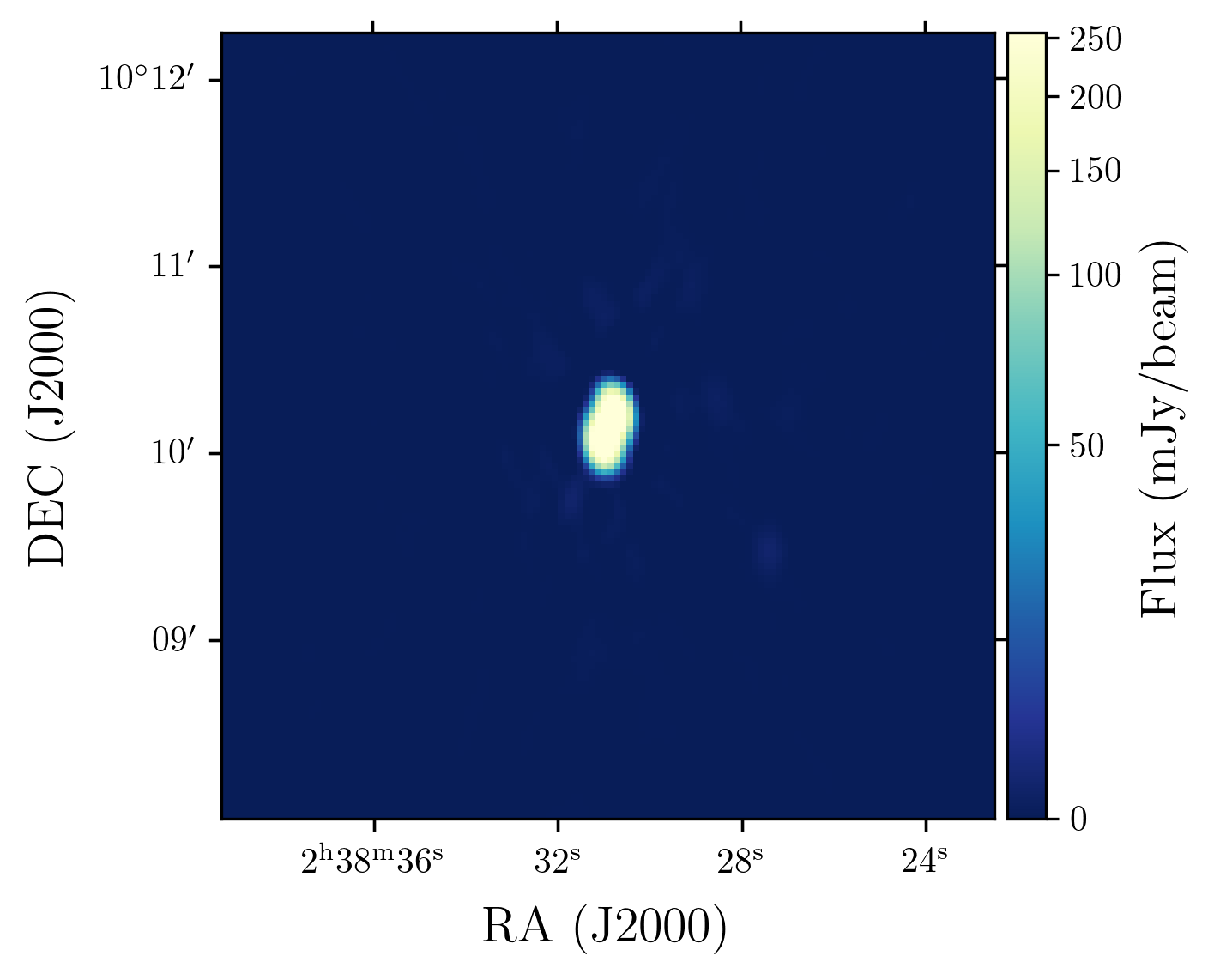}
  \label{subfig:frII_3}
\end{minipage}
\begin{minipage}[c]{0.33\textwidth}
  \vspace*{\fill}
  \centering
  \includegraphics[width=\textwidth]{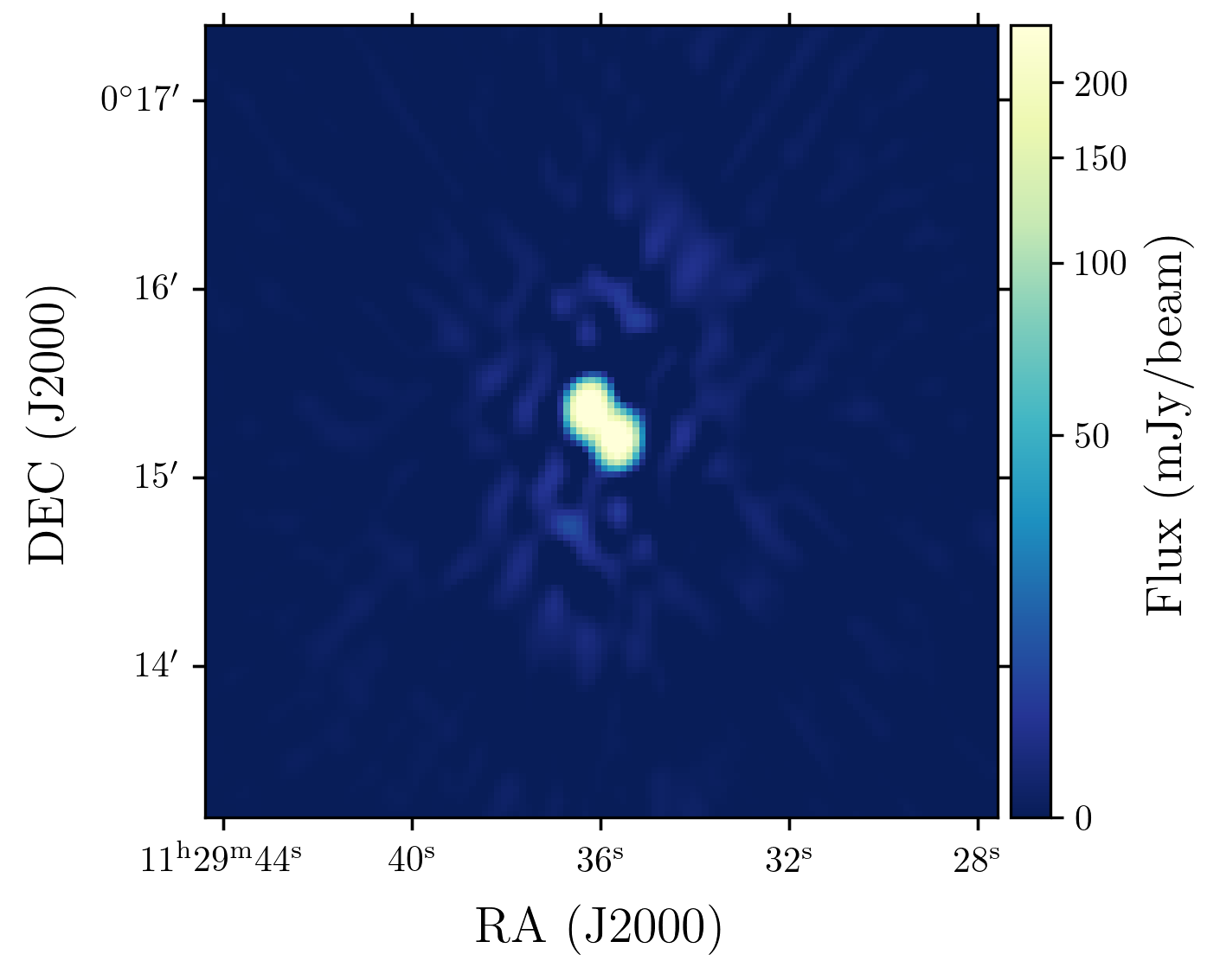}
  \label{subfig:frII_4}
\end{minipage}
\begin{minipage}[c]{0.33\textwidth}
  \vspace*{\fill}
  \centering
\includegraphics[width=\textwidth]{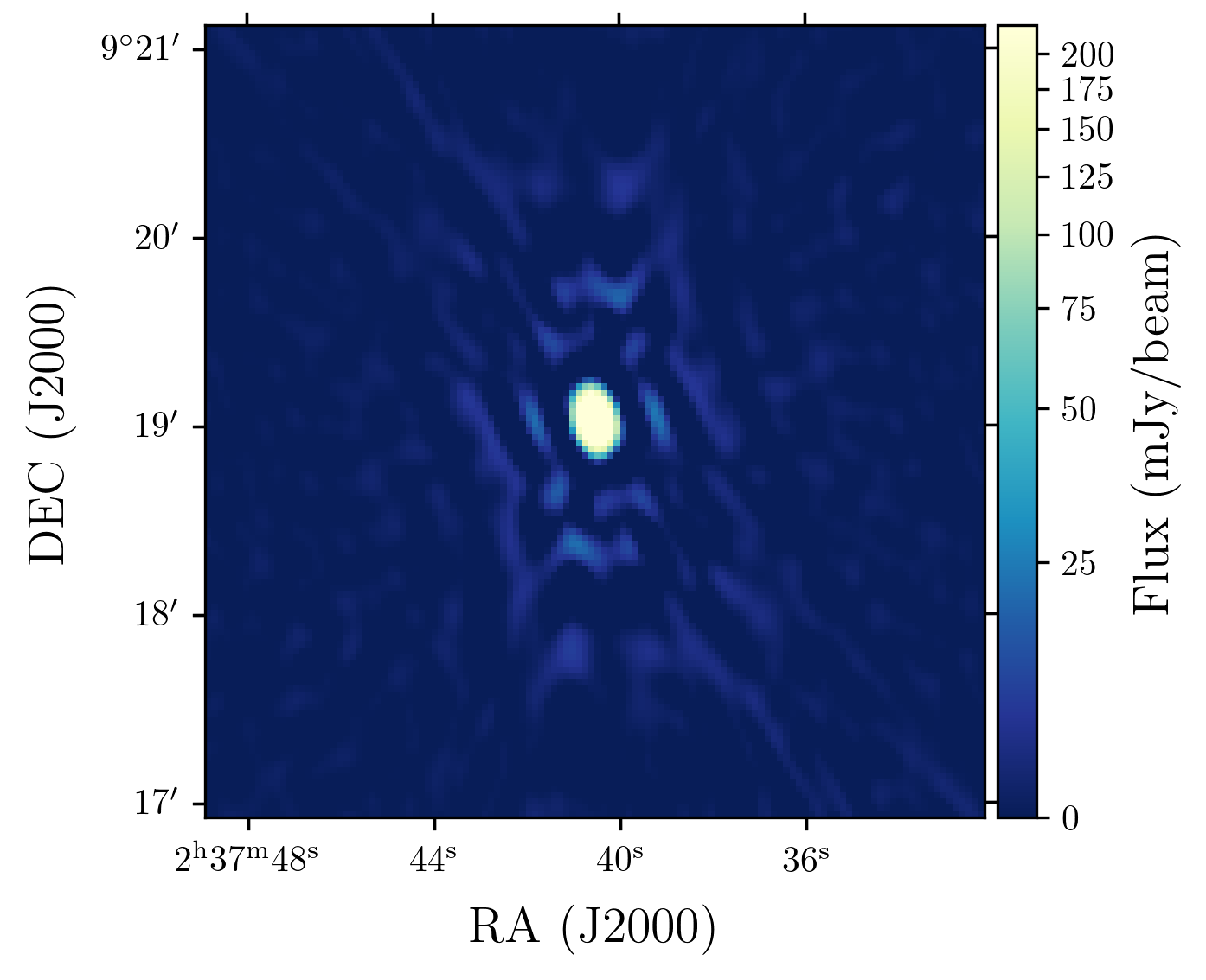}
  \label{subfig:frII_5}
\end{minipage}

\begin{minipage}[c]{0.33\textwidth}
  \vspace*{\fill}
  \centering
  \includegraphics[width=\textwidth]{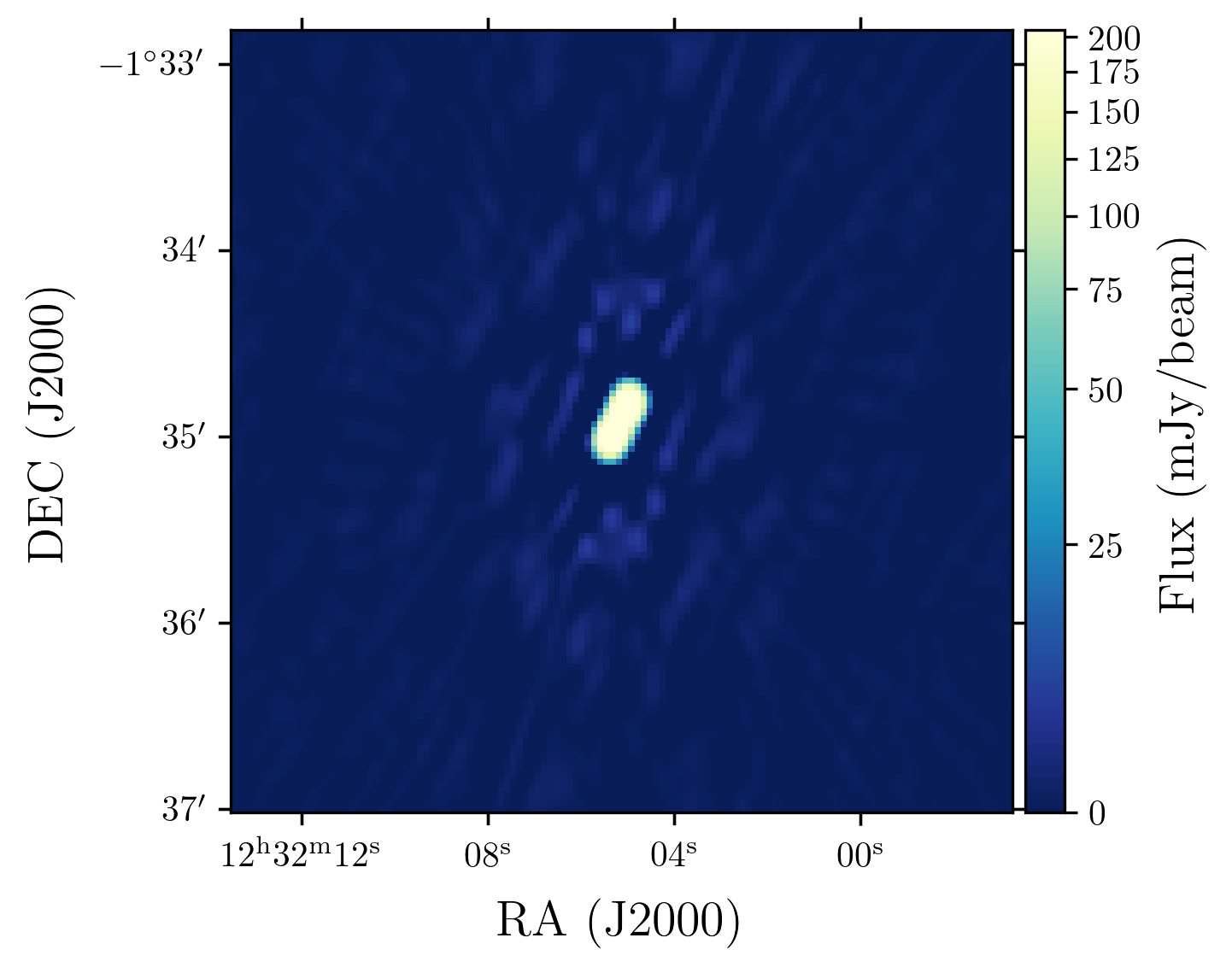}
  \label{subfig:frII_6}
\end{minipage}
\begin{minipage}[c]{0.33\textwidth}
  \vspace*{\fill}
  \centering
  \includegraphics[width=\textwidth]{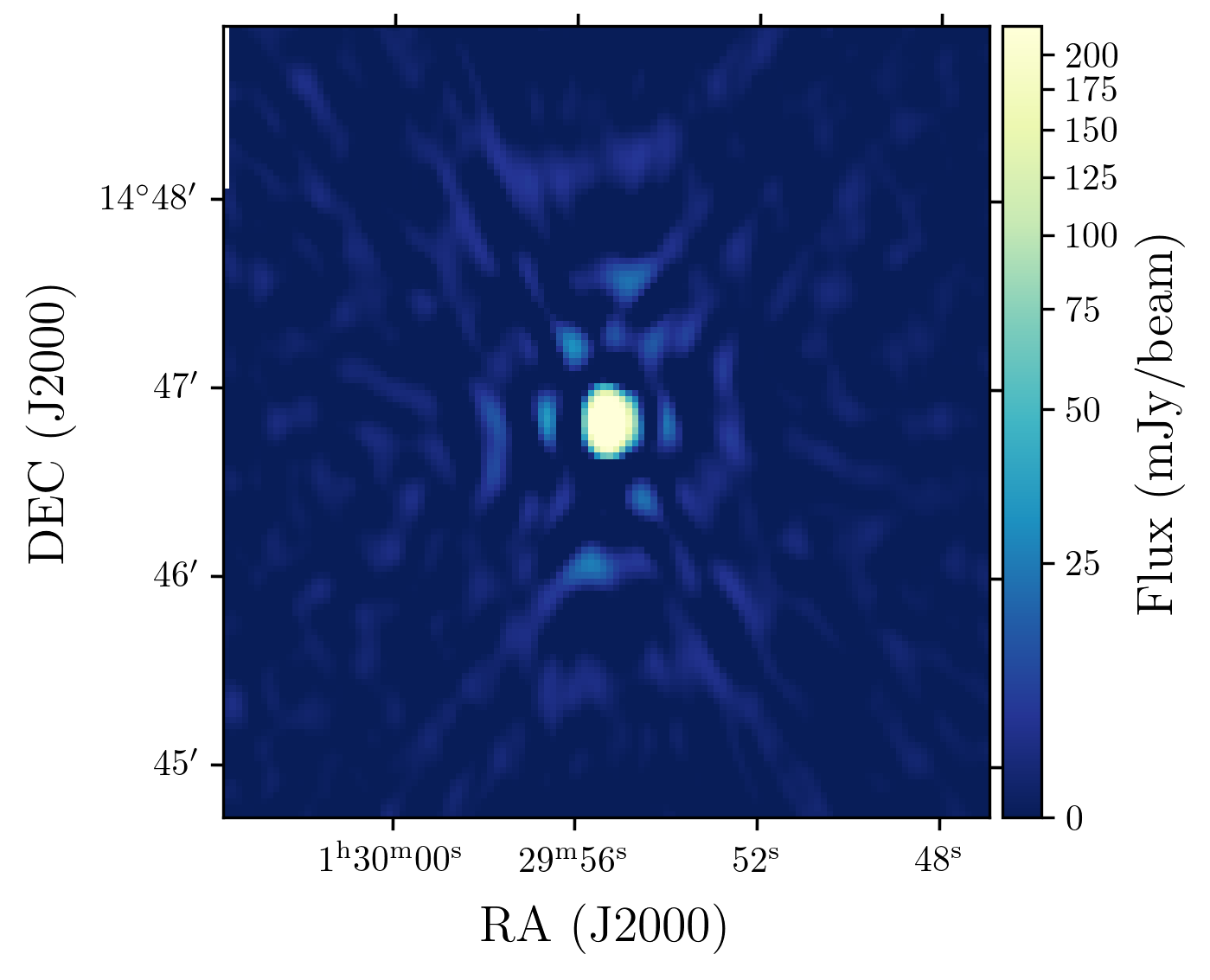}
  \label{subfig:frII_7}
\end{minipage}
\begin{minipage}[c]{0.33\textwidth}
  \vspace*{\fill}
  \centering
  \includegraphics[width=\textwidth]{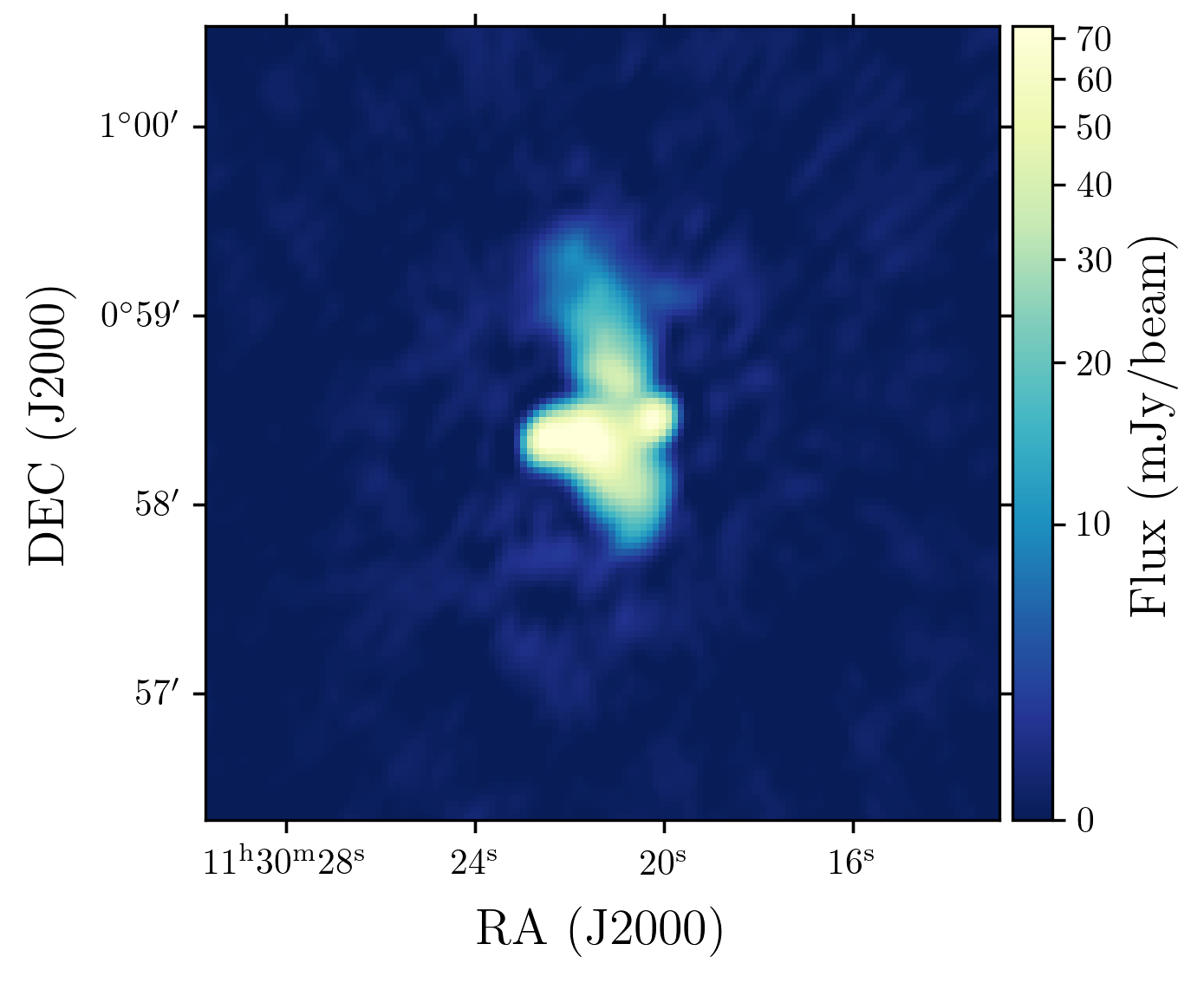}
  \label{subfig:frII_8}
\end{minipage}

\begin{minipage}[c]{0.33\textwidth}
  \vspace*{\fill}
  \centering
  \includegraphics[width=\textwidth]{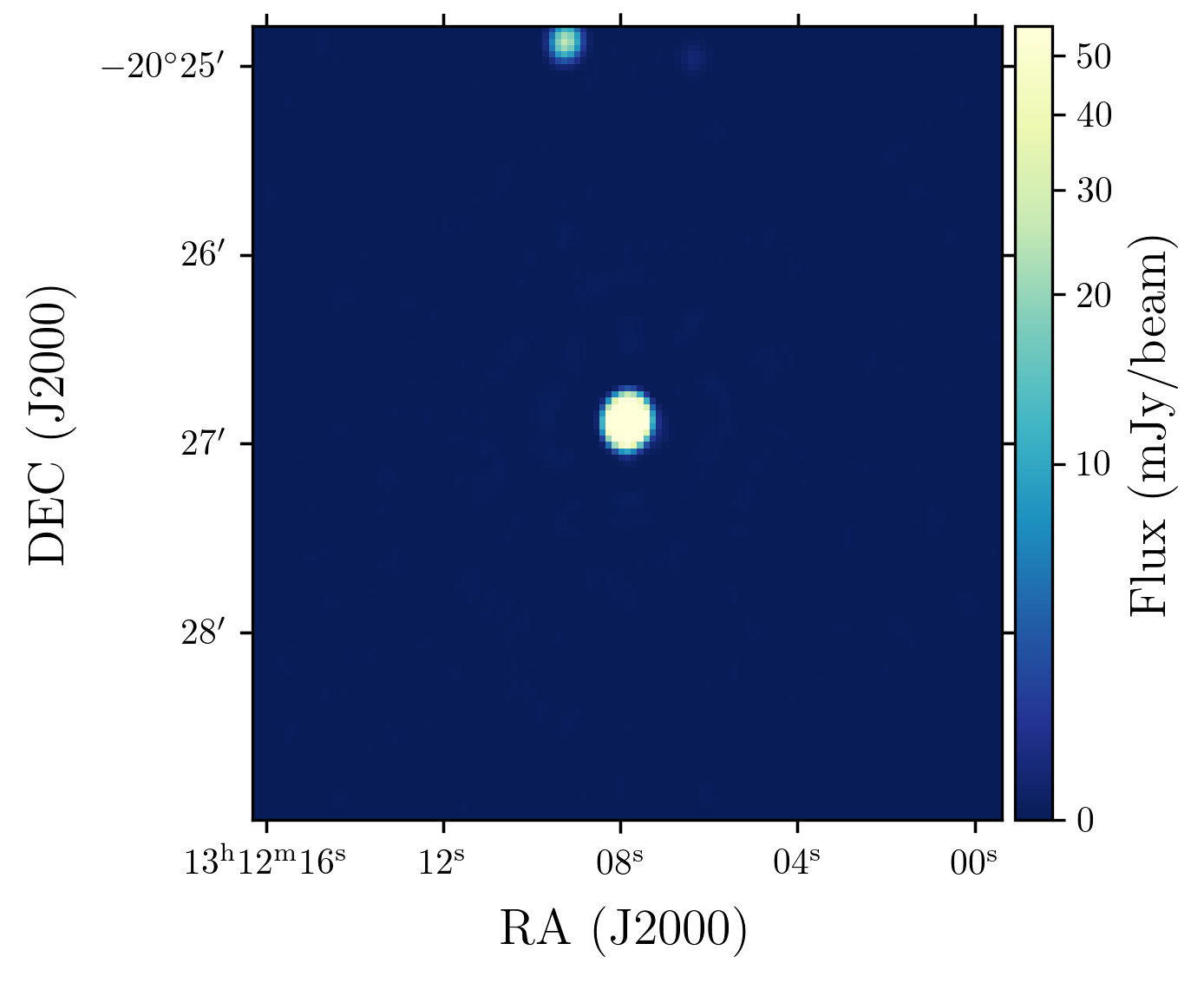}
  \label{subfig:frII_9}
\end{minipage}
\begin{minipage}[c]{0.33\textwidth}
  \vspace*{\fill}
  \centering
  \includegraphics[width=\textwidth]{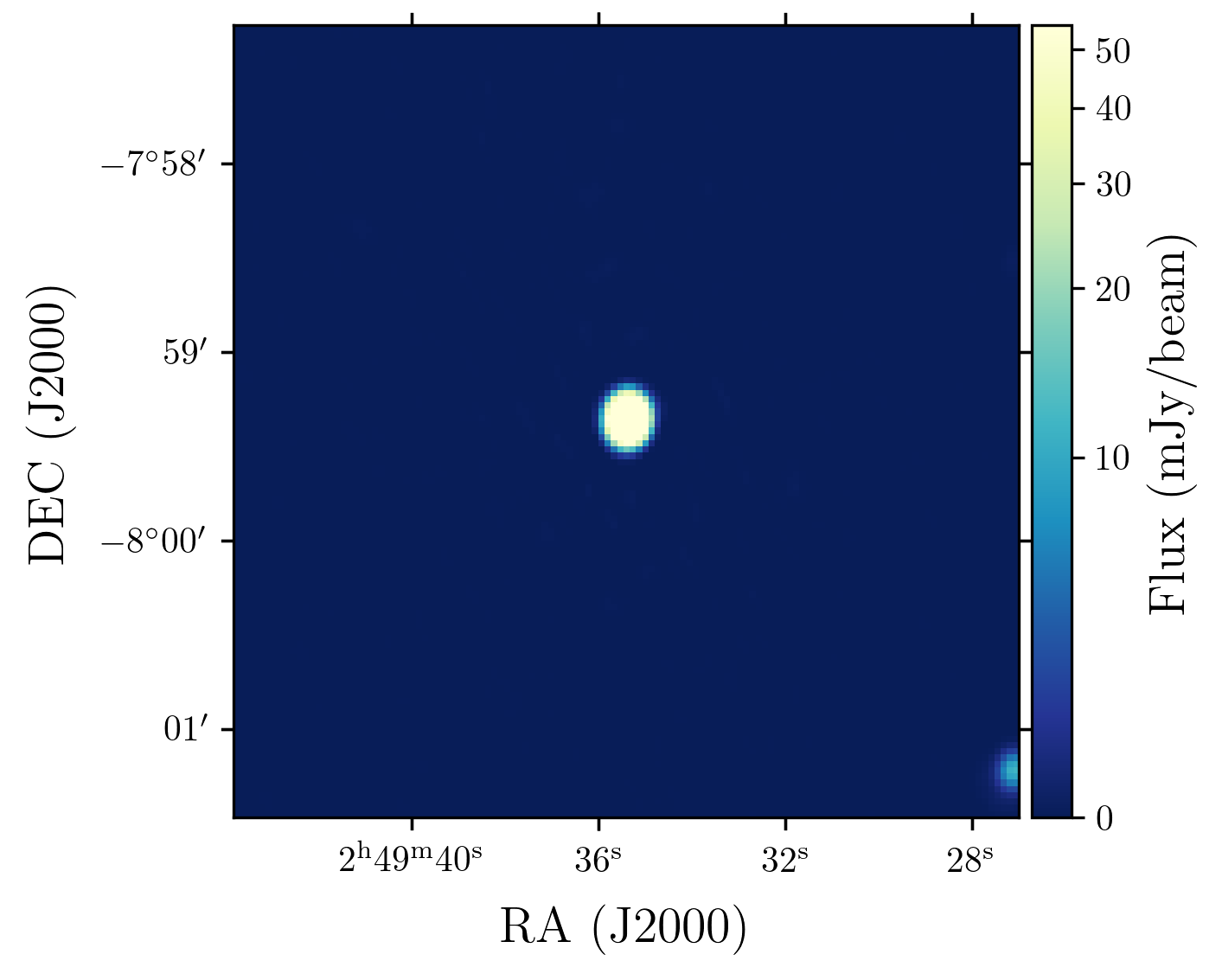}
  \label{subfig:frII_10}
\end{minipage}
\begin{minipage}[c]{0.33\textwidth}
  \vspace*{\fill}
  \centering
  \includegraphics[width=\textwidth]{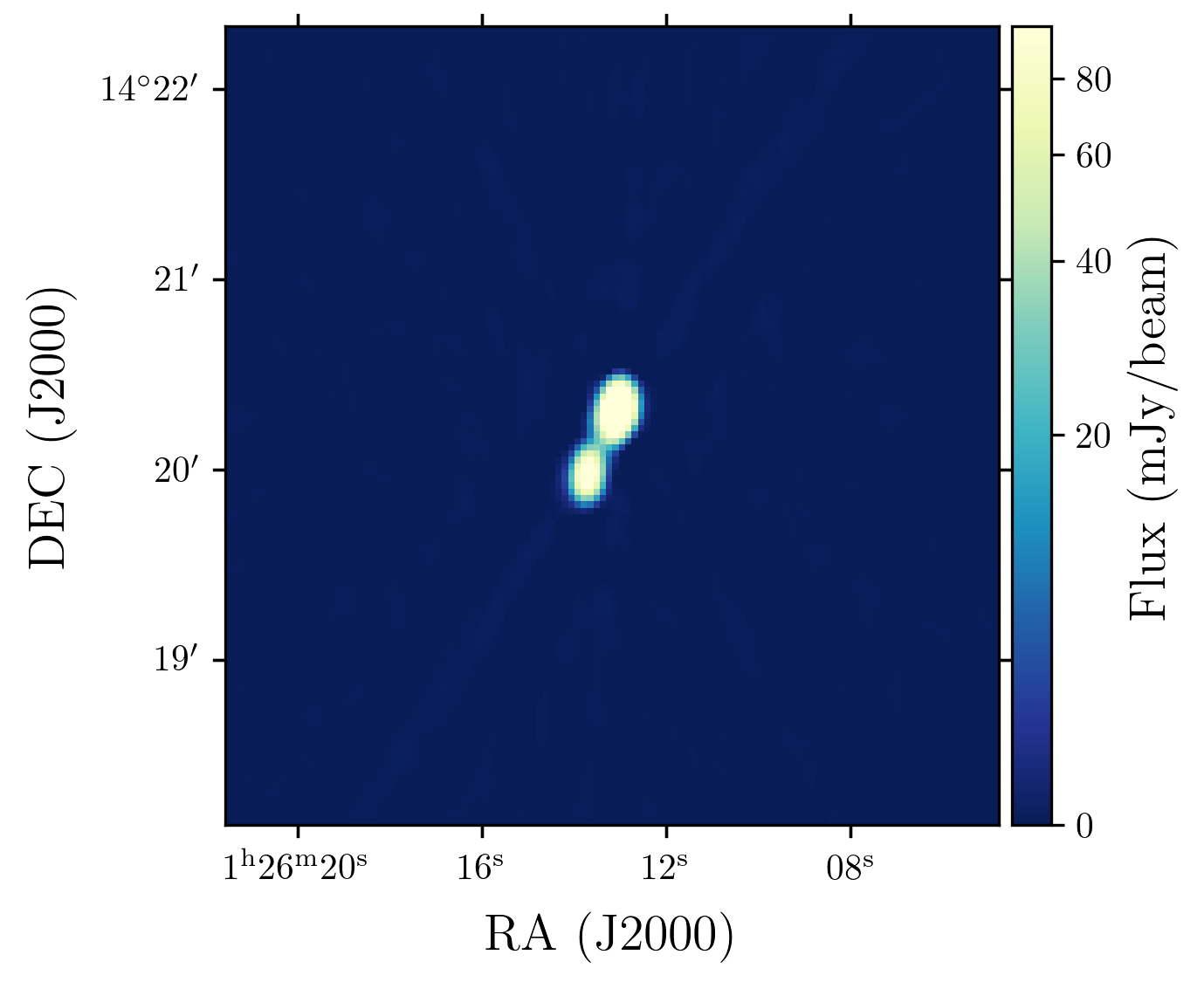}
  \label{subfig:frII_11}
\end{minipage}

\caption{Selection of brightest sources from the combined catalogue with $\alpha < -0.5$. As discussed in Section~\ref{sec:spectral_indices}, this range of spectral indices is generally associated with synchrotron, which is the dominant emission mechanism in radio lobes. Though a number of sources in this sample are unresolved, many show the two component structure characteristic of FRIIs or even more complex extended structure.}
\label{fig:bright_FRII}
\end{figure*}

\FloatBarrier

\newpage

\onecolumn
\section{Table of sources}
\begin{table}[h]
    \centering
    \caption{Example of the final source catalogue structure. Columns are as described in Section~\ref{sec:columns}.}
    \label{tab:catalog_lines}
    \begin{tabular}{lllllllll}
    \hline
    Pointing\_id  & Source\_name & Source\_id & Isl\_id & RA & E\_RA & DEC & E\_DEC  & Sep\_PC \\ 
    & & & & J2000, $\degree$ & $\degree$ & J2000, $\degree$ & $\degree$ & $\degree$ \\
    \hline \hline
    PT-J0001-1540 & J000012.63-154312.0 & 1607 & 1618 & 0.052609 & 6.82e-05 & -15.720001 & 9.42e-05 & 0.36  \\
    PT-J0006+1728 & J000409.33+163842.5 & 1285 & 1331 & 1.038859 & 8.34e-05 & 16.645139 & 1.36e-04 & 1.04   \\
    PT-J0126+1420 & J012317.38+133309.1 & 1504 & 1530 & 20.822411 & 1.70e-04 & 13.552529 & 1.65e-04 & 1.06  \\
    PT-J0240+0957 & J023823.97+094637.3 & 888 & 904 & 39.599861 & 3.05e-05 & 9.777035 & 5.93e-05 & 0.54     \\
    PT-J0249+0440 & J024648.18+035955.5 & 1469 & 1483 & 41.700764 & 8.47e-05 & 3.998738 & 1.61e-04 & 0.98   \\
    PT-J0249-0759 & J024633.91-073732.5 & 2518 & 2529 & 41.641290 & 1.64e-04 & -7.625700 & 2.18e-04 & 0.83  \\
    PT-J1133+0015 & J113059.76+000708.0 & 708 & 717 & 172.749000 & 2.06e-04 & 0.118892 & 2.53e-04 & 0.53    \\
    PT-J1232-0224 & J122945.15-021418.1 & 507 & 517 & 187.438138 & 1.29e-04 & -2.238372 & 1.77e-04 & 0.59   \\
    PT-J1312-2026 & J130900.75-204924.6 & 2326 & 2334 & 197.253139 & 4.04e-05 & -20.823501 & 5.03e-05 & 0.82\\
    PT-J2023-3655 & J202008.96-371033.3 & 2055 & 2067 & 305.037343 & 2.38e-04 & -37.175915 & 1.73e-04 & 0.77\\
    \hline \\
    \end{tabular}
    \begin{tabular}{llllllll}
    \hline
    Total\_flux & E\_Total\_flux & Peak\_flux & E\_Peak\_flux & Spectral\_index & Spectral\_index\_correction & E\_Spectral\_index \\ 
    mJy & mJy & mJy/beam & mJy/beam & & & \\
    \hline \hline
    0.409 & 0.050 & 0.273 & 0.022 & -1.90 & 0.10 & 0.21 \\
    4.564 & 0.458 & 2.101 & 0.151 & -0.10 & 0.80 & 0.09 \\
    19.007 & 1.561 & 2.288 & 0.121 & -0.53 & 0.83 & 0.07\\
    4.692 & 0.368 & 2.575 & 0.062 & -0.74 & 0.21 & 0.05 \\
    1.418 & 0.231 & 0.985 & 0.102 & 0.46 & 0.72 & 0.16  \\
    0.421 & 0.113 & 0.285 & 0.049 & -0.94 & 0.51 & 0.27 \\
    0.423 & 0.114 & 0.254 & 0.046 & -1.51 & 0.21 & 0.10 \\
    0.530 & 0.121 & 0.394 & 0.057 & -2.36 & 0.25 & 0.10 \\
    1.761 & 0.117 & 1.136 & 0.050 & 0.09 & 0.50 & 0.07  \\
    0.511 & 0.116 & 0.306 & 0.047 & -1.56 & 0.43 & 0.23 \\
    \hline \\
    \end{tabular}
    \begin{tabular}{llllllllll}
    \hline
    RA\_max    & E\_RA\_max & DEC\_max   & E\_DEC\_max & Maj   & E\_Maj & Min   & E\_Min & PA & E\_PA \\ 
    J2000, $\degree$ & $\degree$ & J2000, $\degree$ & $\degree$ & $''$ & $''$ & $''$ & $''$ & $\degree$ & $\degree$ \\
    \hline \hline
    0.052609 & 6.82e-05 & -15.720001 & 9.42e-05 & 9.35 & 0.80 & 7.78   & 0.57 & 6.66 & 19.55  \\
    1.038859 & 8.34e-05 & 16.645139 & 1.36e-04 & 14.55 & 1.16 & 10.66  & 0.70 & 171.83 & 11.22\\
    20.824116 & 1.70e-04 & 13.551953 & 1.65e-04 & 19.38 & 1.50 & 17.70 & 1.33 & 38.22 & 36.76 \\
    39.601394 & 3.05e-05 & 9.776665 & 5.93e-05 & 14.21 & 0.51 & 9.25   & 0.25 & 80.33 & 3.76  \\
    41.700764 & 8.47e-05 & 3.998738 & 1.61e-04 & 11.36 & 1.38 & 7.74   & 0.69 & 10.10 & 13.35 \\
    41.641290 & 1.64e-04 & -7.625700 & 2.18e-04 & 10.19 & 1.97 & 7.82  & 1.22 & 154.03 & 30.80\\
    172.749000 & 2.06e-04 & 0.118892 & 2.53e-04 & 11.83 & 2.42 & 8.42  & 1.34 & 146.13 & 26.09\\
    187.438138 & 1.29e-04 & -2.238372 & 1.77e-04 & 9.72 & 1.53 & 7.90  & 1.04 & 162.61 & 31.48\\
    197.253139 & 4.04e-05 & -20.823501 & 5.03e-05 & 9.35 & 0.43 & 8.03 & 0.33 & 17.92 & 12.99 \\
    305.037343 & 2.38e-04 & -37.175915 & 1.73e-04 & 12.52 & 2.08 & 9.91 & 1.38 & 72.24 & 30.76\\
    \hline \\
    \end{tabular}
    \begin{tabular}{lllllll}
    \hline
    DC\_Maj & E\_DC\_Maj & DC\_Min & E\_DC\_Min & DC\_PA & E\_DC\_PA \\
    $''$ & $''$ & $''$ & $''$ & $\degree$ & $\degree$ \\
    \hline \hline
    5.60 & 0.80 & 4.16 & 0.57 & 31.11 & 19.55  \\
    9.03 & 1.16 & 8.62 & 0.70 & 161.03 & 11.22 \\
    17.71 & 1.50 & 15.05 & 1.33 & 64.76 & 36.76\\
    0.00 & 0.51 & 0.00 & 0.25 & 0.00 & 3.76    \\
    6.93 & 1.38 & 3.53 & 0.69 & 22.80 & 13.35  \\
    6.49 & 1.97 & 3.62 & 1.22 & 141.44 & 30.80 \\
    8.15 & 2.42 & 4.45 & 1.34 & 133.16 & 26.09 \\
    4.97 & 1.53 & 3.85 & 1.04 & 135.68 & 31.48 \\
    6.02 & 0.43 & 4.12 & 0.33 & 44.66 & 12.99  \\
    8.28 & 2.08 & 4.65 & 1.38 & 29.77 & 30.76  \\ 
     \hline
    \end{tabular}
\end{table}
\begin{table}
    \ContinuedFloat
    \centering
    \caption{continued}
    \begin{tabular}{llllllll}
    \hline
    Isl\_Total\_flux & E\_Isl\_Total\_flux & Isl\_rms & Isl\_mean & Resid\_Isl\_rms & Resid\_Isl\_mean & S\_Code & N\_Gaus \\
    mJy/beam & mJy/beam & mJy/beam & mJy/beam & & & \\
    \hline \hline
    0.369 & 0.034 & 0.020 & -0.004 & 0.007 & -0.004 & S & 1 \\
    4.343 & 0.276 & 0.141 & -0.021 & 0.121 & -0.002 & S & 1 \\
    11.940 & 0.416 & 0.121 & -0.010 & 0.121 & -0.010 & M & 2\\
    3.954 & 0.160 & 0.062 & 0.005 & 0.021 & 0.004 & M & 2   \\
    1.231 & 0.136 & 0.098 & -0.020 & 0.019 & -0.019 & S & 1 \\
    0.349 & 0.060 & 0.047 & -0.000 & 0.009 & -0.000 & S & 1 \\
    0.384 & 0.064 & 0.043 & 0.003 & 0.003 & 0.003 & S & 1   \\
    0.426 & 0.065 & 0.054 & -0.001 & 0.006 & -0.001 & S & 1 \\
    1.613 & 0.082 & 0.046 & -0.019 & 0.035 & -0.014 & S & 1 \\
    0.373 & 0.056 & 0.043 & -0.008 & 0.007 & -0.008 & S & 1 \\
    \hline \\
    \end{tabular}
    \begin{tabular}{llllllll}
    \hline
    RA\_mean  & DEC\_mean  & Cutout\_Spectral\_index & Cutout\_Total\_flux & Cutout\_flag & Cutout\_class & Resolved & Flag\_artifact \\
    J2000, $\degree$ & J2000, $\degree$ & & mJy & & & & \\
    \hline \hline
    -- & -- & -- & -- & -- & -- & True & False                     \\
    -- & -- & -- & -- & -- & -- & True & False                     \\
    20.822423 & 13.552525 & 10.482 & -0.24 & C & I & True & False   \\
    39.599830 & 9.777032 & 3.583 & -0.75 & C & P & True & False     \\
    -- & -- & -- & -- & -- & -- & True & False                     \\
    -- & -- & -- & -- & -- & -- & False & False                    \\
    -- & -- & -- & -- & -- & -- & True & False                     \\
    -- & -- & -- & -- & -- & -- & False & False                    \\
    -- & -- & -- & -- & -- & -- & True & False                     \\
    -- & -- & -- & -- & -- & -- & False & False                    \\
    \hline
    \end{tabular}
\end{table}

\end{appendix}

\end{document}